\documentclass[fleqn,usenatbib]{mnras}

\usepackage[T1]{fontenc}

\DeclareRobustCommand{\VAN}[3]{#2}
\let\VANthebibliography\thebibliography
\def\thebibliography{\DeclareRobustCommand{\VAN}[3]{##3}\VANthebibliography}

\usepackage{graphicx}	
\usepackage{amsmath}	

\usepackage{amssymb}	
\usepackage{subfigure}
\usepackage{csquotes}
\usepackage{longtable}
\usepackage{tabularx}
\usepackage{caption, booktabs}
\usepackage{setspace}
\usepackage{threeparttable}
\usepackage{threeparttablex}
\usepackage[utf8]{inputenc}
\usepackage{xcolor}
\newcolumntype{C}[1]{>{\centering\arraybackslash}p{#1}}

\title[Different environments of supernova IIP and Ibc]{Metallicity differences between Type IIP and stripped-envelope supernova environments.}

\author[R. Ganss et al.]{
R. Ganss,$^{1}$
J. L. Pledger,$^{1}$ \thanks{E-mail: jpledger@uclan.ac.uk}
A. E. Sansom,$^{1}$
P. A. James,$^{2}$
J. Puls,$^{3}$
and S. M. Habergham-Mawson$^{2}$
\\
$^{1}$Jeremiah Horrocks Institute, University of Central Lancashire, Preston, PR1 2HE, UK \\ 
$^{2}$Astrophysics Research Institute, Liverpool John Moores University, 146 Brownlow Hill, Liverpool, L3 5RF, UK \\
$^{3}$LMU München, Universitätssternwarte, Scheinerstr. 1, D-81679 München, Germany \\
}

\date{Accepted 2025 September 05. Received 2025 September 05; in original form 2025 May 21}

\pubyear{2025}

\begin{document}
\label{firstpage}
\pagerange{\pageref{firstpage}--\pageref{lastpage}}
\maketitle

\begin{abstract}

This work presents measurements of local H\,{\sc ii} environment metallicities of core-collapse supernovae (SNe) in hosts with redshifts up to z$\sim$0.025. 139 SNe environments were observed at the Isaac Newton Telescope and data of an additional 268 SNe environments were found in archival data of MUSE, MaNGA and PISCO. The project focuses on SNe with clean Type IIP, Type Ib and Type Ic classifications. We present the largest spectroscopic sample to date, evaluating environment metallicities of 79 Type Ib, 66 Type Ic and 93 Type IIP by N2 and O3N2 strong emission line methods. The cumulative distribution functions (CDFs) of the SN environment metallicities show Type Ib and Type Ic SNe tending towards higher metallicity than Type IIP. We test the null hypothesis that Type Ib/Ic/IIP progenitors are drawn from the same parent population. There is no statistically significant difference between progenitors of Type Ib and Type Ic SNe. However, when comparing Ib/Ic with IIP SNe, the tests indicate strong statistical significance (significance level better than 1\%) to reject our null hypothesis suggesting that the samples are not drawn from the same parent population. The significance is even higher (level better 0.1\%) when testing Type IIP vs. the combined Type Ib+Ic sample. These results support a different physical nature of Type IIP and Types Ib/Ic progenitors. It challenges stellar evolution and SNe explosion models to reproduce the distinct CDFs found.

\end{abstract}



\color{black}
\section{Introduction}

Core-collapse supernovae (CCSNe) result from the death of massive stars ($>$8M$_{\odot}$) and are responsible for the production of heavy chemical elements that enrich the interstellar medium (ISM) and contribute towards subsequent generations of star formation. Over the past decades, the supernova classification scheme has expanded from simply hydrogen-rich Type II and hydrogen-poor Type I SNe. Type II subtypes include those SN classified from the plateau in their light curves (Type IIP; \citealt{barbon1979}), those with linearly fading light curves (Type IIL; \citealt{filippenko1997}) and those with spectroscopic classifications such as Type IIb, where the hydrogen lines disappear at late times \citep{filippenko1988, filippenko1993}, or Type IIn which have long-lasting narrow emission lines \citep{schlegel1990}. Similarly, Type I SN have thermonuclear Type Ia (not discussed here), hydrogen-poor, helium-rich Type Ib SN and H+He-poor Type Ic.

To understand the diversity of CCSNe their progenitors have become of interest. Direct confirmation of the supernova progenitor using deep, high spatial resolution, archival imaging is the "gold standard" of progenitor detection (e.g. hydrogen-rich Type IIP SN2008bk in NGC 7793 \citealt{mattila2010}). However, such imaging is often unavailable, particularly for more distant SNe or the detection limits are poor \citep{smartt2009, vandyk2017, Smartt2009b}. Indeed, even where such pre-SN imaging exists, there is frequently tension between single star and binary models, both of which can fit the observational data of the progenitor candidate (e.g. Type Ic SNe iPTF13bvn \citep{cao2013,bersten2014}), Type Ib SN2019yvr \citep{kilpatrick2021, sun2022}. \citet{Morozova2018} has also shown that the presence of circumstellar material (CSM) can lead to an underestimate of the initial mass from pre-explosion imaging if not accounted for.

In lieu of direct detections, other approaches to investigate the nature of the progenitor have been explored. For example, \citet{Lyman2016} investigate the bolometric light curves of 38 stripped-envelope SNe, determining the nickel masses, ejecta masses and explosion energies. Comparison of their observational results to stellar evolutionary models support a binary scenario for the SNe. Indeed, models from \citet{Dessart2024} suggest the diversity can result from varying the period of binary systems rather than other parameters such as metallicity. However, \citet{martinez2022b} find that the diversity of Type II SNe is largely due to the explosion energy of a single star.

To infer a progenitor and initial mass for a SN it is possible to explore the stellar population around the SN since the surrounding stellar population should reflect the age and mass of the star that exploded \citep{maund2016}. \citet{sun2023} investigate the ultra-violet/optical properties of the stellar populations surrounding 41 stripped envelope SNe to infer an age and initial mass of the progenitors. They find that Type Ic SNe progenitors are systematically younger and more massive than Ib and IIb which have similar progenitor masses; IIb SNe exhibit hydrogen in their early-time spectra before becoming H-poor like a classical Type Ib SNe. This suggests a hybrid model in which hydrogen envelopes are stripped through interactions that are not dependent on initial mass (e.g. binary) for all subtypes, whilst helium envelopes are stripped via mass-sensitive processes such as stellar winds, resulting in more Ic SNe. Stellar wind strength also increases with metallicity, which results in more stripping of the outer envelopes \citep{mokiem2007} and an increased number of Wolf Rayet (WR) stars in higher metallicity regions \citep{vink2005}.

Given the importance of metallicity it has become increasingly popular to explore both the local environment as well as the global properties of SNe environments for different SN subtypes. Since massive stars have relatively short lifetimes of 3-50\,Myr, they have little time to stray from their natal environment, thus the properties of the explosion site, should also be representative of the progenitor. For example, the metallicity of the surrounding ISM should be consistent with the ISM that the progenitor was made from. The work of \citet{brazzini2024} shows global variations in metallicity across galaxies, but no strong evidence for local variations (e.g. see their figure 11). 

Investigations of local SNe environments have included analysis of the H\,{\sc ii} region for Star Formation Rate (SFR) \citep{anderson2008, anderson2012, galbany2014} and metallicity (see \citealt{ganss2022} and references therein). The latter of these is of interest because it impacts, along with luminosity, the strength of the stellar winds and hence influences progenitor mass-loss \citep{vink2021}. Most recently, \citet{pessi2023} used data from the Very Large Telescope (VLT) with the Multi Unit Spectroscopic Explorer (MUSE) instrument as part of the All-Sky Automated Survey for Supernovae (ASAS-SN, \citealt{Shappee2014}) of 112 CCSNe (78 II; 9 IIn; 7 IIb; 4 Ic; 7 Ib; 3 Ibn; 2 Ic-BL) detected between 2014 and 2018 to look at properties of SNe. They found that stripped-envelope SNe (Ib and Ic) are located in environments with a higher median SFR and oxygen abundance than Type II. The latter is confirmed at a statistical significance level with p\,=\,0.05 from Kolmogorov–Smirnov (KS) tests using the calibration in \citet{dopita2016} but this is not confirmed from the N2 and O3N2 calibrations of \citet{marino2013}. \citet{pessi2023} also find that Type Ic SNe have a higher mean oxygen abundance than Ib (but not at a statistically significant level), whilst Type IIb have similar oxygen abundances to Type Ic but similar SFRs to Ib. However, their low sample size makes these results difficult to interpret. 

With a larger sample of 77 narrow-lined Type IIn SN, \citet{ransome2022} analysed H$\alpha$ at SNe locations with the expectation that Type IIn would mimic ongoing star-formation if the dominant progenitors were high-mass Luminous Blue Variables (LBV). They found that only a subset follow the spatial distribution of H$\alpha$ emission with $\sim$40\% having no association with H$\alpha$ emission at all. One interpretation of this is a bimodal evolution with $\sim$60\% of Type IIn SNe resulting from high mass progenitors such as LBVs (e.g. SN2005gl, \citealt{galyam2007}) whilst the remaining progenitors are in binary systems \citep{smith2015}. \cite{moriya2023} use a sample of 21 Type IIn SNe from MUSE, Potsdam MultiAperture Spectrophotometer (PMAS) and the Mapping Nearby Galaxies at APO (MaNGA) surveys. They find IIn in environments with metallicities of 12$+$log(O/H)\,$\sim$\,8.1 to 8.8 (based on the N2 calibration), and that Type IIn SNe with a higher peak luminosity are preferentially located in environments with lower metallicity. This could suggest that Type IIn in lower metallicity environments have higher explosion energies.

\citet{pessi2023b} investigate a global analysis of all star-forming regions within the SNe host galaxy using the same ASAS-SN dataset for 99 CCSNe. They conclude there is a strong dependence of CCSNe production on metallicity, finding that there is a strong decrease in the number of CCSNe (per unit of star-formation) as oxygen abundance increases which is inconsistent with theory \citep{castor1975, kudritzki1987, puls2000, mokiem2007}. To explain this, the authors suggest that metallicity could play a significant role in the "explodability" of a massive star with more massive stars producing CCSNe at low metallicities. While this shows the importance of metallicity, without splitting their sample into subtypes we cannot assess the role of stripping with respect to subtype ratios as all massive stars would produce a CCSNe of some type. Similarly, \citet{laplace2020} show that low metallicity could be the crucial factor in producing more SNe from binary systems because at high metallicity all the H-rich envelope is stripped, whereas at low metallicity lower-mass progenitors still have their envelopes that can extend to larger radii of $>$400R$_{\odot}$, encouraging Roche-Lobe overflow and additional stripping of the hydrogen envelope.

\citet{aguilera-dena2023} apply MESA (Modules for Experiments in Stellar Astrophysics, \citealt{paxton2011}) models and find that explosion energies, ejecta masses and nickel masses for Ib SNe are not well matched by single, high mass progenitors with WR-like mass loss, but are reproduced well by lower mass progenitors with negligible mass loss and a luminosity below the WR threshold. For Ic SNe they show that higher progenitor metallicities (Z$=$0.04) are required to match ejecta masses and explosion energies from observations of Type Ic SN. However our previous work \citep{ganss2022} found Ic SNe at all metallicities with little to no preference for super-solar metallicities. 

When looking at SNe and their progenitors the effect of dust and mass-loss episodes has been seen. Prior to exploding, the high-mass LBV progenitor of Type IIn SN2009ip was observed to undergo multiple outbursts \citep{mauerhan2013}. Similarly, \citet{fraser2013} identified two outbursts prior to SN2011ht, also a Type IIn, however pre-SN imaging fails to reveal the progenitor. Most recently, Type IIL SN2023ixf has been found to be surrounded by a C-rich dust shell and an increased mass-loss rate $\sim$3\,years before the SN of $\sim$10$^{-4}$M$_{\odot}$yr$^{-1}$ \citep{niu2023}. Similarly, \citet{chugai2022} model Type Ibn SN 2010al spectra and the circumstellar shell concluding that the progenitor star had a mass-loss episode $\sim$0.4\,years prior to the SNe explosion. \citet{xiao2023} conducted a wider survey using mid-IR Spitzer data to investigate dusty SNe over a broad range of subtypes. They find that dusty SNe have a larger proportion located within regions of higher SFR compared to normal (non-dusty) SNe but find little to no dependence on metallicity. 

Understanding the effect of metallicity on the SN subtype is vital to understanding their progenitors, whether single or binary, but there is not yet a clear consensus about this. This paper builds on our previous work (\citealt{ganss2022} hereafter Paper I) where we concluded that there was a lack of Type Ib SNe at low metallicities based on a sample of 65 CCSNe (13 Type Ib). However, the low number statistics limited the results. In this paper we present a much larger sample of Type IIP, Type Ib and Type Ic CCSNe to investigate the metallicity of their environments as a function of SN subtype. We focus on SNe within a distance of $\sim$100\,Mpc and crucially, we present the largest spectroscopic sample to date of Type Ib and Ic SNe metallicities, thus reducing the unreliability of results due to low number statistics. In this paper we use the metallicity calibration of \citet{marino2013}, however we note that we are mainly concerned with the relative differences in metallicity between the SN subtypes rather than the absolute values. 

The paper is structured as follows: Section \ref{sect_obs} presents the SNe sample selection criteria, the new target observations and the archival data used. Section \ref{sect_dat} describes the data reduction process and the method of SNe environment metallicity measurement. Section \ref{sect_res} presents the results followed by the discussion in Section \ref{sect_dis}. Finally, Section \ref{sect_con} summarizes the conclusions.

\section{Observations} \label{sect_obs}

\subsection{Target selection}    \label{sub_trgts}

The most important step to enlarge the target sample is the extension of maximum luminosity distance from $\sim$30\,Mpc in Paper I to $\sim$100\,Mpc. We chose the 100\,Mpc limit as a compromise of a significantly larger number of targets and the inevitably worse spatial resolution of observed SN environments.

Due to the larger uncertainty of the luminosity distance with increasing distance, the luminosity distance criterion creates a certain arbitrariness in the target selection. To avoid this, we switched from luminosity distance to a maximum redshift z$\lesssim$0.025 for the host galaxy of the target as a more objective selection criterion. We chose a maximum redshift of z$\sim$0.025 roughly corresponding in the Cosmic Microwave Background (CMB) reference frame to a Hubble distance of $\sim$100\,Mpc. We took the redshifts from the NASA/IPAC Extragalactic Database (NED\footnote{https://ned.ipac.caltech.edu}, \citealt{helou1991}). We used various catalogues to identify SNe within this redshift limit (Rochester Supernovae Archives (RAS\footnote{https://www.rochesterastronomy.org/snimages/index.html}, \citealt{galyam2013}), Transient Name Server (TNS\footnote{https://www.wis-tns.org}, \citealt{galyam2021}), Weizmann Interactive Supernova Data Repository (WISeREP\footnote{https://www.wiserep.org},  \citealt{yaron2012}), Open Supernovae Catalogue (OSC\footnote{https://github.com/astrocatalogs/supernovae},  \citealt{guillochon2017}), UC Berkeley Filippenko Group's Supernova Database (SNDB\footnote{http://heracles.astro.berkeley.edu/sndb/}, \citealt{silverman2012}),  Asagio Supernova Catalogue(ASNC\footnote{https://sngroup.oapd.inaf.it/asnc.html}, \citealt{barbon1999})). In total, we found 1395 CCSNe within redshift z$\lesssim$0.025 which have at least one classification of Type IIP, Ib or Ic.

As in Paper I, we have only considered SNe with reliable classification as Type IIP, Type Ib or Type Ic as potential targets for the project. We restricted Type II SNe to the Type IIP subtype as this is the only subtype with unambiguous, confirmed progenitors (single red supergiants, see \citealt{smartt2009,vandyk2017} and references therein) and because work by \cite{graham2019} indicates that different Type II subtypes have different metallicity distributions. Additionally, we excluded Type Ic-BL SNe from the Type Ic class because they are often associated with gamma-ray bursts indicating a different physical nature of Type Ic-BL progenitors. We discuss the implications of subtype and classification issues in more detail in Sections \ref{sub_ic_bl} and \ref{sub_sn_mis}.

\begin{table}
   \begin{threeparttable}
    \caption{Overview of all SNe observed by INT/IDS and/or found in archival data. The columns are the instrument used, number of SNe within z${\lesssim}$0.025 and the number of SNe with reliable classifications. The column `misc Ib/c' implies all Types Ib/c, Ca-rich Ib or Ic events, Ic-BL, Ibn and Icn; the column `misc. II' implies generic Type II, IIL, ambiguous IIb and ambiguous IIn. The difference between `Number of Individual SNe' and the sum of the instrument targets is caused by targets observed with multiple instruments.}
    \setlength{\tabcolsep}{0.1cm}
    \begin{tabular}{cccccccccc}
      \hline
      \hline
      Instrument & Number & Ib & Ic & IIP & IIb & IIn & misc.& misc. \\
                 & of SNe  &    &     &     &    &     &  Ib/c & II \\
    \hline
      IDS & 139 & 58 & 29 & 31 & 2 & 0 & 9 & 10 \\
      MUSE & 166 & 18 & 21 & 54 & 20 & 13 & 13 & 27 \\
      MaNGA & 22 & 5 & 8 & 2 & 3 & 2 & 1 & 1 \\
      PMAS/PPak & 113 & 19 & 20 & 29 & 12 & 13 & 11 & 9 \\
      \hline 
      Number of & 407 & 90 & 71 & 107 & 35 & 27 & 30 & 47 \\
      Individual SN \\
      \hline
    \end{tabular}%
    \label{tab_observations}%
    \end{threeparttable}
\end{table}%

\begin{table*}
   \begin{threeparttable}
   \caption{Brief overview of the resolution capabilities of the instruments used in the current paper. For more technical details see Paper I for INT/IDS, \citet{henault2003,bacon2010} for MUSE, \citet{roth2005,kelz2006} for PMAS/PPak and \citet{smee2013,drory2015} for MaNGA. Survey information is available in \citet{galbany2018} for PISCO, \citet{sanchez2012} for CALIFA, \citet{bundy2014} for MaNGA and references herein.}
    \setlength{\tabcolsep}{0.1cm}
    \begin{tabular}{ccccc}
        \hline
      \hline
       & IDS & MUSE & PMAS/PPak & MaNGA \\
      \hline
      field of view & 1.5$\arcsec$$\times$3.3$\arcsec$ (slit) & 1$\arcsec$$\times$1$\arcsec$ & 1.3$\arcsec$$\times$1.3$\arcsec$ & 3$^{\circ}$ diameter \\
      spatial sampling & 0.4$\arcsec$/px & 0.2$\arcsec$$\times$0.2$\arcsec$/px & 0.2$\arcsec$/px & 0.5$\arcsec$$\times$0.5$\arcsec$/px \\
      FWHM & 2.8 {\AA} & 2.4 {\AA} & 6 {\AA} & 2.5 {\AA} \\
      resolving power & $\backsim$1600 & 1770-3590 & $\backsim$850 & 1560-2650 \\
      spatial resolution & n.a. & 0.46$\arcsec$ & 2.7$\arcsec$ & 2$\arcsec$ \\
      \hline
    \end{tabular}%
    \label{tab_instruments}%
    \end{threeparttable}
\end{table*}%

Because it was not within the scope of the project to classify each SN, we used the various catalogues outlined above to check the classifications. SNe classified in all references exclusively as Type Ib, Type Ic or Type IIP were always accepted as target candidates. For Type IIP SNe this meant they had to have a well-sampled light curve extending to the plateau. The procedure was significantly more complex for targets which had multiple classifications in the catalogues (unclear Type Ib or Ic, generic Type II, etc.). In these cases we searched the references given in the catalogues for classifications in The Astronomer's Telegram (ATel\footnote{https://astronomerstelegram.org/},  \citealt{rutledge1998}) and/or Central Bureau for Astronomical Telegrams\footnote{http://www.cbat.eps.harvard.edu/} (CBETs) or International Astronomical Union Circulars (IAUC). Most importantly, we searched the literature for re-classifications (e.g. \citealt{harutyunyan2008, modjaz2014, matheson2001, shivvers2017, shivvers2019}), light curves of SNe (e.g. \citealt{jones2009, galbany2016,martinez2022a}) as well as dedicated papers on specific SNe. If the published literature reached a consensus on the SN subtype then the SN was accepted as part of our sample. Any SNe that remained ambiguous were removed. For clarity, all references on which our classification assessment is based are given in the tables in the appendix (Tables \ref{tab_results} and \ref{tab_iibiin_results}). 

In Paper I, we had a sufficiently low (<75°) host galaxy inclination as a further selection criterion to avoid ambiguities in the SNe H\,{\sc ii} region identification. We decided not to select by galaxy inclination in this work because of the large range in morphologies, sizes and distances of the host galaxies, which render the interpretation of measured inclinations unclear. All SNe were visually checked and only removed from the sample if the angular distance of the SNe site to the galaxy centre was smaller than the extraction aperture, meaning that the extracted SNe spectrum could be contaminated by Active Galactic Nuclei (AGN) or other nuclear activity. We applied this check during the data reduction process.

The total size of the selected sample meeting our redshift and classification criteria was 732 potential SNe consisting of 196 Type Ib, 233 Type Ic and 303 Type IIP SNe. We closed the target list on 31-Dec-2022. We note that the work presented in this paper is of those SNe for which we obtained our own data or could access archival data, details of which are given below. 

\subsection{INT/IDS observations}    \label{sub_intobs}

From previous observations (see table 1 of Paper I) with the Intermediate Dispersion Spectrograph (IDS) on the Isaac Newton Telescope (INT) we had a total of 100 long-slit observations of 78 individual supernovae. From October 24, 2022 to November 2, 2022, we were able to make another 62 observations in 10 nights with the INT/IDS (INT proposal ID I/2022B/4). The same instrumental setup as our first observing runs was used (see table 2 in Paper I for details). In total, 162 long-slit observations of 139 different targets were available with INT/IDS. See Table \ref{tab_observations} for details of the different SNe types observed with INT/IDS.

\subsection{Archival data}    \label{sub_arch_dat}

To further increase the number of observations we searched for SNe site observations by Integral Field Unit (IFU) in the following archives: The ESO Science Archive\footnote{https://archive.eso.org/scienceportal/home} (\citealt{romaniello2022}) for suitable MUSE (\citealt{bacon2010}) data, the Calar Alto Legacy Integral Field Area (CALIFA) survey\footnote{https://califa.caha.es} (\citealt{sanchez2012}) of PMAS/PPaK (\citealt{roth2005,kelz2006}) instrument observations and the MaNGA survey\footnote{https://www.sdss4.org/dr17/manga} (\citealt{bundy2014}) for suitable data from the MaNGA IFU (\citealt{smee2013,drory2015}). See Table \ref{tab_instruments} for the most important technical data of the three IFUs; for more technical details of the instruments we refer to the studies quoted.

In the ESO Science Archive we searched the data of the MUSE instrument mounted on the VLT (UT4). In total, we found MUSE data of 166 CCSNe environments within z$\lesssim$0.025, 93 of these had a reliable classification including 18 Type Ib, 21 Type Ic and 54 Type IIP.

From the CALIFA survey using the PMAS/PPak instrument, we searched several sources: the CALIFA H\,{\sc ii} Regions Catalogue\footnote{https://ifs.astroscu.unam.mx/CALIFA/HII$\_$regions/} (\citealt{espinosa2020}), the SNe environmental metallicities published in \cite{galbany2018} and the PISCO (PMAS/PPak Integral-field Supernova Hosts Compilation) data published online\footnote{https://zenodo.org/records/1172732} by \cite{galbany2018}. In these data, we found observational data and environmental metallicities of 113 CCSNe environments with z${\lesssim}$0.025 among them 19 Type Ib, 20 Type Ic and 29 Type IIP.

In the MaNGA Survey we found IFU observational data from 22 CCSNe environments with z$\lesssim$0.025 (among them 5 Type Ib, 8 Type Ic, 2 Type IIP).

In total, we obtained archival environment observations of 294 CCSNe of all types (among them 39 Type Ib, 48 Type Ic, 83 Type IIP) from the three IFU instruments (see Table \ref{tab_observations} for details). In total with the INT/IDS data, we had observations of the environments of 407 targets, of which 268 met our selection criteria (90 Type Ib, 71 Type Ic and 107 Type IIP targets).

\begin{figure*}
\includegraphics[trim={0cm 0cm 0cm 0cm}, clip, width=1.92\columnwidth, angle=0]{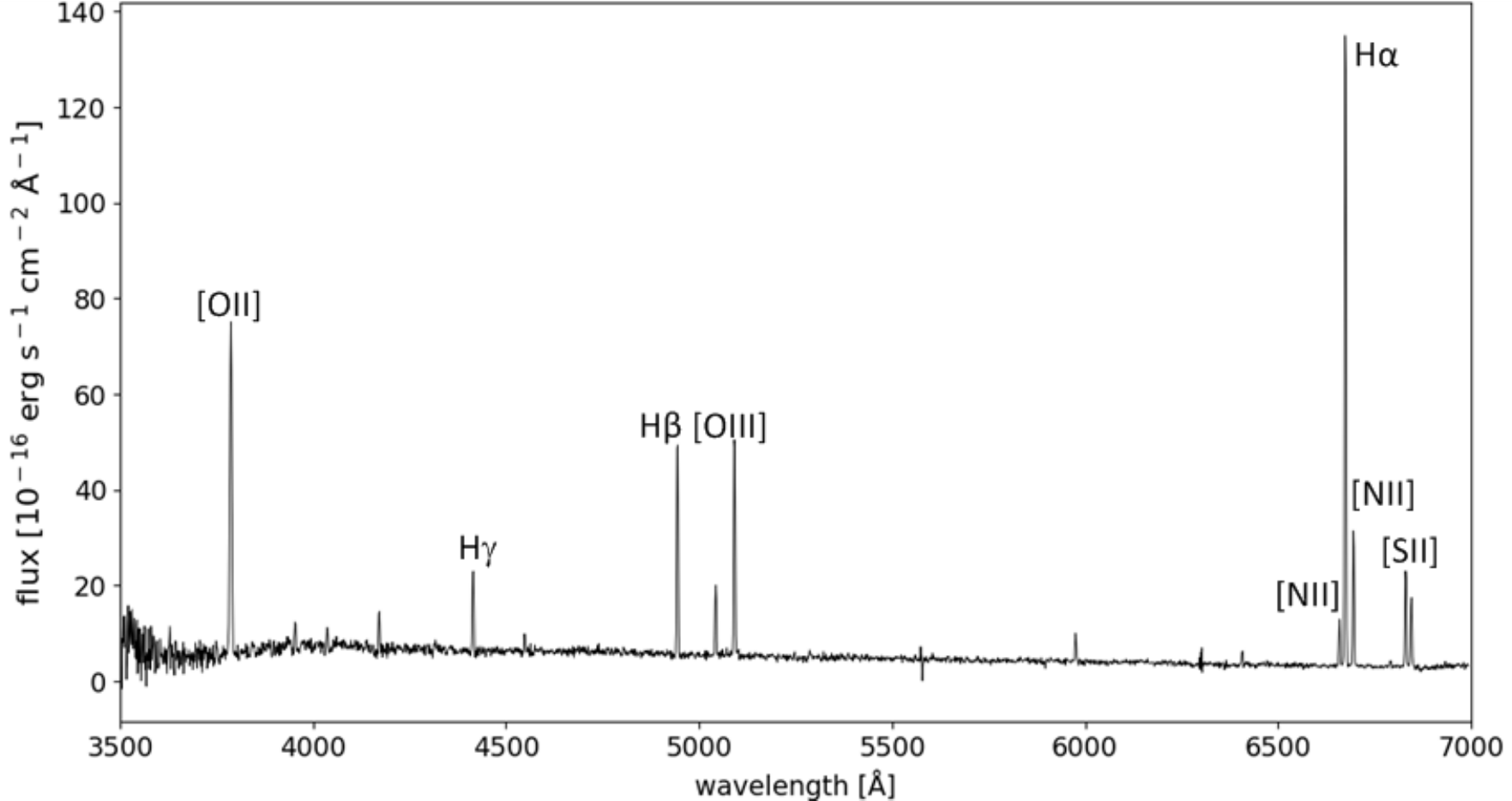}
\caption{Obtained INT/IDS spectrum of Type Ib SN2006dn environment as an example for the extracted 1D environment spectra. The figure shows the most prominent emission lines of which H$\alpha$, [N\,{\sc ii}]$\lambda$6583, H$\beta$ and [O\,{\sc iii}]$\lambda$5007 have been used for the metallicity estimation.}
\label{fig_sn2006dn}
\end{figure*}

\section{Data Reduction and Analysis}  \label{sect_dat}

\subsection{INT data reduction} \label{sect_dat_int}

Reduction of the new INT/IDS data was performed as in Paper I: Using the standard software packages IRAF (Image Reduction and Analysis Facility; \citealt{tody1986}) and Starlink (\citealt{currie2014}), we processed the observational data with bias subtraction, flat correction and S-distortion correction; wavelength and flux calibration were performed using arc frames and standard star observations. The INT/IDS data were trimmed to the range 3500-7000 {\AA}, in which all required diagnostic lines lie. Determination of the pixel position of the SNe environment was done as described in Paper I and extraction of the spectrum was done with the IRAF \textsc{apall} routine with 4\arcsec extraction aperture as a reasonable choice to account for seeing and imperfections of the optical path of the equipment. The 4\arcsec aperture size scales with 19.4\,pc/Mpc and corresponds to linear size of 1.94\,kpc for a host galaxy at distance 100\,Mpc.

The Balmer decrement has been used for interstellar extinction correction of the SNe site spectrum based on an H$\alpha$/H$\beta$ ratio of 2.86, assuming case B recombination \citep{hummer1987} and empirical extinction curves from \citet{osterbrock2006}. Targets lacking H$\beta$ emission were only corrected for Galactic extinction\footnote{We tested the impact of the extinction correction on the resulting metallicities and for the N2 calibration there was no difference in the metallicity results, for O3N2 the difference was 0.01-0.03 dex depending on the signal-to-noise ratio of the observation.} with values taken from NED (based on \citealt{schlafly2011}) and assuming a \citet{fitzpatrick1985} reddening law.  Extinction correction was performed using the Starlink package DIPSO \textsc{dred} routine. A typical extracted spectrum obtained by the INT/IDS long-slit observations is shown in Figure \ref{fig_sn2006dn} for target SN2006dn as an example. 

\begin{figure*}
\includegraphics[trim={0cm 0cm 0cm 0cm}, clip, width=1.92\columnwidth, angle=0]{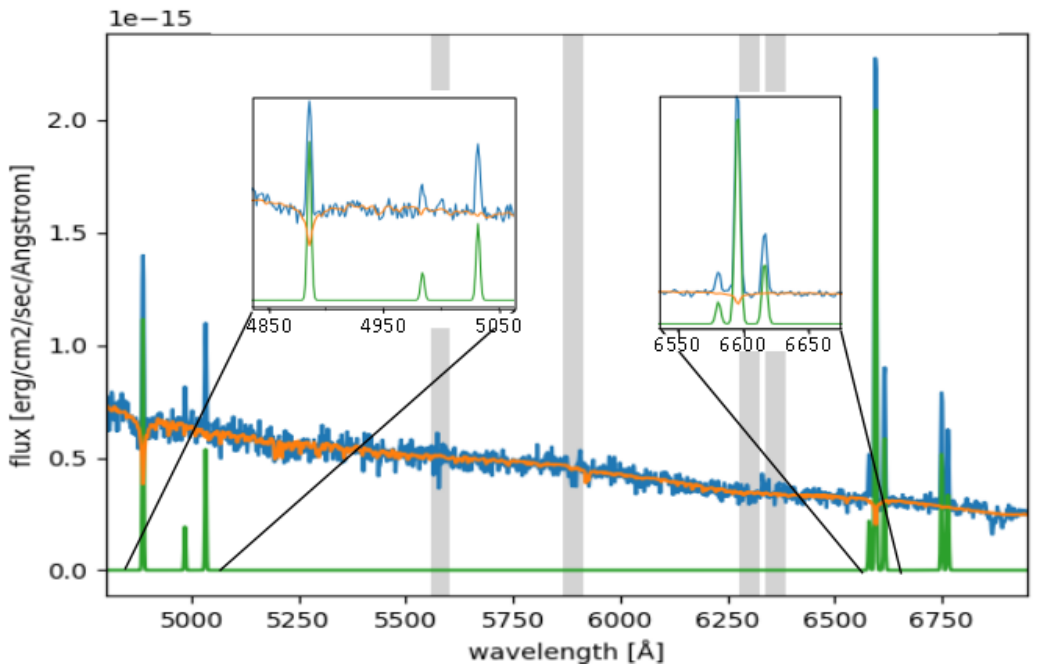}
\caption{Example of PPXF emission line fit. The figure shows the PPXF fit of the explosion site spectrum of SN2001B obtained by INT/IDS. The obtained spectrum is shown in blue. The orange line is the best PPXF fit to the continuum including underlying stellar absorption lines. Green is the PPXF emission line fit. The light gray regions are masked to avoid any fitting issues by sky line residuals. The two inlays are zoomed H$\beta$ and H$\alpha$ region, respectively. The chosen example is one of the targets with strong stellar absorption underlying the H$\beta$ line, which is much lower for H$\alpha$.}
\label{fig_sn2001b}
\end{figure*} 

\subsection{Archival data reduction} \label{sect_dat_arc}

The downloaded MUSE and MaNGA data are wavelength and flux calibrated but not corrected for interstellar extinction (\citealt{weilbacher2020,law2016}). The interstellar extinction correction of extracted MUSE and MaNGA spectra was performed in same way as described above for the INT/IDS data.

The extraction of SNe site data from the IFU data was done by the Python package MUSE Python Data Analysis Framework (MPDAF\footnote{https://mpdaf.readthedocs.io/en/latest/index.html}, \citealt{bacon2016,piqueras2019}). MPDAF has been developed mainly for work with MUSE data but is suited for general use with spectra, images and cubes. We used MPDAF for all cubes – independent of the archival source - to avoid any issues with different software packages. The SNe site spectra found in MUSE, MaNGA and the emission lines of PISCO cubes were all extracted with a circular aperture of 2\arcsec radius. The resulting aperture area is about twice the INT aperture area. We did test different aperture sizes, but since the metallicity calibrations rely on flux ratios rather than absolute flux the aperture size makes little to no difference to our results.

The archival data of CALIFA H\,{\sc ii} regions catalogue and PISCO survey are wavelength and flux calibrated data which are corrected for interstellar extinction as described in \citet{sanchez2012} and \citet{galbany2018}. No additional data reduction was required. We searched the CALIFA H\,{\sc ii} regions catalogue for emission line fluxes by the SNe RA-Dec coordinates with search radius 6 arcsec and took the H\,{\sc ii} region emission line fluxes of the H\,{\sc ii} region closest to the SNe site. PISCO data are available as flux data in the online version of table 3 of \cite{galbany2018}, as spectra of SNe sites and as IFU cubes containing the emission line fluxes of H$\alpha$, H$\beta$, [NII]$\lambda$6583, [OIII]$\lambda$5007 and [SII]$\lambda\lambda$6717,6731 per spaxel (both provided by \citealt{galbany2018} and downloaded from  \url{https://zenodo.org/records/1172732}). We used all three of these PISCO data sources for evaluation of the environment metallicities and averaged the results. No data reduction was necessary for the SNe site spectra of the PISCO survey and the data were taken for emission line fitting as they are. The PISCO cubes were processed with MPDAF as described above.
 
\subsection{Emission line fitting} \label{sect_line_fit}
We use the strong emission line method to estimate SNe environment metallicities \citep{pettini2004}. This method needs only flux ratios of strong nebular emission lines for the metallicity determination, which are much easier to detect than the auroral [O\,{\sc iii}]$\lambda$4363 line required for the direct method. Many different ratios of strong emission lines have been proposed to determine gas-phase metallicities (see e.g. \citealt{kewley2008, kewley2019} and references therein). This work uses the N2 and O3N2 indicators with the calibration as defined in \citet{marino2013}.

The N2 and O3N2 indicators themselves have been chosen because they use closely spaced emission line ratios making the methods robust to reddening and flux calibration corrections. 

The emission lines H$\alpha$, H$\beta$, [NII]$\lambda$6583 and [OIII]$\lambda$5007 of the extracted SNe site spectra were fitted by Gaussian profiles for flux estimation. Two independent tools with different fitting approaches were used to check the reliability of fluxes. We use the Penalized PiXel-Fitting (PPXF)\footnote{https://pypi.org/project/ppxf/7.0.0} (version 7.0.0, \citealt{cappellari2017}) and the DIPSO Emission Line Fit (ELF) routine; comparison between the results is discussed in Section \ref{sect_dis}

The PPXF tool fits the spectrum by an optimized linear combination of stellar spectra templates (for details see \citealt{cappellari2017} and references therein). Additionally, PPXF allows for fits of emission lines in the spectra. We fitted all extracted SNe site spectra with PPXF in the wavelength interval 4800-6950 {\AA} using simple stellar population template spectra from the MILES library\footnote{http://research.iac.es/proyecto/miles/pages/ssp-models.php} \citep{vazdekis2010}. We used a Gaussian fit template for the emission lines (see Figure \ref{fig_sn2001b} for an example).

With the DIPSO ELF command we fitted emission lines by Gaussian profiles. ELF fits were done in two small wavelength intervals ($\Delta\lambda$ typically 240 {\AA}  and 360 {\AA}, respectively) in the H$\alpha$ and H$\beta$ region independently. However, all emission lines within each interval were fit simultaneously.

The emission line fluxes of the hydrogen lines may be contaminated by underlying stellar absorption along the line of sight underestimating the true emission line flux of the SNe environment. While PPXF intrinsically takes into account stellar contamination of the emission line fluxes, the simple DIPSO Gaussian fit is not able to correct for possible stellar absorption. Therefore we applied the workaround as described in Paper I, figure 2, to all targets exhibiting significant stellar contamination (122 targets) of the H$\beta$ line to compensate for the stellar absorption. In all cases we neglect the much lower relative stellar absorption at the H$\alpha$ line.   

Despite these different approaches, the two tools provide very similar metallicity values (differences between PPXF and DIPSO results are addressed in section \ref{sub_ppxf}). All subsequent results are using the PPXF results, supplemented by DIPSO results of three targets (SN2004A, SN2016ccm and SN2017ewx) for which PPXF failed to fit.

From the observational data of the 268 Type Ib, Ic and IIP targets of Table \ref{tab_observations}, the data reduction process provided emission line fluxes for the N2 indicator of 238 SN. For 205 of these SNe we also got H$\beta$/[OIII] fluxes, for application of the O3N2 indicator. 18 spectra exhibited either only hydrogen lines or no emission lines at all. The check of potential nucleus contamination excluded 7 SNe due to angular centre distance less than the extraction aperture (see section \ref{sub_trgts}). We also excluded 4 targets which exhibit a broadened H$\alpha$ line indicating active interaction with circumstellar matter (e.g. \citealt{smith2017,smith2024}) and blending with [NII] lines making uncertainties high or [NII] flux estimation impossible. One target failed the validity check of the \cite{marino2013} N2 calibration (see Section \ref{sub_indicators}). In sum we had fluxes for N2 calibration of 238 targets (79 Type Ib, 66 Type Ic, 93 Type IIP; see Table \ref{tab_targets} in the appendix) and fluxes for O3N2 calibration of 205 targets (63 Type Ib, 61 Type Ic, 81 Type IIP).  

\section{Results}   \label{sect_res}

Metallicity estimation by the strong emission line method requires a calibration derived from empirical metallicity data estimated by direct methods. The N2/O3N2 calibrations were initially introduced by \citet{pettini2004} but were updated by \cite{marino2013} who used over 450 and over 3000 H\,{\sc ii} regions, respectively to calculate the linear regression line fits. Throughout the paper we will use the updated calibrations of \cite{marino2013} for N2 and O3N2, their equations 2 and 4, respectively. In this paper we label the calibrations M13-N2 and M13-O3N2, respectively. For comparison with previous works that use the \citet{pettini2004} calibration we convert the values of M13-N2 and M13-O3N2 using equations \ref{m13topp04_n2} and \ref{m13topp04_o3n2}. 

\begin{equation}
\text{PP04-N2} = -1.887 + 1.234 \times \text{M13-N2} 
\label{m13topp04_n2}
\end{equation}
and
\begin{equation}
\text{PP04-O3N2} = -4.03 + 1.495 \times \text{M13-O3N2}
\label{m13topp04_o3n2}
\end{equation}

Table \ref{tab_results} presents the measured metallicities of the SNe environments in our sample, inferred by the N2 and O3N2 (where available) calibrations from \cite{marino2013} calibration.

\subsection{Validity ranges of the indicators}    \label{sub_indicators}

The validity ranges for the M13-N2 and M13-O3N2 calibrations are given by \citet{marino2013} as --1.6 < N2 < --0.2 and --1.1 < O3N2 < 1.7. All N2 and O3N2 values of our sample are larger than the lower limits whilst the upper limits are exceeded by the values of SN2012cw, SN2014C (both N2 only) and SN2001ch, SN2012A, SN2013ce, SN2016I (all O3N2). The indicator values of these targets except SN2012cw exceed the upper limit within the estimated uncertainty interval of the indicators. They have been accepted taking into account that the given range limits are not strict limits due to their statistical nature. SN2012cw exceeds the upper limit significantly and given that it exhibits emission line radiation characteristic of a low-ionization nuclear emission-line region (LINER), (see section \ref{sub_sn2012cw}) it has been removed from our sample.

\begin{figure}
\includegraphics[trim={0.5cm 0.6cm 0.5cm 0.3cm}, clip, 
width=1.0\columnwidth]{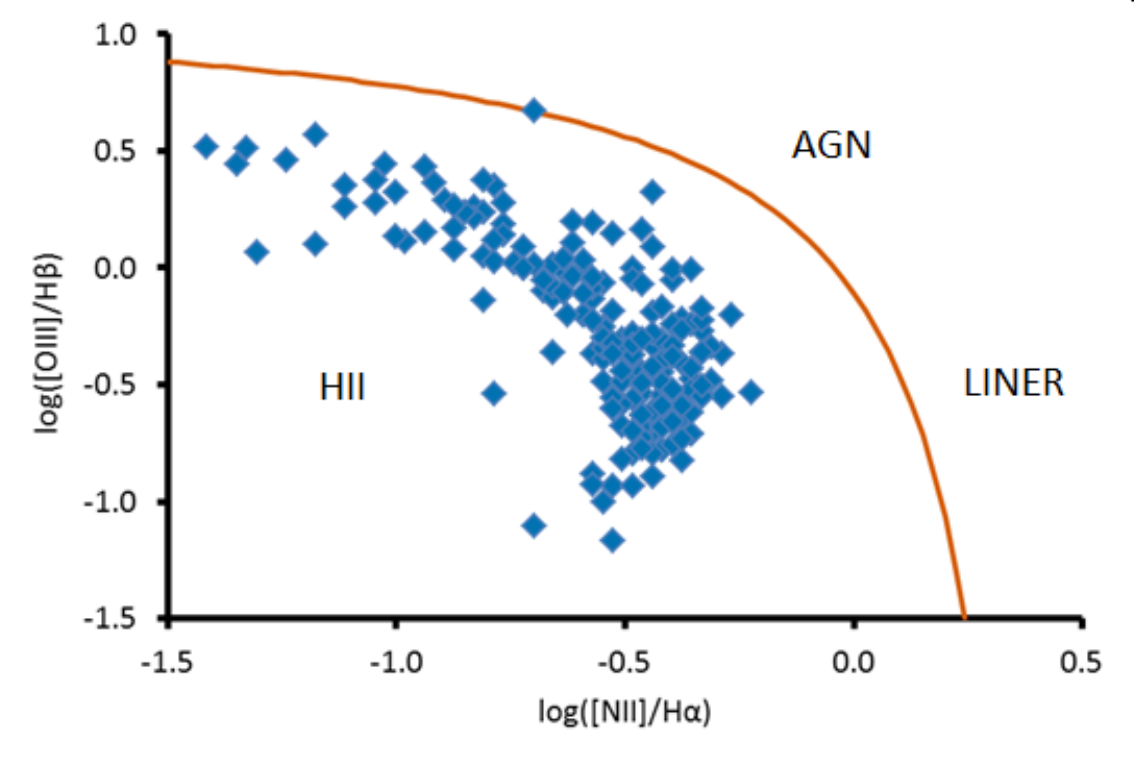}
\caption{BPT-diagram of all targets with O3N2 calibration (blue diamonds). The red solid line is the decision line between the H\,{\sc ii} region and the AGN/LINER region as given by \protect\cite{kewley2001}, equation (5). All targets but one (SN1961V) are well within the H\,{\sc ii} region of the \protect\cite{kewley2001} decision line}
\label{fig_bpt_diagram}
\end{figure}

\subsection{AGN activity}    \label{sub_agn}

Application of the strong emission lines method is restricted to gas phase emission caused by stellar ionisation. We checked this precondition by means of the BPT-diagram (Baldwin, Phillips \& Terlevich, \citealt{baldwin1981}) shown in Figure  \ref{fig_bpt_diagram}. All targets but one (SN1961V) with O3N2 results are well within the H\,{\sc ii} region; for targets with N2 results only, the BPT diagram is not applicable.

\begin{table}
\caption{Number of targets (N), mean values and standard deviations ($\sigma$) of the M13-N2 and M13-O3N2 metallicities split into the three SNe types and for the total sample.}
    \begin{tabular}{c|ccc|ccc}
    \hline
    \hline
    SN    & \multicolumn{1}{c}{N(N2)} & \multicolumn{2}{c|}{M13-N2} & \multicolumn{1}{c}{N(O3N2)} & \multicolumn{2}{c}{M13-O3N2} \\
    type  &       & \multicolumn{2}{c|}{[12+log(O/H)]} &       & \multicolumn{2}{c}{[12+log(O/H)]} \\
    \multicolumn{1}{c|}{} &       & \multicolumn{1}{c} {mean} & \multicolumn{1}{c|}{$\sigma$} &       & \multicolumn{1}{c} {mean} & \multicolumn{1}{c}{$\sigma$} \\
    \hline
    Ib    & 79    & 8.51  & 0.07  & 63    & 8.50  & 0.10 \\
    Ic    & 66    & 8.51  & 0.09  & 61    & 8.50  & 0.10 \\
    IIP   & 93    & 8.46  & 0.12  & 81    & 8.43  & 0.13 \\
    \hline
    all   & 238   & 8.49  & 0.10  & 205   & 8.47  & 0.12 \\
    \hline
    \end{tabular}%
  \label{tab_statistics}%
\end{table}%

\begin{figure*}
\subfigure{\includegraphics[trim={0.3cm 0cm 1.35cm 1.1cm}, width=1.0
\columnwidth]{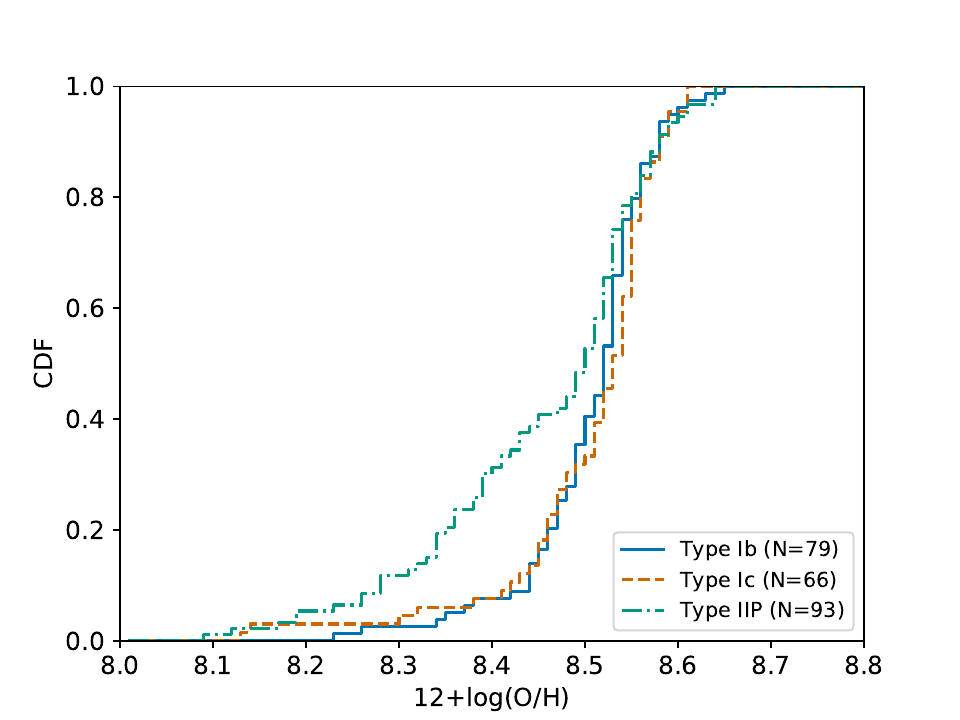} \label{fig_n2_cdf}}
\subfigure{\includegraphics[trim={0.3cm 0cm 1.35cm 1.1cm}, 
width=1.0\columnwidth]{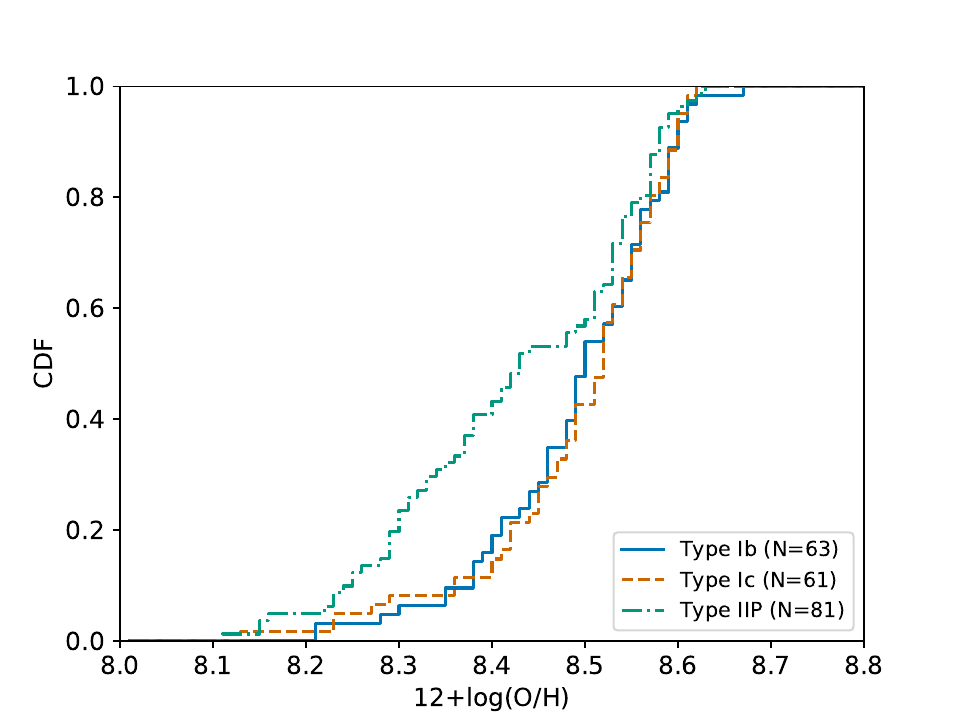} \label{fig_o3n2_cdf}}
\caption{CDFs of the SNe environment metallicities measured with the M13-N2 (left) and M13-O3N2 (right) calibration. Binning width for CDF calculation: 0.0125 dex.}
\label{fig_cdfs}
\end{figure*}

\subsection{Uncertainties}

The observational uncertainties of our observed data by photon noise and data reduction process have been estimated to be between $\pm$0.01 and $\pm$0.05\,dex (median value $\sim$0.035\,dex) mainly depending on the signal-to-noise ratio (SNR).

The error on the metallicity calibrations must be considered in a number of ways. The calibrations result from linear regression line fits to a large number of data points and result in a 1$\sigma$ (68\%) dispersion around the line of $\pm$0.16\,dex for M13-N2, $\pm$0.18\,dex for M13-O3N2. However, the regression fit itself has a much smaller formal error of $\pm$0.035 dex for both the M13-N2 and M13-O3N2 fits, respectively. Whilst the larger uncertainties must be taken into account when considering absolute metallicity values, we note that in this work we are mainly interested in the relative differences between SN subtypes, rather than the absolute values of metallicity. We address the errors on the calibration in more detail in Section \ref{sub_mcs}.

\subsection{Environment metallicities}    \label{sub_mets}

Table \ref{tab_statistics} lists the mean and standard deviation of metallicities for the three SNe types and for the total sample. Differences in the mean metallicities between subtypes are small, with the largest difference being between Ib/Ic and IIP SNe at 0.07\,dex for the M13-O3N2 calibration. A simple t-test for the O3N2 results suggests a statistically significant difference between the mean of Ib and IIP SNe if the maximum observational uncertainty of $\pm$0.05 is assumed, however this becomes (just) not significant (p = 0.053) if the calibration uncertainty of $\pm$0.18 is applied to all data points. The scatter of Type Ib metallicities around the mean is significantly smaller (factor 1.7) than for Type IIP for the N2 calibration, but this reduces (factor 1.3) for M13-O3N2 calibration. The smaller range of Type Ib metallicities compared to IIP observed in Paper I, is still observed in this larger sample, but the difference between Ib and IIP has decreased.

\subsection{Cumulative distribution functions}    \label{sub_cdfs}

Figure \ref{fig_cdfs} shows the normalised cumulative distribution functions (CDFs) of the results for M13-N2 and M13-O3N2 calibrations, respectively. The figure reflects the standard deviation results of Table \ref{tab_statistics} by a narrower CDF of Type Ib compared with Types Ic and IIP. The difference in CDFs at low metallicities seems to be significant: while Types Ic and IIP are already present at metallicities from about 8.1\,dex, Type Ib metallicities only start above 8.2 dex. In addition, the fraction of Type IIP at low metallicities up to about 8.5\,dex is significantly greater than for Types Ib and Ic. At metallicities above about 8.5\,dex, no difference in the CDFs can be seen.

\begin{table}
\begin{threeparttable}
\caption{P-values of the two-sample KS-test and two-sample AD-test for the M13 results. P-values of less than 0.01 support a statistically significant difference between the parent 
populations of the given supernova subtypes (see section  \ref{sub_test} for full interpretation of the results).}

\setlength{\tabcolsep}{0.151cm}
   \begin{tabular}{c|cc|cc}
   \hline
   \hline
   & \multicolumn{2}{c|}{KS-test} & \multicolumn{2}{c}{AD-test} \\
   \hline
   SN type & M13-N2 & M13-O3N2  & M13-N2 & M13-O3N2 \\
       & p-value & p-value  & p-value & p-value \\
   \hline
   Ib-Ic  & 0.1215 & 0.7614 & 0.3470 & 0.9686 \\
   Ib-IIP & 0.0012 & 0.0057 & 0.0057 & 0.0041 \\
   Ic-IIP & 0.0078 & 0.0019 & 0.0035 & 0.0046 \\
   \hline
   (Ib+Ic)-IIP & 0.00032 & 0.00032 & 0.00060 & 0.00046 \\
   \hline
   \end{tabular}%
  \label{tab_m13testresults}%
  \end{threeparttable}
\end{table}%

\subsection{Statistical tests}    \label{sub_tests}

We applied the two-sample Kolmogorov-Smirnov test (KS-test, \citealt{press1988}) and the two-sample Anderson-Darling test (AD-test, \citealt{pettitt1976})  to check if the differences between the three SN subtypes are statistically significant. The tests have different approaches (see paper I, section 4, for a short description of the test differences) to test the null hypothesis that two discrete samples are drawn from same parent population. In this work we use the two-sample KS-test and AD-test as implemented in the package `twosamples'\footnote{https://twosampletest.com/}\citep{dowd2020}, version 2.0.0,  of the statistical software called `R'\footnote{https://www.r-project.org} \citep{Rproject2021}. The statistical tests of the package `twosamples' are not based on tables to calculate the p-value of the given samples but perform real permutations between samples to evaluate the p-value. The null hypothesis, that both samples have the same parent population, must be rejected if the p-value is less than the chosen significance level $\alpha$.

Table \ref{tab_m13testresults} lists the p-value results of the sample combinations Ib-Ic, Ib-IIP and Ic-IIP for M13-N2 and M13-O3N2 metallicity calibrations for both the two-sample KS-test and the two-sample AD-test. The p-value for a significance level of 1\% must be <0.01 to reject the null hypothesis with 99\% confidence which is the case for the Ib/Ic vs. IIP both the KS and AD test results. For the combined Ib+Ic vs. IIP results the level is even lower than 0.1\%.

The main result of the statistical tests is statistically significant evidence to reject the null hypothesis that the Ib/Ic and IIP samples are drawn from the same parent population with significance level better than 1\%. This suggests that IIP SNe arise from different progenitor stars to Ib and Ic SNe, whilst the p-values of SNe Ib vs Ic indicate no statistical significance of different parent populations of Ib and Ic SN.

\begin{figure}
\includegraphics[trim={0.6cm 0.25cm 0.6cm 0cm}, clip, width=1.0\columnwidth, 
angle=0]{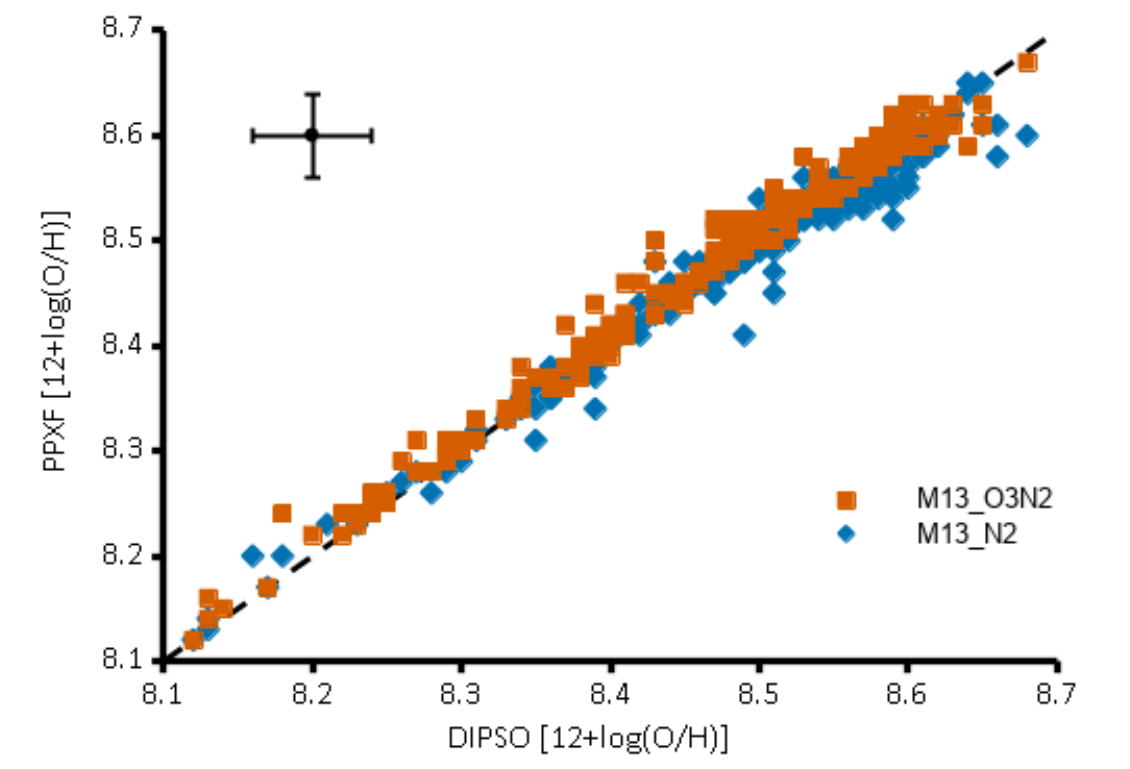}
\caption{Comparison of metallicities derived from PPXF fitting versus metallicities derived by DIPSO fitting for M13-N2 (blue diamonds) and M13-O3N2 (red squares). The cross indicates the typical observational uncertainty of $\pm$0.04 dex; the dashed line represents the 1:1 line. Despite that both methods have completely different approaches to fit the emission lines (Section \ref{sect_line_fit}), the values agree remarkably well.}
\label{fig_ppxf_vs_dipso}
\end{figure}

\section{Discussion}  \label{sect_dis}

In this section we consider the reliability of our results and the impact of the errors by using Monte Carlo (MC) simulations. We compare long-slit and IFU observations to assess the consistency between the two different observing methods. We also review the results from PPXF with those from DIPSO to investigate whether our results are dependent on the analysis tools. 

\begin{figure}
\includegraphics[trim={0.6cm 0.15cm 0.6cm 0cm}, clip, width=1.0\columnwidth, 
angle=0]{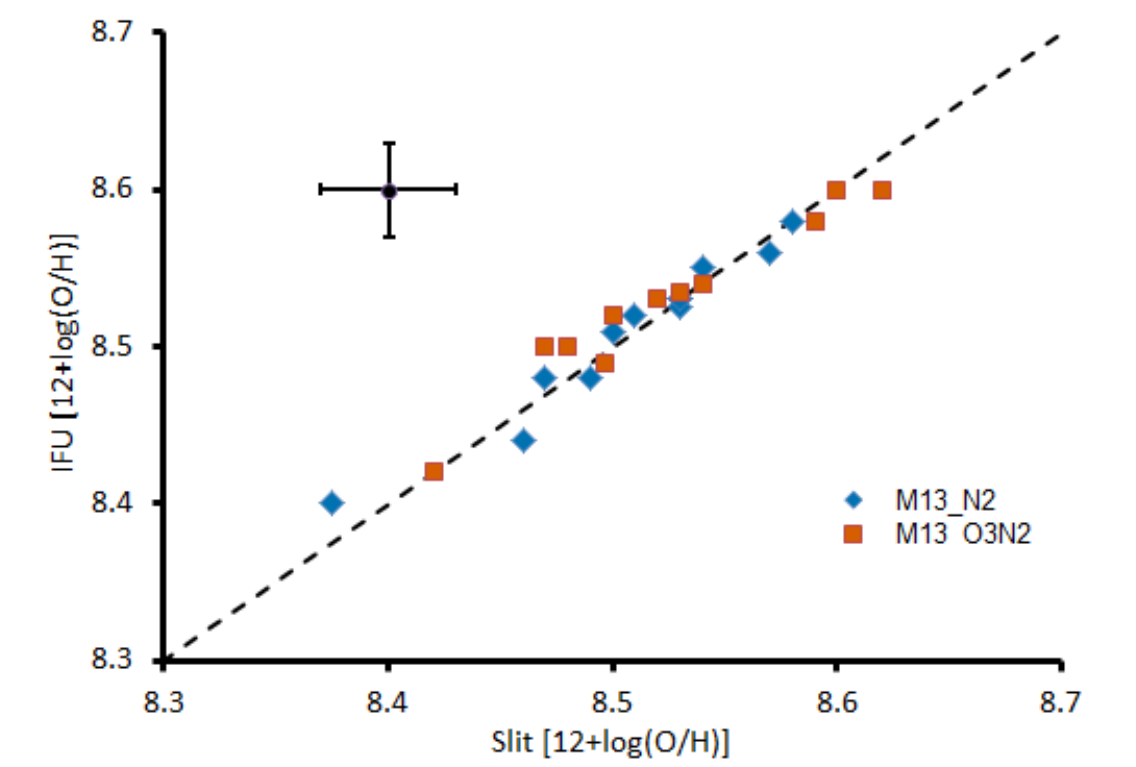}
\caption{Comparison of metallicities of targets observed by both IFU and by INT/IDS instruments for M13-N2 (blue diamonds) and M13-O3N2 (red squares). The dashed line represents the 1:1 line. There is no significant difference in the metallicities between the results obtained by the two instrument types (and different telescopes at different locations).}
\label{fig_ifu_vs_slit}
\end{figure}

\begin{table*}
\begin{threeparttable}
\caption{Statistical properties of the SNe host galaxies. The table shows the means of absolute B-magnitudes M\textsubscript{B}, of the projected centre distance and of the host inclinations, their standard deviations $\sigma$ and the standard errors of the mean ($\Delta$mean) of the hosts for the different SNe types. The numbers of hosts are given as well and they differ slightly for the three quantities due to different availability of host data. Data of absolute magnitudes were taken from GLADE+ catalogue \citep{dalya2022}, projected centre distance were calculated by angular distance of SNe site to host centre and host luminosity distance taken from GLADE+. The host inclination values were taken from Hyperleda \citep{makarov2014}. See text in Section \ref{sub_host} for interpretation.}
\setlength{\tabcolsep}{0.11cm}
   \begin{tabular}{c|cccc|cccc|cccc}
   \hline
   \hline
   & \multicolumn{4}{c|}{absolute magnitude} & \multicolumn{4}{c|}{projected centre distance} & \multicolumn{4}{c}{inclination} \\
   \hline
   SN type &  N(hosts) & mean M\textsubscript{B} & $\sigma$ & $\Delta$(mean M\textsubscript{B}) & N(hosts) & mean d\textsubscript{c} & $\sigma$ & $\Delta$(mean d\textsubscript{c}) & N(hosts) & mean I & $\sigma$ & $\Delta$(mean I) \\
   &  & [mag]  & [mag] & [mag] &  & [kpc]  & [kpc] & [kpc] &  & [\textdegree]  & [\textdegree] & [\textdegree] \\
   \hline
   Ib   & 79 & -20.4 & 0.94 & 0.11 & 79 & 4.4 & 3.05 & 0.34 & 77 & 55.0 & 18.39 & 2.10 \\
   Ic   & 66 & -20.4 & 0.92 & 0.11 & 66 & 4.4 & 4.14 & 0.51 & 64 & 50.5 & 19.96 & 2.50 \\
   IIP  & 91 & -20.0 & 1.19 & 0.12 & 91 & 5.2 & 5.83 & 0.61 & 85 & 56.3 & 19.70 & 2.14 \\
   \hline
   all  & 236 & -20.2 & 1.06 & 0.07 & 236 & 4.7 & 4.90 & 0.32 & 226 & 54.2 & 19.40 & 1.29 \\
   \hline
   \end{tabular}%
  \label{tab_host_statistics}%
  \end{threeparttable}
\end{table*}%

\subsection{PPXF vs. DIPSO results}    \label{sub_ppxf}

As mentioned in Section \ref{sect_line_fit} we used PPXF as well as DIPSO for emission line fitting to estimate effects of stellar contamination on the metallicities. Both tools use different approaches for flux estimation of the emission lines. The main advantage of DIPSO is the numerical stability and the use of local spectral data around the emission line(s) only. PPXF fits the complete spectrum obtained by stellar templates and thus provides information on stellar contamination of the emission lines. On the other hand, PPXF is more sensitive to the input parameters and needs the data over a broad spectral range, requiring an accurate masking of residual sky lines and/or other artificial spectral issues. However, PPXF can easily fit the underlying hydrogen absorption features which is more complicated in DIPSO. 

Despite the different approaches to fitting the spectra, the results of the two tools agree remarkably well as shown in Figure \ref{fig_ppxf_vs_dipso} within the typical overall observational and data reduction process uncertainty of about $\pm$0.04 dex as indicated by the error bar. The few outliers for M13-N2 calibration generally had poorer fits and lower SNR observations. The good agreement between different methods of fitting the emission lines gives us confidence in our measured emission lines fluxes.           

\subsection{Long-Slit vs. IFU results}    \label{sub_long_slit}

We checked the reproducibility of the results using targets with multiple observations from both IFU and long-slit observations. The reproducibility of our INT observations is unchanged from paper I (see Figure 5) as we did not obtain any repeat observations in our 2022 INT observing run. However we have 12 new targets observed by both INT/IDS long-slit and IFU observations (7 MUSE, 1 MaNGA, 3 PMAS, 1 observed by both MUSE and MaNGA). Figure \ref{fig_ifu_vs_slit} compares the metallicities for both observational techniques based on observations of different instruments at different observatories. The consistency of the results is significantly better than the observational uncertainty. For the only target with three observations (SN2006fo observed by INT, MUSE and MaNGA) the difference between the three results is 0.01 dex. The excellent agreement between the two observational techniques, for multiple targets, provides confidence in our extraction of 1D-spectra and in the comparability of long-slit and IFU results.     

\subsection{Check of SNe host biases}    \label{sub_host}

We selected our targets strictly by the redshift of the host galaxy. With the exception of a few targets, excluded by close host centre distance to avoid host centre contamination (see Section \ref{sub_trgts}), our sample is volume limited and so there should be no observational bias. To confirm this, we checked several host properties (host type, absolute luminosity, projected SNe centre distance and host inclination) for differences between the SNe types potentially biasing the statistical results (see Table \ref{tab_host_statistics}). The large difference in the mean of the distance from centre of Type IIP (5.2 kpc) compared to Type Ib (4.4\,kpc) and Ic (4.4\,kpc) is caused by a few outliers especially SN2016ccm. The projected mean distance of Type\, IIP reduces to 4.80 without SN2016ccm and the standard deviation reduces from 5.83\,kpc to 4.01\,kpc. The mean of inclination is in the same range for all SN types with a very large scatter.
 
Potential bias based on targeted versus untargeted SN surveys has been discussed by \citet{sanders2012} whereby pre-2015 surveys targeted brighter (and potentially more metal-rich) galaxies. Since we only look for relative differences in the metallicities we don't expect our results to be impacted by host luminosity as any such bias should be applicable to all SN subtypes. The relative metallicities would only be impacted if one SN subtype was dominated by pre-2015 observations and the other by post-2015 observations; this is not the case in our survey. For completeness, we present histograms of the host galaxy luminosity separated by SN subtype in Figure \ref{fig_host_luminosities}. Type IIP SN hosts do extend to (slightly) fainter host galaxies however, fainter IIP hosts relative to Ib/Ic hosts could be a natural consequence of a genuine difference in the environment metallicities between IIP and Ib/c. 

\subsection{Comparison of O3N2 and N2 results}   \label{sub_o3n2}

\begin{figure}
\includegraphics[trim={0.1cm 0.15cm 0.7cm 0cm}, clip, width=1.0\columnwidth, 
angle=0]{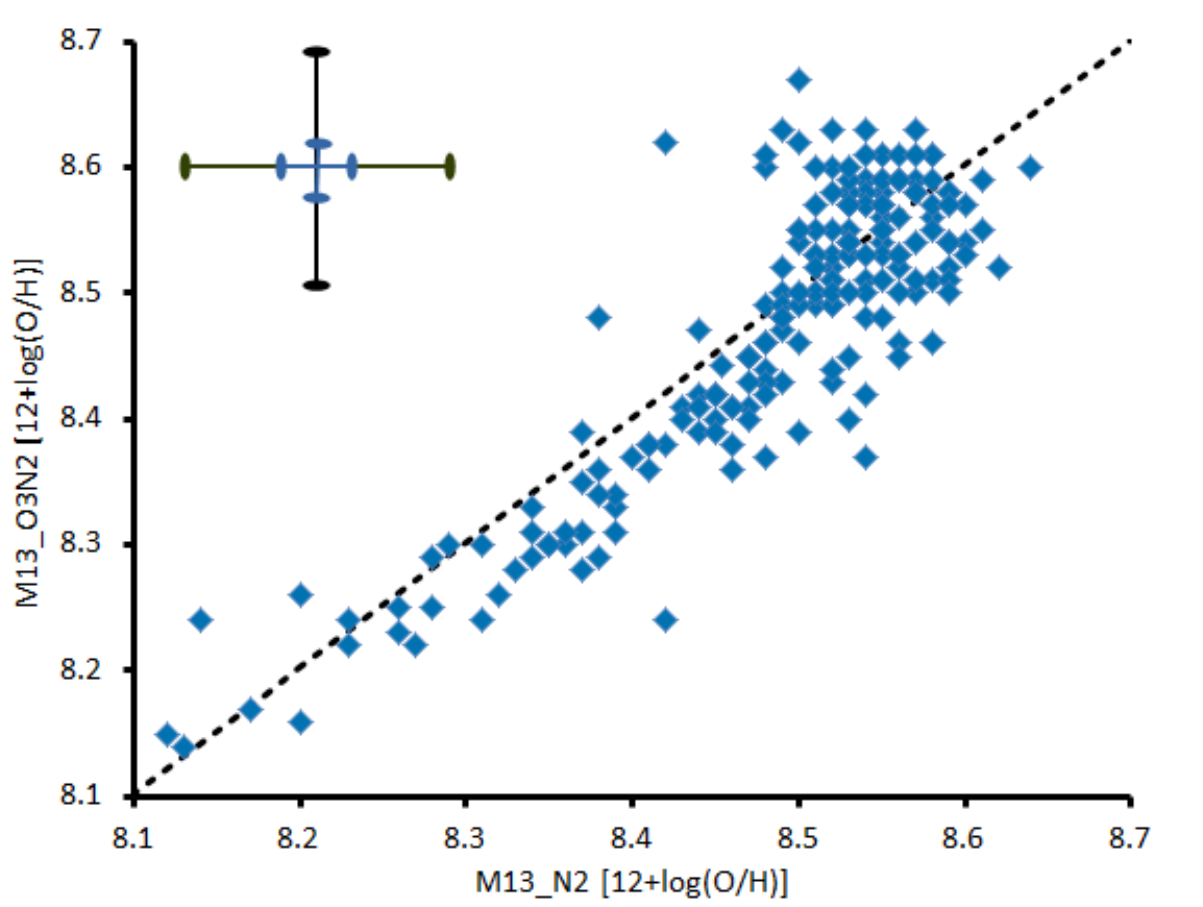}
\caption{Comparison  of M13-O3N2 vs. M13-N2 calibration for all targets where both results are available. The black and blue overlaid crosses represent the calibration uncertainties and the typical observational uncertainty, respectively. The dashed line is the 1:1 line. This figure confirms the increased scatter seen in Paper I for M13-N2 values > 8.5\,dex. It reveals also that M13-O3N2 provides lower metallicity values in the range 8.2 to 8.5 dex.}
\label{fig_n2_vs_o3n2}
\end{figure}

\begin{table}
   \begin{threeparttable}
    \caption{Smallest metallicities of Type Ib, Ic and IIP SNe found in previous and this work. The number in brackets behind is the number of targets found in previous work fulfilling our target selection criteria. All metallicities but \citet{modjaz2011} are M13-N2 values; the PP04-N2 values of \citet{anderson2010}, \citet{leloudas2011} and \citet{sanders2012} have been converted to M13-N2 values by rearranged equation (\ref{m13topp04_n2}). The values of \citet{modjaz2011} are PP04-O3N2 values. See text Section \ref{sub_prev} for discussion.}
    \setlength{\tabcolsep}{0.1cm}
    \begin{tabular}{cccc}
        \hline
      \hline
      work & lowest Ib & lowest Ic & lowest IIP \\
      \hline
      \cite{anderson2010}     & 8.11(14) & 8.41(10) & 8.26(5) \\
      \cite{galbany2018}      & 8.43(17) & 8.14(20) & 8.10(27) \\
      \cite{kuncarayakti2018} & 8.26(11) & 8.02(9) & 8.13(23) \\
      \cite{leloudas2011}     & 8.50(6) & 8.37(3) & -(0) \\
      \cite{modjaz2011}       & 8.29(9) & 8.49(9) & -(0) \\
      \cite{sanders2012}      & 8.34(5) & 8.30(4) & -(0) \\
      \cite{pessi2023}        & 8.13(6) & 8.16(5) & 8.09(23) \\
      this work               & 8.23(79) & 8.13(66) & 8.09(93) \\
      \hline
    \end{tabular}%
    \label{tab_previous}%
    \end{threeparttable}
\end{table}%

\begin{figure*}
\includegraphics[trim={0cm 0cm 0cm 0cm}, clip, width=1.92\columnwidth, 
angle=0]{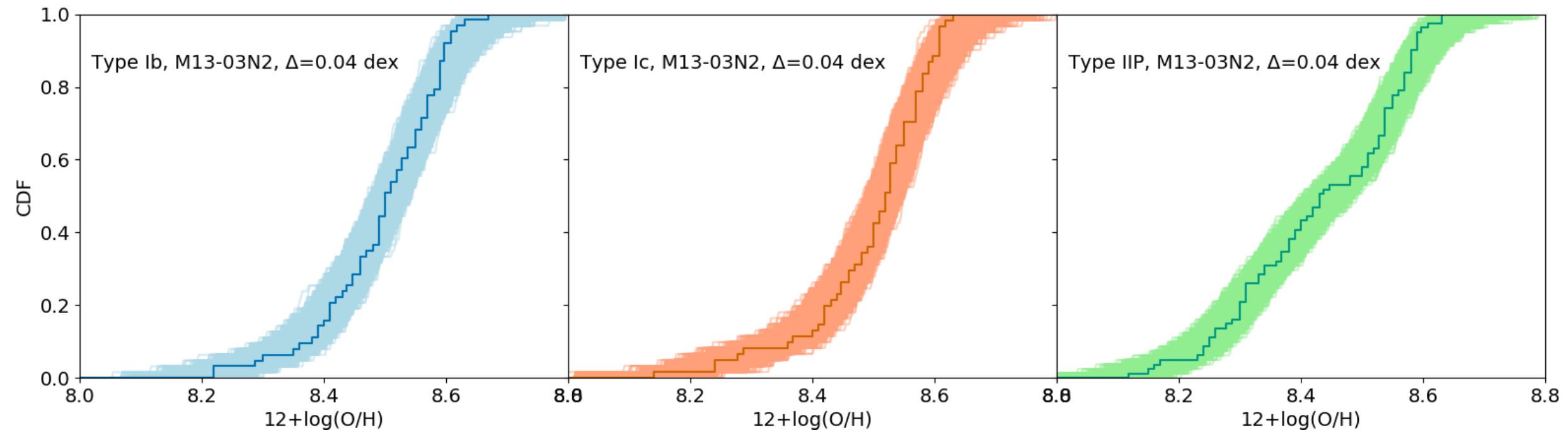}
\caption{CDF scatter range of the Monte-Carlo simulations (20000 runs) for uncertainty $\Delta$ = $\pm$0.04 dex of M13-O3N2 samples. The solid lines are the undisturbed CDFs of Figure \ref{fig_cdfs}, right panel.}
\label{fig_cdf_scatter}
\end{figure*}

\begin{figure*}
\subfigure{\includegraphics[trim={0.5cm 0cm 1.1cm 1cm}, width=.95
\columnwidth]{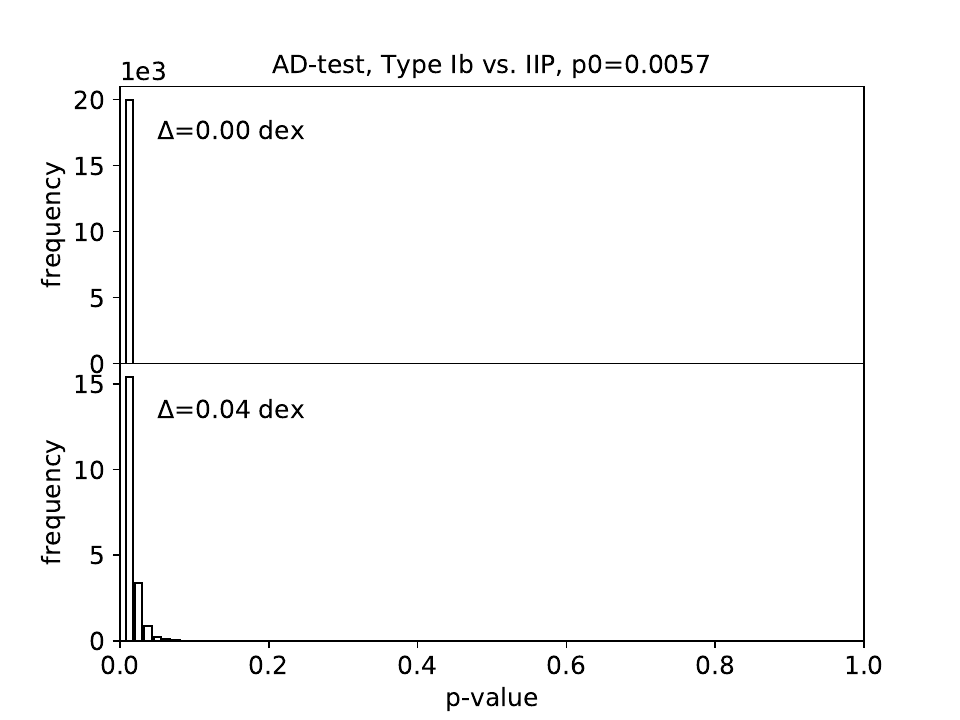} \label{fig_mn2_ib_iip_1}}
\vspace{0cm}
\subfigure{\includegraphics[trim={0.5cm 0cm 1.1cm 0.8cm}, width=0.95
\columnwidth]{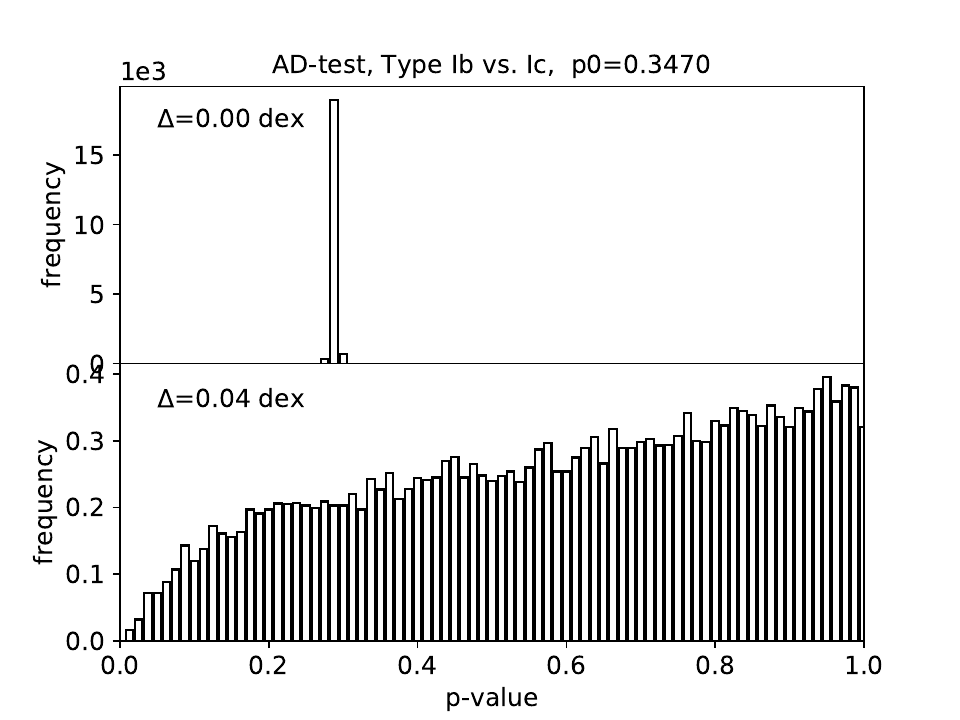} \label{fig_mn2_ib_ic_1}}
\caption{Example of the effect of metallicity uncertainty on the p-value distribution based on Monte-Carlo (MC) simulations. The figures shown are the MC AD-test p-value distributions for the M13-N2 Ib sample vs. M13-N2 IIP sample (left) and vs. M13-N2 Ic sample (right). The upper panels show the initial p-values of the samples without metallicity uncertainty, the lower panels show the p-value distributions whne applying an uncertainty of 0.04\,dex to all sample targets. The behaviour shown is representative and has been seen for all MC results independent of statistical test and sample: the smaller the initial p0 value the lower the sensitivity to uncertainties and vice versa. For our largest samples (combined Ib+Ic sample vs. IIP sample), we have seen almost no sensitivity of the p-values at 0.04\,dex uncertainty.}
\label{fig_pvalues}
\end{figure*}

We repeat the comparison of M13-O3N2 and M13-N2 calibrations done in Paper I with the larger sample (Figure \ref{fig_n2_vs_o3n2}). As in Paper I, an increasing scatter for metallicities >$\sim$8.5\,dex is visible. The increased scatter may be caused by saturation of the [N\,{\sc ii}] line for metallicities greater than solar (12 + log(O/H) = 8.69, \citealt{asplund2009}) as discussed by e.g. \cite{pettini2004} and \cite{marino2013}. However, confirming the results of Paper I, the spread starts at about 8.5\,dex, well below solar metallicity. Figure \ref{fig_n2_vs_o3n2} also reveals a systematic difference of the metallicity by M13-O3N2 compared with M13-N2 in the range 8.2 to 8.5\,dex by about 0.04\,dex. This systematic difference was not obvious in Paper I because of the smaller sample. Consequently, a mixture or comparison of M13-O3N2 and M13-N2 metallicities provides questionable results at this level. This result is unexpected and reinforces the importance of only comparing metallicities derived from the same calibration (see \citealt{kewley2008} for full discussion).

\subsection{Comparison to other work}   \label{sub_prev}

In Paper I we found a M13-N2 cutoff metallicity of about 8.3\,dex for Type Ib SNe, which was clearly distinct to the lower metallicities of Type Ic and IIP (M13-N2 cutoff 8.15\,dex and 8.18\,dex, respectively). With the larger sample, the Type Ib cutoff lowers to 8.23\,dex but is still above the Ic and IIP cutoffs (8.13\,dex and 8.09\,dex, respectively). Recently, \cite{pessi2023} reported a Type Ib metallicity as low as 8.13\,dex for SN2016ccd. This target is within a projected distance of 260\,pc from the host galaxy centre and a 90$^\circ$ host inclination so was excluded from our sample by project design. But we can confirm the low metallicity we derived from MUSE data (our value: 8.15\,dex). We checked previous work for low metallicity results of Type Ib and present a summary in Table \ref{tab_previous}. We note that previous works do not correct for the underlying Balmer absorption we outline in section \ref{sect_line_fit}. The impact of this correction corresponds to, on average, a difference in metallicity of 0.018\,dex ($\sigma$=0.016), although in extreme cases the difference can be up to 0.1\,dex.

For the table we searched the previous work for all Ib, Ic and IIP targets fulfilling the target criteria of our project (beside classification essentially the redshift condition). With exception of \cite{modjaz2011} values, all values in Table \ref{tab_previous} are M13-N2 values, either directly from the published data or calculated from published PP04-N2 values by rearranging equation (\ref{m13topp04_n2}). From \cite{modjaz2011} we have only PPO4-O3N2 values not transferable to M13-N2. But by means of rearranging equation (\ref{m13topp04_o3n2}) we get a M13-O3N2 value of 8.24\,dex, which should be, according to Figure \ref{fig_n2_vs_o3n2}, a lower limit for M13-N2. Table \ref{tab_previous} shows the only other low metallicity Type Ib, the SN2002jz (formerly classified as Type Ic in the IAU circulars, \citealt{shivvers2019}) with environment metallicity of M13-N2 = 8.11\,dex by \cite{anderson2010}. This target is, with M13-N2 = 8.23\,dex, our lowest Type Ib target as well. The different metallicity value is likely caused by the slightly different approach by \cite{anderson2010} to estimate the environment metallicity (they evaluated the H\,{\sc ii} region closest to the SNe site not the SNe site itself, see \citealt{anderson2010}). The other previous work confirms the higher Type Ib cutoff compared with Type IIP SNe. For Type Ic, the lower cutoff metallicity varies between studies from 8.02\,dex to 8.49\,dex. However, all but this work and \citet{galbany2018} have low number statistics and/or take the metallicity closest to the SN, not directly at the SNe site. This work and \citet{galbany2018} are in good agreement. 

\subsection{Monte-Carlo simulations}    \label{sub_mcs}

We checked the sensitivity of the p-values of the statistical tests using MC simulations and applied the same algorithm described in section 5.4 of Paper I. As an example, the results for the typical observational uncertainty of about 0.04\,dex (which was assumed to be the same for all targets in the MC simulations) are shown in Figure \ref{fig_cdf_scatter} and Figure \ref{fig_pvalues}. Figure \ref{fig_cdf_scatter} shows the scatter in the CDFs for different SNe types. At uncertainty 0.04\,dex the differences between Ib and Ic CDFs to the IIP CDFs is still visible. This is reflected in the p-value distributions which are shown in Figure \ref{fig_pvalues}. However, there is a significant change in the sensitivity of the p-value distributions to uncertainties compared to Paper I due to the significantly larger samples: if the initial p-value (p0) is small (empirically: less than 0.1), the sensitivity to the typical observational uncertainty 0.04\,dex is very low and almost not visible for initial p-values <0.001. If the initial p-value (p0) is high (empirically: greater than 0.2), the p-value distributions tend to be almost uniform for typical observational uncertainty of 0.04\,dex. This behaviour has been seen for all sample combinations and independently of the statistical test used. The qualitative difference in the MC p-value distributions between KS- and AD-test we have seen in Paper I disappears completely with the larger samples as it was caused only by the relatively small sample in Paper I. 

\begin{figure}
\includegraphics[trim={0cm 0cm 0cm 0cm}, clip, width=1.0\columnwidth, angle=0]{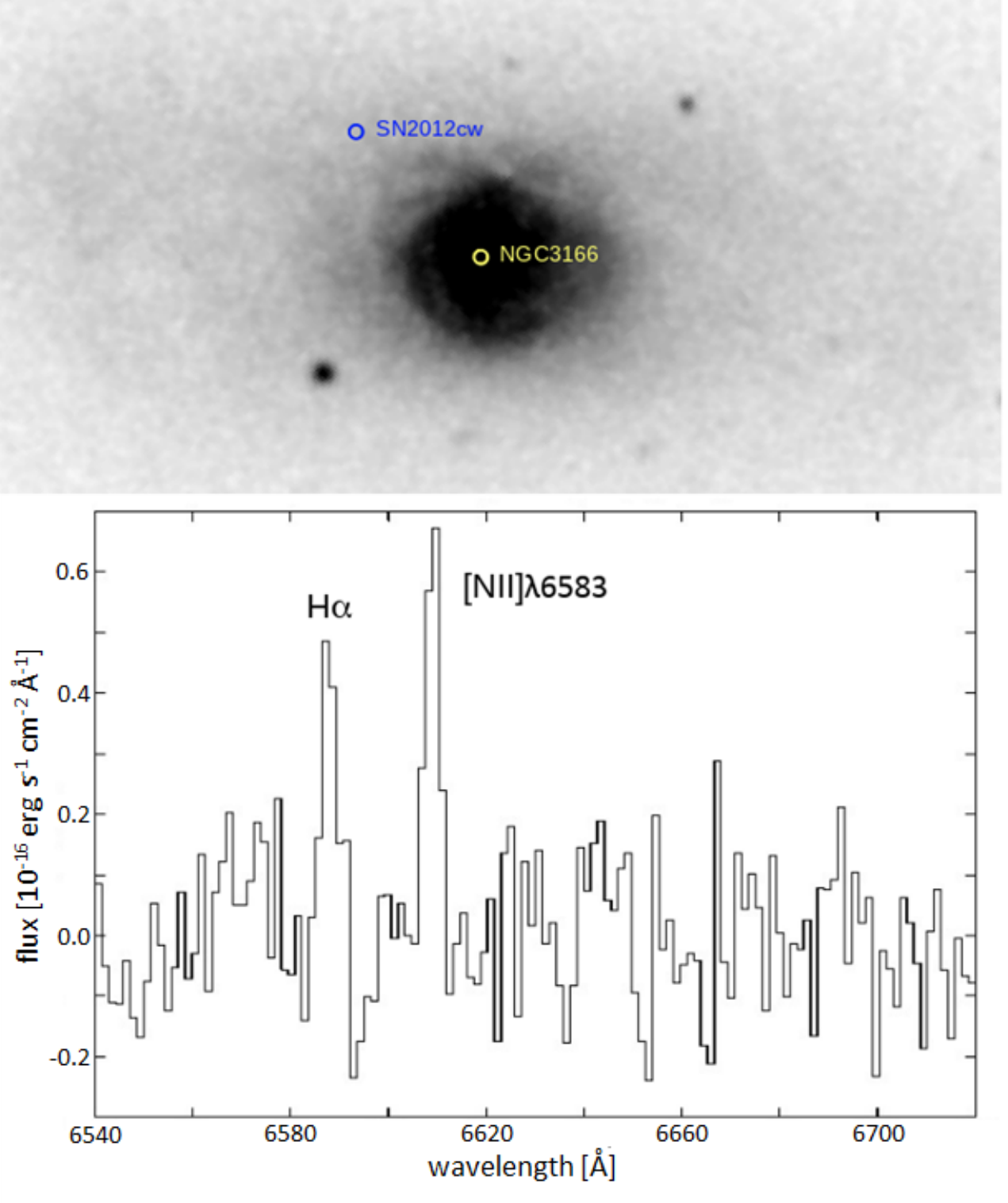}
\caption{Explosion site of SN2012cw in NGC 3166 and the obtained emission lines at H$\alpha$ (observer frame). The [NII] flux is stronger than H$\alpha$ indicating low-ionisation LINER-like emission at a large distance from the host centre (projected distance: $\sim$5.1 kpc).}
\label{fig_sn2012cw}
\end{figure}

\subsection{SN2012cw}   \label{sub_sn2012cw}

We excluded SN2012cw from our sample because it violates the validity range of the M13-N2 calibration significantly. However, SN2012cw is an interesting object because it is the single spectrum of our collection exhibiting stronger [NII] flux than H$\alpha$ flux (see Figure \ref{fig_sn2012cw}). This is characteristic for low-ionisation excitation known from LINER galaxies, but SN2012cw is at a large projected distance (about 5.1 kpc) from the centre of its lenticular host and we can exclude an AGN excitation. We observed SN2012cw twice at INT with a time interval of about 15 months and thus we can exclude a rapid transient process. Unfortunately, the H$\beta$ and [OIII] emission is hidden in the noise in our observations and we cannot evaluate the position of SN2012cw in the BPT diagram directly. By means of PPXF we revealed H$\beta$ and [OIII] fluxes out of the noise and get averaged SN2012cw coordinates (-0.1,0.21) being in the LINER region of BPT diagram (recall Figure \ref{fig_bpt_diagram}). We propose to call the SN2012cw environment a "LIER" (Low-Ionisation Excitation Region) observed at large distance from the host centres (e.g. \citealt{singh2013,hviding2018,byler2019}). The excitation mechanism may be shock ionisation (e.g. \citealt{congiu2023}) of CSM or ISM by the SNe ejecta.  

\begin{figure*}
\subfigure{\includegraphics[trim={0.3cm 0cm 1.35cm 1.1cm}, width=1.0
\columnwidth]{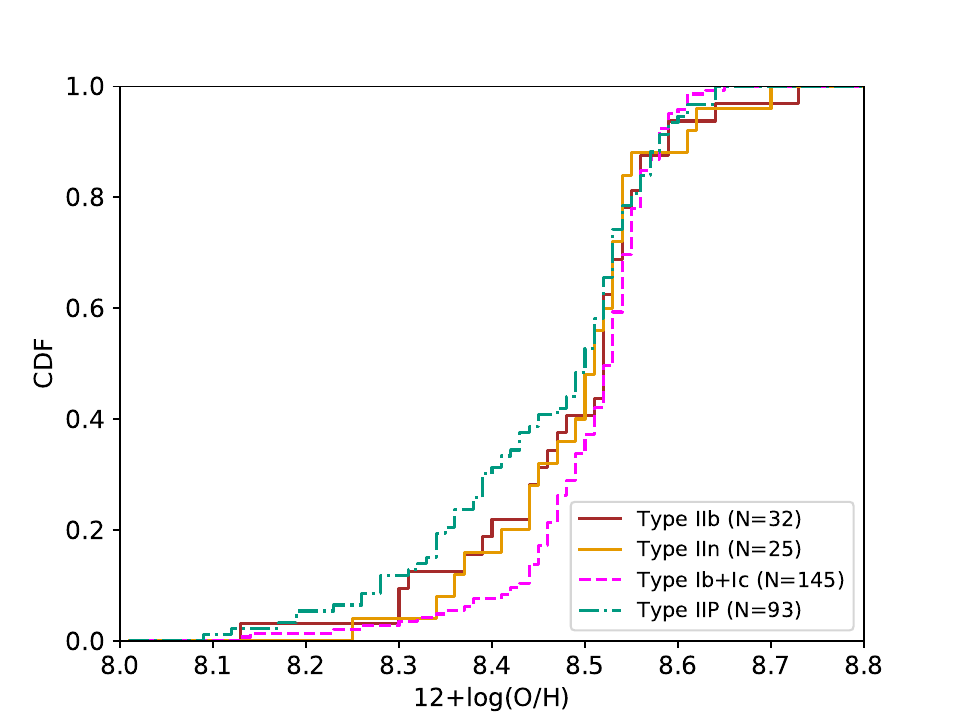} \label{fig_iibn_n2_cdf}}
\subfigure{\includegraphics[trim={0.3cm 0cm 1.35cm 1.1cm}, 
width=1.0\columnwidth]{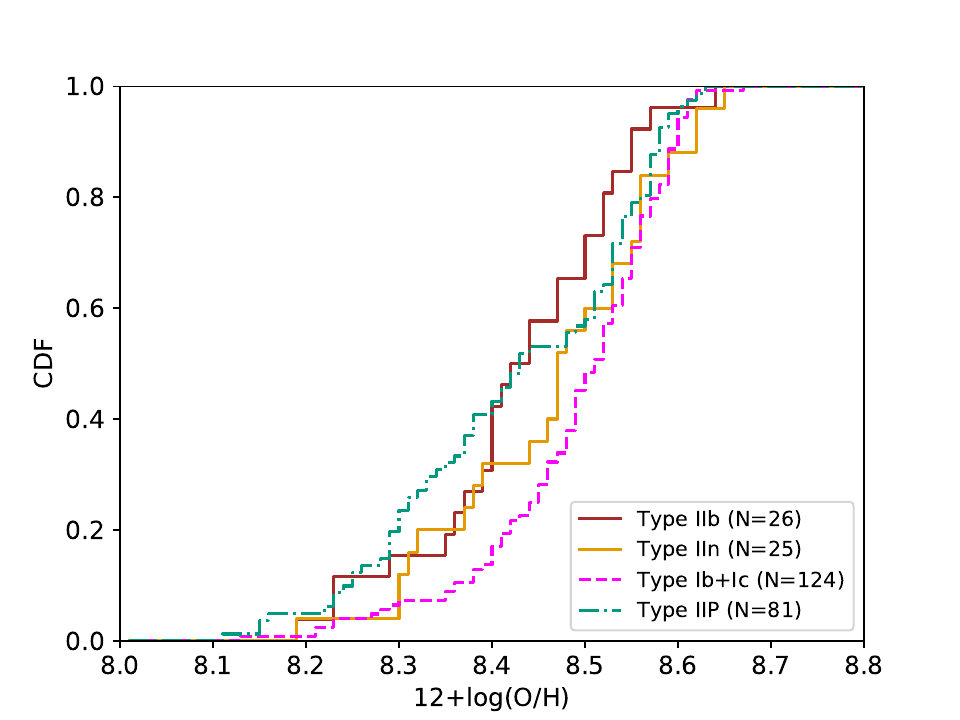} \label{fig_iibn_o3n2_cdf}}
\caption{Supplementary CDFs of the Type IIb (dark red) and Type IIn (gold) environment metallicities derived from archival environment spectra. The CDFs are created with M13-N2 (left) and M13-O3N2 (right) calibrations. The dashed lines are the CDFs of our combined Type Ib+Ic (magenta) and Type IIP (green) sample, respectively, for comparison. Binning width for CDF calculation: 0.0125 dex.}
\label{fig_iib_iin_cdfs}
\end{figure*}

\subsection{Archival Type IIb and Type IIn data} \label{sub_iib_iin}

We found spectral data of 32 Type IIb and 25 Type IIn CCSNe (see Table \ref{tab_iibiin_targets}) in the used archives. Beyond the initial scope of the project we provide the metallicities of these SNe in Table \ref{tab_iibiin_results}. Data reduction, emission line fitting and calibration for the evaluation of environment metallicities were the same as described in Section \ref{sect_dat}. Figure \ref{fig_iib_iin_cdfs} shows the CDFs of Type IIb and Type IIn SNe compared with CDFs of our Type IIP and combined Type Ib+Ic sample. The environment metallicities of IIb and IIn SNe are in between the IIP and Ib+Ic SNe for metallicities less than about 8.5 dex. No differences are visible for higher metallicities in case of M13-N2 calibration. For the M13-O3N2 calibration, the Type IIb metallicities tend to be lower than Type IIP metallicities starting at metallicities of about 8.45 dex. Taking into account the relatively low Type IIb number it is not clear that this tendency would hold with a larger sample of Type IIb SN. The application of statistical tests (Table \ref{tab_iib_iin_tests}) reveals no statistical significance to reject the null hypothesis that the samples are drawn from the same parent population.      

\subsection{Type Ic-BL SNe} \label {sub_ic_bl}
There are only 8 Ic-BL SN environment spectra within our survey parameters, so including them as a separate class would not yield any statistical results due to the low number statistics. Both Ic and Ic-BL are predicted to result from the death of massive stars and previous work has shown that Ic-BL prefer more metal-poor regions compared to Type Ic SN \citep{modjaz2011}, so a misclassification of a Ic as Ic-BL (or vice versa) could impact our results. To test this we compare the metallicities of our Ic SNe to the Ic-BL SN. Metallicity determination was possible for 7 out of the 8 Ic-BL SN and are presented in Figure \ref{fig_ic_bl} which shows that all of the Ic-BL lie between log(O/H)+12$\sim$8.3 to 8.6. We therefore conclude that Ic-BL do not represent Ic SNe at lower metallicity.

\subsection{Misclassification of SN} \label{sub_sn_mis}

Clean SNe classifications are crucial for this project. Our efforts to provide a clean sample are highlighted in Section \ref{sub_trgts} so we are confident that the classifications used here are accurate. The requirement of a clean Type IIP classification (rather than Type II only) removes any issues of Type IIb SN being misclassified as Type IIP as their light curves would not show the necessary plateau. Similarly, Type II SN with short-lived flash ionization could be misclassified as Type IIn SNe but their light curves would reveal their true nature and again this would not impact our Type IIP distribution. However, we would like to discuss three inherent classification issues debated in the literature, which may impact the distributions: a) the possibility of hidden helium in SNe spectra resulting in a Type Ic classification instead of a Type Ib, b) Type IIb to Ib transition and c) the distinction of Type IIP versus other Type II.

\begin{figure*}
\subfigure{\includegraphics[trim={0.3cm 0cm 1.35cm 1.1cm}, width=1.0
\columnwidth]{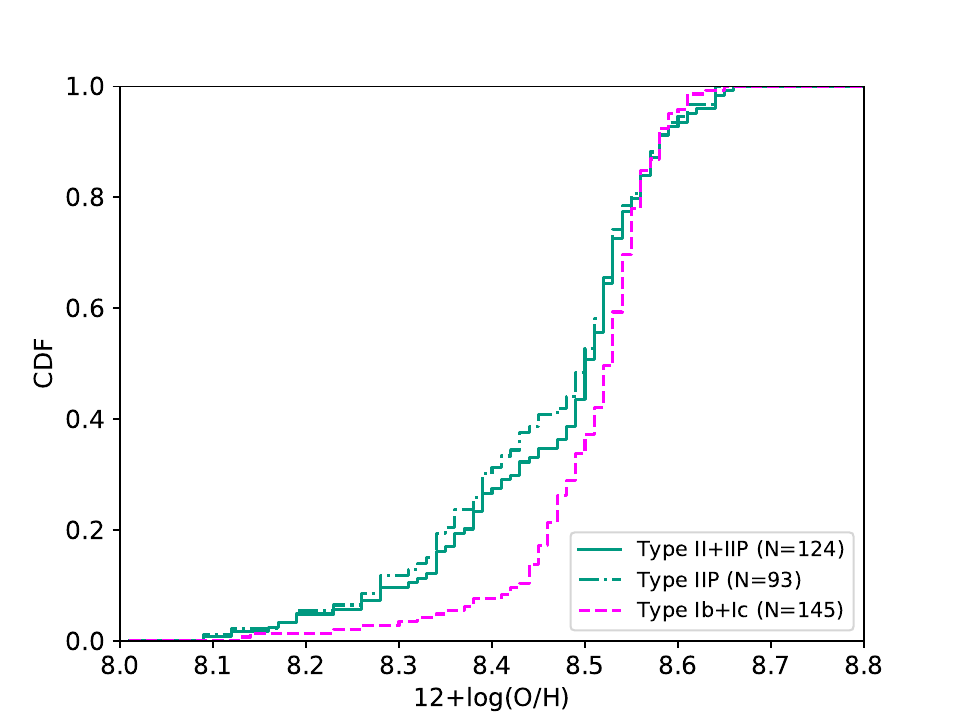} \label{fig_ext_n2_cdf}}
\subfigure{\includegraphics[trim={0.3cm 0cm 1.35cm 1.1cm}, 
width=1.0\columnwidth]{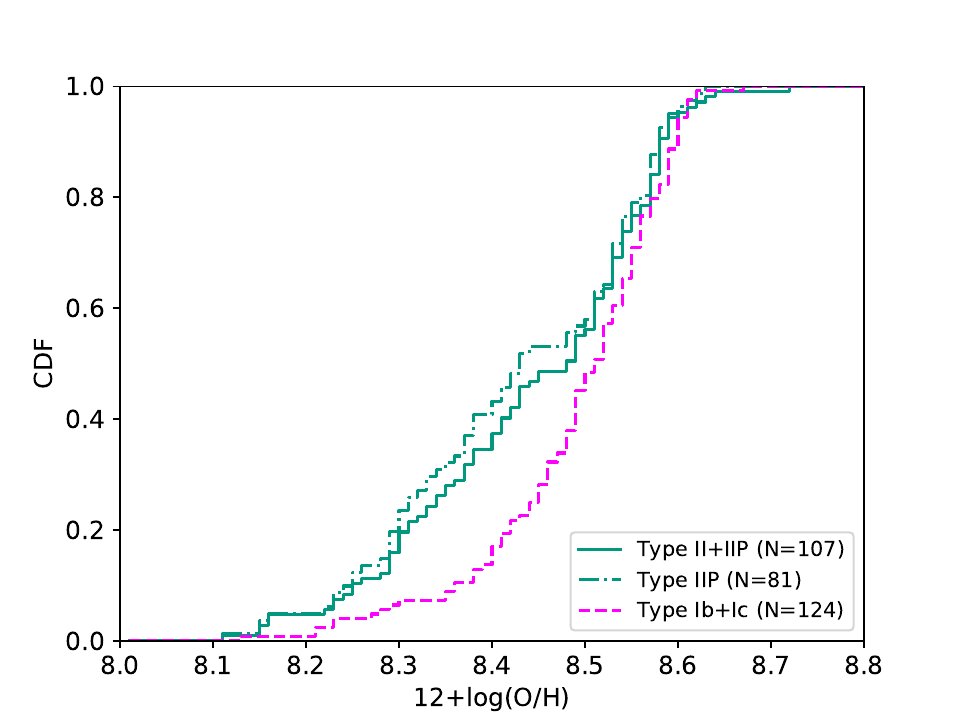} \label{fig_ext_o3n2_cdf}}
\caption{CDFs of the SNe environment metallicities measured with the M13-N2 (left) and M13-O3N2 (right) calibration for the extended Type IIP+II sample (green solid line) and the clean Type IIP CDF (green dashdot line). The magenta dashed line is the combined Type Ib plus Type Ic sample CDF for comparison. Binning width for CDF calculation: 0.0125 dex.}
\label{fig_ext_cdfs}
\end{figure*}

\subsubsection{Hidden Helium} \label{sub_iip_ii}

\citet{dessart2012} suggests that weak mixing of \textsuperscript{56}Ni into He-rich regions can hide the helium in the spectrum and mask a Type Ib SNe as a Type Ic SN. \citet{hachinger2012} uses non-LTE models with different mass He envelopes and create synthetic spectra to investigate the detection of the He\,{\sc i} 6678\AA~ and He\,{\sc i} 7065\AA~ lines. They conclude that with 0.14\,M$_{\odot}$ of helium present in the envelope, the optical helium lines can be detected. Their spectra suggest that He\,{\sc i} lines become stronger in later-time spectra ($\sim$ 30 days post-explosion) for Ib SN, whilst none are detected at any point in the Ic spectra. Based on this work we suggest that optical spectra for SNe Ib classification are best obtained $\sim$20 days post explosion and later. 

\begin{table}
\begin{threeparttable}
\caption{Supplementary p-values of the two-sample KS-test and two-sample AD-test for M13 results of Type IIb and Type IIn samples vs. Type Ib, Type Ic and Type IIP samples; for a description of p-value calculation see Section \ref{sub_cdfs}. There is no statistical significance in any way to reject the null hypothesis that any samples are drawn from the same parent population.}
\setlength{\tabcolsep}{0.151cm}
   \begin{tabular}{c|cc|cc}
   \hline
   \hline
      & \multicolumn{2}{c|}{KS-test} & \multicolumn{2}{c}{AD-test} \\
      \hline
      SN type & M13-N2 & M13-O3N2  & M13-N2 & M13-O3N2 \\
       & p-value & p-value  & p-value & p-value \\
      \hline
         IIb-Ib & 0.4376 & 0.0317 & 0.3327 & 0.0149 \\
         IIb-Ic & 0.1579 & 0.0182 & 0.1874 & 0.0070 \\
         IIb-IIP & 0.5309 & 0.5846 & 0.6053 & 0.4234 \\
         IIb-IIn & 0.9599 & 0.2977 & 0.9952 & 0.3409 \\
         IIn-Ib & 0.5822 & 0.5419 & 0.4710 & 0.3856 \\
         IIn-Ic & 0.1233 & 0.3231 & 0.1361 & 0.2763 \\
         IIn-IIP & 0.5057 & 0.3093 & 0.5042 & 0.3269 \\
      \hline
   \end{tabular}%
  \label{tab_iib_iin_tests}%
  \end{threeparttable}
\end{table}%

We checked the dates of the classification spectra of our Type Ib and Type Ic SNe where available and found a median value of 7.5 days post discovery with a range from 0 to 60 days. Taking into account some additional days between explosion and discovery, the classification spectra of many targets are in the range of the suggested 20 days but there is a large spread. Whilst hidden helium may incorrectly classify Ib as Ic SNe in a small minority of cases, our statistical tests provide no statistical significance to reject the null hypothesis that Ib and Ic progenitor metallicities are drawn from the same parent sample and a few misclassified Type Ib or Type Ic will not change this result. Importantly, the issue of hidden helium is not a factor in the analysis of Type IIP and Ib/Ic SNe as we combine the Ib and Ic samples. 

\subsubsection{Type IIb to Type Ib Transition}

Type IIb SNe are characterised by hydrogen lines in the early phase disappearing usually in timescale of months. Some Type IIb SNe (e.g. SN2011ei, \citealt{milisavljevic2013}, or SN2022crv, \citealt{gangopadhyay2023,dong2024}) exhibit very quick disappearance of the hydrogen features after the maximum. It is believed that the IIb progenitors have kept small amounts of hydrogen (e.g. \citealt{filippenko1997}) which is thinned with the rapid ejecta expansion and the underlying He-rich layers become visible. On the other hand, the progenitor of a Type Ib SNe does not have to lose its hydrogen completely to be classified as Type Ib (e.g. \citealt{deng2000, parrent2007,hachinger2012}). Consequently, a continuum between the Type IIb and Ib with a decreasing amount of residual hydrogen in the progenitor envelope has been suggested (e.g. \citealt{hachinger2012,yoon2017,prentice2017,sravan2019,gilkis2022}).

Type Ib with residual hydrogen and the transitional character of Type IIb SNe create the risk of mis-classifications. Some Type Ib classified SNe may be actually Type IIb SNe (late classification spectrum in the case of a quickly transitioning Type IIb) or some Type IIb may be actually Type Ib SNe (early classification spectrum of Type IIb with long lasting hydrogen phase). This can only be investigated by longer observation times with multiple classification spectra, which is not the case for our Type Ib sample. As mentioned above, the classification spectra of our sample are younger than 60 days after discovery and some of our Type IIb SNe may actually be Type Ib. This will not change our principal result of a strong statistical significance of different parent populations of Type IIP and Type Ib/Ic progenitors, but we can not exclude the possibility that a low-metallicity Type IIb is actually a low-metallicity Type Ib (see Figure \ref{fig_iib_iin_cdfs}). More work is needed to assess this possibility.   

\subsubsection{Ambiguous Type II classifications} \label{sub_iip_ii}

Type IIP and Type IIL are characterised by photometric features with a plateau (IIP) phase during declining and a rapid linear (IIL) declining light curve, respectively (\citealt{barbon1979,filippenko1997}), and they require a longer time monitoring (up to $\sim$100 days) to determine this. However, Type IIP is not well defined in terms of the period of the plateau phase, in which photometric band, how "constant" the plateau is and so on. Consequently, some studies (\citealt{anderson2014,sanders2015,valenti2016}) argue for a continuous distribution of Type II SNe properties and that no separate Type IIP / IIL classes exist. 

\begin{table}
\begin{threeparttable}
\caption{Change of p-values by including targets with ambiguous Type II (N=31 for M13-N2, N=26 for M13-O3N2) classification to the IIP sample. The p-values increase drastically with the ambiguous targets included; for a description of p-value calculation see Section \ref{sub_cdfs}.}
\setlength{\tabcolsep}{0.151cm}
   \begin{tabular}{cc|cc|cc}
   \hline
   \hline
   & & \multicolumn{2}{c|}{M13-N2} & \multicolumn{2}{c}{M13-O3N2} \\
   \hline
   SN type & test & vs. IIP & vs. IIP+II  & vs. IIP &  vs. IIP+II \\
   & & p-value & p-value & p-value & p-value \\
   \hline
      Ib & KS & 0.0012 & 0.0082 & 0.0057 & 0.0324 \\
      Ic & KS & 0.0078 & 0.0078 & 0.0019 & 0.0101 \\
      Ibc & KS & 0.00032 & 0.00259 & 0.00032 & 0.00322 \\
      Ib & AD & 0.0057 & 0.0263 & 0.0041 & 0.0226 \\
      Ic & AD & 0.0035 & 0.0091 & 0.0046 & 0.0187 \\
      Ibc & AD & 0.00060 & 0.00318 & 0.00046 & 0.00337 \\
   \hline
   \end{tabular}%
  \label{tab_ext_testresults}%
  \end{threeparttable}
\end{table}%

A continuous Type II distribution challenges the project design based on a clean Type IIP sample. Maybe such a strict requirement is needless because there exists no clean Type IIP population at all. We address this issue by extending our clean IIP sample and including environment metallicities of SNe with ambiguous Type II classifications for which we found archival environment spectra. These SN have a IIP classification alongside other classifications (e.g. Type IIb/IIn/IIL). This extended sample did not include any Type II SNe with solid classifications of IIb, IIn or IIL.

In total, this II+IIP sample extends the Type IIP sample by additional 29 (M13-N2) and 24 (M13-O3N2) targets, respectively. Figure \ref{fig_ext_cdfs} shows the change in the CDF; the CDF of the extended sample comes closer to the combined Type Ib+Ic sample CDF reducing the statistical "distance". This is echoed in the statistical test results in Table \ref{tab_ext_testresults} with p-values increasing typically by a factor of a few. We cannot draw any physical conclusion from this simple test, but it shows that it could be important to keep a clean SN subtype sample to test the behaviours of different subtypes with metallicity. This is echoed by the different distributions for Type IIb and IIn SNe in Figure \ref{fig_iib_iin_cdfs}. We note that, even with the extended Type IIP sample the statistical tests indicate a statistical significance (at a significance level better than 1\%) to reject the null hypothesis that the extended Type IIP sample and the combined Type Ib+Ic sample are drawn from the same parent population (see Table\,\ref{tab_ext_testresults}).

\subsection{Interpretation of statistical results}    \label{sub_test}

All statistical test results of this paper indicate at a high significance level of usually better than 1\% to reject the null hypothesis that the Type IIP samples and Type Ib and Ic samples are drawn from the same parent population. This is the first statistically significant result that supports the commonly discussed assumption of a different nature of Type IIP and Type Ib/Ic progenitors. 

Regarding the nature of Type Ib/Ic progenitors, two scenarios are under discussion: the single massive star and the close binary scenario with current preference for the close binary scenario based on stellar evolution models and observational evidence revealing a high binary fraction of massive stars (see e.g \citealt{sana2009, duchene2013, de_mink2014}). In Paper I we argued for higher fraction of single massive stars as Type Ib progenitors based on the significantly narrower Type Ib CDFs and lack of low metallicity Type Ib. Both arguments have weakened with the larger samples: the Type Ib CDFs are still narrower and the lowest metallicities are still higher than for IIP and Ic, but the differences have considerably dropped. Taking into account the (still) low number statistics in the low metallicity range even with the clearly larger samples, it is questionable if there is a difference between the CDFs in the low metallicity range at all. More data in the low metallicity range are required to answer this.

The overall shapes of the CDFs clearly indicate the preference of Type Ib and Type Ic to higher metallicity environments. This is consistent with the single massive star scenario, where the star loses its outer shell(s) by strong stellar winds (see e.g. \citealt{puls2008,vink2001,vink2021} and references therein), but is inconsistent with the lack of direct-detections of Ibc progenitors.  \citet{Smartt2009b} suggests this lack of detections support lower mass binary progenitors as the dominant channel of Type Ibc SNe. However, if there are only binary progenitors, it is difficult to explain the preference of Type Ib/Ic SNe to higher metallicity environments, which is simpler to explain by the higher mass single star scenario. In addition, the observed frequency of CCSNe subtypes, requires binary and single-star scenarios \citep{Smith2011}.

Models by \citet{Dessart2024} show that some SN subtypes can be a function of the initial separation of a binary system rather than initial mass or metallicity. This suggests that the shapes of the CDFs do not provide any distinction between single or binary scenarios. However \citet{Dessart2024} do not produce Ic SNe in their binary analysis and only run solar-metallicity models. Metallicity could potentially be the key to producing Type Ic supernovae and highlights the complexity of modelling required to understand the role of both binarity and metallicity simultaneously.
 
\citet{Xiao2019} compared SNe environment metallicities from 107 Type II and 45 Type Ibc SNe from the PISCO survey with predicted oxygen abundance from best-fitting BPASS stellar population models \citep{Eldridge2017} (that accounted for ionizing photon leakage; figure 9 in \citet{Xiao2019}). Whilst they do not produced CDFs, comparison of the observed and theoretical results suggest no clear monotonic relation between the range of metallicities, initial mass or ages for Type II and Ibc SNe. However, their Type II SNe sample include all subtypes. We show in Figure \ref{fig_iib_iin_cdfs} that Type IIb and IIn SNe have different distributions to IIP SNe, and demonstrate in Section \ref{sub_iip_ii} how inclusion of ambiguous Type II SNe can influence the IIP CDF. It would be interesting to see the results of a similar analysis using our own sample, without the contamination by ambiguous Type II SNe.

\section{Summary and Conclusions}   \label{sect_con}

This work evaluates the metallicities of environments of Type IIP, Type Ib and Type Ic SNe in host galaxies with redshift up to z$\sim$0.025 (corresponds roughly with CMB related Hubble distance of 100 Mpc). From our INT/IDS observations supplemented by archival MUSE, PISCO and MaNGA data we obtained the environment metallicities of 79 Type Ib, 66 Type Ic and 93 Type IIP SNe by the M13-N2 calibration (63 Type Ib, 61 Type Ic, 81 Type IIP for M13-O3N2 calibration). 

The CDFs of the environment metallicities show a clear distinction between Type IIP CDF and Type Ib/Ic CDFs up to a metallicity of about 8.5\,dex, which almost vanishes for higher metallicities. Mean values and standard deviations of the three SNe types are significantly different. The narrower distribution of Type Ib SNe seen in our Paper I does not hold with the significantly larger sample sizes. The Type Ib distribution still starts at higher metallicities than Type IIP/Ic distributions but the difference is small and, taking into account the low number of SNe in the low-metallicity range, not significant. 

We tested the null hypothesis that progenitors of different SNe types are taken from the same parent population, using KS- and AD-test. The test of Type IIP vs. Type Ib and Type Ic rejects the null hypothesis with a statistical significance level of better than 1\%. The significance level of Type IIP vs. combined Ib+Ic sample is even better with 0.1\% (see Table \ref{tab_m13testresults}). This work is the first to present statistically significant test results with a sufficient sample size. They give high statistical confidence of a different physical progenitor nature of Type IIP and Type Ib/Ic SNe.

The result testing Type Ib vs. Type Ic does not allow to reject the null hypothesis. We cannot derive any differences in the Type Ib and Type Ic progenitors from our results. The small difference between their CDFs suffers from low numbers of targets in the low metallicity range and more data of low-metallicity SNe environments are required to assess this difference.

We extend the study to include Type IIb and IIn SNe and analysis of the CDFs suggest that Type IIb SN could explain the missing Type Ib SN at low metallicity. However, due to low number statistics this is not a robust result and a larger sample of Type IIb SN needs to be collated for this to be tested further.

Our results challenge the stellar evolution and SNe explosion models to reproduce the clear distinction between Type IIP and Types Ib/Ic CDFs (see Figure \ref{fig_cdfs}) as a function of metallicity. Given our larger sample size, the results of the statistical tests are too robust to be ignored even when taking into account all deficiencies of strong emission line method and clean SNe classifications. 

\section*{Acknowledgements}

The work presented is based on observations made with the Isaac Newton Telescope. The Isaac Newton Telescope is operated on the island of La Palma by the Isaac Newton Group of Telescopes in the Spanish Observatorio del Roque de los Muchachos of the Instituto de Astrofísica de Canarias. We would like to thank Isaac Newton Group of Telescopes staff and our co-observers A. Brocklebank, T. Davison, A. de Burgos, L. Holden, D. Nespral, S.Percival, T. Wilson, and T. Zegmott for their contributions to the observation data. Thanks to Abigail Bell, student at University of Central Lancashire, for her contribution to the SNe classification reviews. J. L. Pledger, A. E. Sansom and S.M. Habergham-Mawson acknowledge financial support through the Panel for the Application of Telescope Time (PATT) travel grant (IDs ST/S005307/1 and ST/M00211X/1). 
We used archival data downloaded from the ESO science portal based on observations made with ESO Telescopes at the La Silla Paranal Observatory under programme IDs 0100.D-0341(A), 0101.B-0706(A), 0101.D-0748(A), 0103.A-0637(A), 0103.D-0440(A), 0104.D-0503(A), 094.B-0321(A), 095.B-0042(A), 095.B-0482(A), 095.B-0532(A), 095.B-0686(A), 095.D-0172(A), 096.B-0230(A), 096.D-0263(A), 096.D-0296(A), 097.B-0165(A), 097.B-0640(A), 097.D-0408(A), 099.B-0242(A), 099.D-0022(A), 106.2104.001, 106.2155.001, 106.21C7.001, 106.21C7.002, 1100.B-0651(A), 60.A-9194(A), 60.A-9301(A). 
We used archival data obtained by CALIFA survey at the Observatorio Astronomico De Calar Alto operated by the Centro Astronomica Hispano en Andalucia. 
We used archival data from MaNGA survey part of the SDSS-IV project. Funding for the SDSS IV has been provided by the Alfred P. Sloan Foundation, the U.S. Department of Energy Office of Science, and the Participating Institutions. SDSS acknowledges support and resources from the Center for High-Performance Computing at the University of Utah. The SDSS web site is www.sdss4.org. SDSS is managed by the Astrophysical Research Consortium for the Participating Institutions of the SDSS Collaboration including the Brazilian Participation Group, the Carnegie Institution for Science, Carnegie Mellon University, Center for Astrophysics | Harvard \& Smithsonian (CfA), the Chilean Participation Group, the French Participation Group, Instituto de Astrofísica de Canarias, The Johns Hopkins University, Kavli Institute for the Physics and Mathematics of the Universe (IPMU) / University of Tokyo, the Korean Participation Group, Lawrence Berkeley National Laboratory, Leibniz Institut für Astrophysik Potsdam (AIP), Max-Planck-Institut für Astronomie (MPIA Heidelberg), Max-Planck-Institut für Astrophysik (MPA Garching), Max-Planck-Institut für Extraterrestrische Physik (MPE), National Astronomical Observatories of China, New Mexico State University, New York University, University of Notre Dame, Observatório Nacional / MCTI, The Ohio State University, Pennsylvania State University, Shanghai Astronomical Observatory, United Kingdom Participation Group, Universidad Nacional Autónoma de México, University of Arizona, University of Colorado Boulder, University of Oxford, University of Portsmouth, University of Utah, University of Virginia, University of Washington, University of Wisconsin, Vanderbilt University, and Yale University.
We thank all known and unknown observers and technical staff of ESO, CALIFA and SDSS for their contributions to made the observations possible. 
Supernova data have been obtained the Bright Supernova pages of the Rochester Academy of Science supported by Purdue university, from Transient Name Server administrated by Ofer Yaron, Avner Sass, Nikola Knezevic, Eran Ofek and Avishay Gal-Yam, supported by The Weizmann Institute of Science, The Israeli Institute for Advanced Studies (IIAS) and The EU via ERC grant No. 725161, from The Astronomer's Telegram edited by R. E. Rutledge and Derek Fox and last but not least from the archival Open Supernova Catalogue created by James Guillochon and Jerod Parrent. Host galaxy data have been obtained from the NASA/IPAC Extragalactic Database (NED), which is operated by the Jet Propulsion Laboratory, California Institute of Technology, under contract with the National Aeronautics and Space Administration. The research has made use of tools created by hard work of the astronomical community, acknowledged here by tool name and URL: IRAF (https://iraf.net), Starlink (http://starlink.eao.hawaii.edu/starlink), Hyperleda (http://leda.univ-lyon1.fr), 'R'-project (http://www.R-project.org), PPXF (https://pypi.org/project/ppxf), MILES stellar library (http://miles.iac.es/pages/stellar-libraries/miles-library.php), MUSE Python Data Analysis Framework (MPDAF, https://mpdaf.readthedocs.io/en/latest/index.html, GLADE+ galaxy catalogue (https://glade.elte.hu/).

\section*{Data Availability}
The data underlying this article will be shared on reasonable request to the corresponding author. Spectroscopic data will be made available through \href{https://uclandata.uclan.ac.uk/}{https://uclandata.uclan.ac.uk/} in due course.



\bibliographystyle{mnras}
\bibliography{sn_ref}

\onecolumn
\section*{Appendix A.1} \label{appndx_a}

\setcounter{table}{0}
\renewcommand{\thetable}{A.\arabic{table}}
\begin{longtable}{@{\extracolsep{\fill}}ccC{3cm}C{2cm}ccccccc@{}}
\caption{Alphabetically ordered list of observed targets. The columns are the target name, target SN type, host galaxy name and morphology, redshift  luminosity distance D\textsubscript{L}, absolute magnitude M\textsubscript{B} of the host galaxies, host inclination, position angle PA of the SN with respect to galaxy centre, distance d\textsubscript{c} of the target from host centre and the instruments from which we obtained data (1=IDS, 2=MUSE ,3=BOSS/MaNGA,  4=PMAS/PPaK). Redshifts and morphologies have been taken from NED (if not marked otherwise), luminosity distances and absolute B-magnitudes have been taken from GLADE+ and inclination data have been taken from Hyperleda.} \\
\hline \hline
 target & type  & host  & host type & redshift & D\textsubscript{L}    &  M\textsubscript{B}   & Incl. &  PA   &d\textsubscript{C} & instr.\\
          &       &   &    &  &  [Mpc]     & [mag] & [\textdegree] & [\textdegree] &  [\arcsec]     &      \\
\hline
\endfirsthead
\caption[]{(continued)}\\
\hline \hline
 target & type  & host  & host type & redshift & D\textsubscript{L}    &  M\textsubscript{B}  & Incl.  &  PA   &d\textsubscript{C} & instr.\\
          &       &   &    &  &  [Mpc]     & [mag] & [\textdegree] & [\textdegree] &  [\arcsec]     &      \\
\hline
\endhead
\tabularnewline
ASASSN14bf & IIP & CGCG 103-030 & ? & 0.022412 & 93.0 & -19.5 & 33 & 287.9 & 5.71 & 2 \\
ASASSN14iz & IIP & ESO 462-G009 & (R')SB(r)a & 0.019277 & 87.1 & -20.3 & 59 & 306.5 & 27.43 & 2 \\
ASASSN14ma & IIP & SDSS J235509.00+101252.9 & ? & 0.013700$^{\star}$ & ? & ? & ? & 55.9 & 2.32 & 2 \\
ASASSN15bb & IIP & ESO 381-IG048 & WR: & 0.015878 & 71.9 & -19.7 & 59 & 77.4 & 10.24 & 2 \\
ASASSN15kz & IIP & IC 4303 & Sb & 0.008022 & 57.6 & -18.9 & 59 & 49.0 & 7.85 & 2 \\
ASASSN15ng & IIP & ESO 221-G012 & SBm? pec & 0.009840 & 41.9 & -19.6 & 90 & 342.5 & 28.28 & 2 \\
iPTF13bvn & Ib & NGC 5806 & SAB(s)b & 0.004493 & 23.1 & -19.9 & 60 & 185.4 & 35.35 & 2 \\
LSQ11jw & Ib & KUG 0202+005 & ? & 0.020002 & 92.1 & -18.0 & 90 & 93.6 & 3.16 & 3 \\
LSQ12fhh & IIP & UGC 02628 & SB(s)bc: & 0.022729 & 98.5 & -20.4 & 84 & 290.7 & 4.81 & 4 \\
LSQ12fnt & Ic & KUG 0330-013 & ? & 0.019000$^{\star}$ & 139.4 & -20.3 & 53 & 159.6 & 10.77 & 3 \\
LSQ15xp & IIP & PGC 035624 & ? & 0.012262 & 55.1 & -18.7 & 40 & 247.8 & 12.42 & 2 \\
PS15afa & IIP & NGC 3404 & SBab? & 0.015364 & 71.2 & -20.8 & 87 & 37.1 & 4.14 & 2 \\
PS15cer & Ib & NGC 7349 & SBb & 0.014950 & 68.2 & -19.7 & 76 & 353.0 & 12.59 & 2 \\
PTF11ixk & Ic & UGC 08399 & SB(r)b & 0.024088 & 103.9 & -20.7 & 56 & 209.6 & 10.92 & 3 \\
SN1961V & IIP & NGC 1058 & SA(rs)c & 0.001735 & 10.9 & -18.2 & 58 & 293.6 & 78.06 & 4 \\
SN1983V & Ib & NGC 1365 & SB(s)b & 0.005457 & 26.9 & -22.3 & 63 & 242.7 & 64.22 & 2 \\
SN1984L & Ib & NGC 0991 & SAB(rs)c & 0.005110 & 16.0 & -18.8 & 28 & 245.4 & 34.86 & 4 \\
SN1985F & Ib & NGC 4618 & SB(rs)m & 0.001815 & 14.7 & -19.6 & 58 & 32.8 & 4.17 & 3 \\
SN1988L & Ic & NGC 5480 & SA(s)c: & 0.006373 & 36.7 & -20.2 & 41 & 17.6 & 10.07 & 4 \\
SN1990aa & Ic & UGC 00540 & S? & 0.016591 & 84.4 & -20.7 & 58 & 125.5 & 14.65 & 1 \\
SN1990U & Ib & NGC 7479 & SB(s)c & 0.007925 & 36.1 & -21.8 & 43 & 202.1 & 67.27 & 1 \\
SN1991ar & Ib & IC 0049 & SAB(s)c & 0.015204 & 68.1 & -20.2 & 37 & 35.5 & 15.22 & 1 \\
SN1991N & Ic & NGC 3310 & SAB(r)bc & 0.003312 & 22.5 & -20.5 & 16 & 148.7 & 8.76 & 4 \\
SN1996an & IIP & NGC 1084 & SA(s)c & 0.004707 & 19.4 & -19.9 & 50 & 31.7 & 26.06 & 2 \\
SN1996aq & Ic & NGC 5584 & SAB(rs)cd & 0.005464 & 20.6 & -19.7 & 42 & 240.9 & 17.86 & 2 \\
SN1996D & Ic & NGC 1614 & SB(s)c & 0.015938 & 71.3 & -21.2 & 42 & 89.8 & 6.67 & 1 \\
SN1997dc & Ib & NGC 7678 & SAB(rs)c & 0.011638 & 53.0 & -21.6 & 43 & 46.5 & 9.76 & 1 \\
SN1997X & Ib & NGC 4691 & (R)SB(s)0/a & 0.003736 & 9.5 & -18.2 & 38 & 94.2 & 9.76 & 2 \\
SN1998dt & Ib & NGC 0945 & SB(rs)c & 0.014964 & 68.0 & -21.0 & 29 & 210.2 & 45.69 & 1 \\
SN1999bg & IIP & IC 0758 & SB(rs)cd: & 0.004256 & 24.6 & -18.5 & 65 & 240.3 & 36.52 & 4 \\
SN1999di & Ib & NGC 0776 & SAB(rs)b & 0.016408 & 82.2 & -21.0 & 18 & 163.4 & 17.84 & 4 \\
SN1999eh & Ib & NGC 2770 & SA(s)c: & 0.006494 & 28.1 & -20.8 & 82 & 239.2 & 15.22 & 1 \\
SN1999em & IIP & NGC 1637 & SAB(rs)c & 0.002392 & 7.7 & -18.0 & 31 & 226.0 & 23.76 & 1,4 \\
SN1999ex & Ic & IC 5179 & SA(rs)bc & 0.011405 & 50.0 & -21.6 & 62 & 233.4 & 27.35 & 2 \\
SN1999gi & IIP & NGC 3184 & SAB(rs)cd & 0.001941 & 13.8 & -20.3 & 14 & 357.9 & 61.65 & 1,4 \\
SN1999gn & IIP & NGC 4303 & SAB(rs)bc & 0.005224 & 16.2 & -20.8 & 18 & 141.2 & 50.64 & 2 \\
SN2000de & Ib & NGC 4384 & Sa & 0.008407 & 43.7 & -19.7 & 42 & 14.7 & 6.50 & 4 \\
SN2000ew & Ib & NGC 3810 & SA(rs)c & 0.003309 & 15.2 & -19.9 & 48 & 189.9 & 20.50 & 1 \\
SN2000F & Ic & IC 0302 & SB(rs)bc & 0.019694 & 77.3 & -21.9 & 51 & 67.0 & 23.38 & 1 \\
SN2000fn & Ib & NGC 2526 & S? & 0.015354 & 69.7 & -20.1 & 64 & 292.2 & 13.48 & 4 \\
SN2001B & Ib & IC 0391 & SA(s)c & 0.005194 & 23.2 & -19.2 & 18 & 240.5 & 6.67 & 1,4 \\
SN2001ch & Ic & PGC 066738 & S? & 0.009773 & 46.8 & -19.7 & 85 & 220.1 & 14.64 & 4 \\
SN2001ci & Ic & NGC 3079 & SB(s)c & 0.003689 & 19.0 & -20.9 & 90 & 351.7 & 27.69 & 1 \\
SN2001du & IIP & NGC 1365 & SB(s)b & 0.005457 & 26.9 & -22.3 & 63 & 266.2 & 88.26 & 2 \\
SN2001em & Ib & UGC 11794 & Sab & 0.019507 & 90.7 & -20.8 & 78 & 103.6 & 12.36 & 1 \\
SN2001fv & IIP & NGC 3512 & SAB(rs)c & 0.004580 & 22.1 & -18.7 & 32 & 225.3 & 23.86 & 2 \\
SN2001is & Ib & NGC 1961 & SAB(rs)c & 0.013122 & 55.5 & -22.8 & 47 & 153.9 & 53.03 & 1 \\
SN2001X & IIP & NGC 5921 & SB(r)bc & 0.004937 & 22.7 & -20.3 & 49 & 207.5 & 35.88 & 2 \\
SN2002cw & Ib & NGC 6700 & SB(rs)c & 0.015304 & 73.8 & -21.0 & 50 & 303.3 & 30.20 & 1 \\
SN2002dz & Ib & PGC 000902 & SA(rs)b pec? & 0.018026 & 84.6 & -21.0 & 74 & 204.5 & 11.55 & 1 \\
SN2002ee & IIP & NGC 5772 & SA(r)b: & 0.016245 & 73.7 & -21.2 & 73 & 25.3 & 44.56 & 4 \\
SN2002hh & IIP & NGC 6946 & SAB(rs)cd & 0.000133 & 14.5 & -22.9 & 18 & 207.5 & 129.81 & 1 \\
SN2002hn & Ic & NGC 2532 & SAB(rs)c & 0.017506 & 73.3 & -21.7 & 36 & 212.5 & 5.30 & 1,3 \\
SN2002ho & Ic & NGC 4210 & SB(r)b & 0.009036 & 45.8 & -20.5 & 46 & 132.8 & 17.81 & 4 \\
SN2002hy & Ib & NGC 3464 & SB(rs)c & 0.012465 & 55.3 & -21.0 & 51 & 328.2 & 22.03 & 2 \\
SN2002J & Ic & NGC 3464 & SB(rs)c & 0.012465 & 55.3 & -21.0 & 51 & 285.3 & 23.94 & 2 \\
SN2002ji & Ib & NGC 3655 & SA(s)c: & 0.005003 & 32.4 & -20.4 & 48 & 236.3 & 25.40 & 1,4 \\
SN2002jz & Ib & UGC 02984 & SBdm: & 0.005157 & 18.4 & -18.2 & 60 & 197.4 & 3.90 & 1 \\
SN2003bl & IIP & NGC 5374 & SB(r)bc? & 0.014483 & 61.7 & -21.2 & 37 & 130.1 & 19.89 & 2 \\
SN2003bn & IIP & PGC 831618 & ? & 0.012769 & 57.1 & -19.6 & 75 & 113.8 & 3.52 & 2 \\
SN2003bp & Ib & NGC 2596 & Sb & 0.019807 & 84.2 & -21.2 & 74 & 57.4 & 21.92 & 4 \\
SN2003E & IIP & ESO 485-G004 & Sc: & 0.014673 & 63.3 & -19.8 & 90 & 323.5 & 14.27 & 2 \\
SN2003el & Ic & NGC 5000 & SB(rs)bc & 0.018655 & 78.1 & -20.5 & 38 & 258.7 & 17.28 & 4 \\
SN2003gd & IIP & NGC 0628 & SA(s)c & 0.002192 & 9.5 & -20.0 & 20 & 175.4 & 160.83 & 1,4 \\
SN2003gk & Ib & NGC 7460 & SB(s)b pec & 0.010624 & 50.3 & -20.7 & 47 & 3.5 & 19.73 & 1 \\
SN2003hg & IIP & NGC 7771 & SB(s)a & 0.014460 & 66.4 & -21.8 & 67 & 247.3 & 10.38 & 2,4 \\
SN2003I & Ib & IC 2481 & S? & 0.017783 & 82.6 & -20.5 & 58 & 95.8 & 10.08 & 4 \\
SN2003ih & Ic & UGC 02836 & S0-: & 0.016702 & 69.9 & -21.1 & 46 & 82.3 & 9.84 & 4 \\
SN2003is & Ib & UGC 11430 & SA(s)c & 0.018286 & 84.5 & -19.8 & 15 & 314.7 & 5.68 & 1 \\
SN2003Z & IIP & NGC 2742 & SA(s)c: & 0.004310 & 21.2 & -20.0 & 61 & 346.1 & 32.85 & 1 \\
SN2004A & IIP & NGC 6207 & SA(s)c & 0.002842 & 25.3 & -20.3 & 65 & 305.5 & 27.27 & 1 \\
SN2004am & IIP & NGC 3034 & I0 & 0.000897 & 7.3 & -20.2 & 77 & 256.4 & 32.80 & 4 \\
SN2004ao & Ib & UGC 10862 & SB(rs)c & 0.005641 & 28.5 & -19.5 & 53 & 166.4 & 25.92 & 1 \\
SN2004aw & Ic & NGC 3997 & SBb & 0.015914 & 69.0 & -20.3 & 63 & 125.1 & 33.34 & 2 \\
SN2004bm & Ic & NGC 3437 & SAB(rs)c: & 0.004260 & 31.5 & -20.3 & 73 & 294.1 & 6.36 & 1 \\
SN2004bs & Ib & NGC 3323 & SB? & 0.017334 & 71.7 & -20.7 & 59 & 119.8 & 4.06 & 4 \\
SN2004dj & IIP & NGC 2403 & SAB(s)cd & 0.000445 & 5.6 & -20.3 & 61 & 94.1 & 159.16 & 1 \\
SN2004dk & Ib & NGC 6118 & SA(s)cd & 0.005247 & 24.0 & -20.5 & 69 & 6.2 & 43.36 & 1,2 \\
SN2004dn & Ic & UGC 02069 & SAB(s)d & 0.014267 & 44.0 & -19.6 & 56 & 181.3 & 26.92 & 1 \\
SN2004eu & Ic & PGC 008915 & ? & 0.021658 & 92.8 & -19.8 & 38 & 191.8 & 8.25 & 4 \\
SN2004fc & IIP & NGC 0701 & SB(rs)c & 0.006104 & 21.4 & -19.1 & 62 & 3.4 & 2.49 & 1 \\
SN2004ge & Ic & UGC 03555 & SAB(rs)bc & 0.015491 & 67.0 & -20.9 & 29 & 106.5 & 6.35 & 4 \\
SN2004gk & Ic & IC 3311 & Sdm: & -0.000457 & 19.7 & -18.0 & 90 & 26.8 & 3.25 & 1, \\
SN2004gn & Ib & NGC 4527 & SAB(s)bc & 0.005791 & 10.4 & -19.2 & 81 & 69.0 & 59.08 & 1 \\
SN2004gq & Ib & NGC 1832 & SB(r)bc & 0.006468 & 26.6 & -20.4 & 72 & 44.3 & 30.61 & 1 \\
SN2004gt & Ic & NGC 4038 & SB(s)m & 0.005417 & 19.4 & -20.5 & 52 & 256.1 & 38.60 & 1,2 \\
SN2004gv & Ib & NGC 0856 & (R’)SA(rs)0/a: & 0.020066 & 95.4 & -21.2 & 41 & 255.6 & 14.55 & 1 \\
SN2005ay & IIP & NGC 3938 & SA(s)c & 0.002695 & 18.7 & -20.5 & 18 & 194.8 & 58.13 & 1 \\
SN2005az & Ic & NGC 4961 & SB(s)cd & 0.008449 & 40.4 & -19.3 & 39 & 304.7 & 9.69 & 4 \\
SN2005bk & Ic & MCG +07-33-027 & S? & 0.024450 & 104.9 & -20.5 & 20 & 141.2 & 7.19 & 3 \\
SN2005bz & IIP & UGC 11162 & S & 0.015414 & 71.8 & -19.4 & 65 & 58.1 & 14.58 & 4 \\
SN2005cs & IIP & NGC 5194 & SA(s)bc & 0.001534 & 12.6 & -21.8 & 33 & 179.4 & 66.91 & 1 \\
SN2005kl & Ic & NGC 4369 & (R)SA(rs)a & 0.003386 & 18.1 & -19.5 & 19 & 308.1 & 7.66 & 1 \\
SN2005V & Ic & NGC 2146 & SB(s)ab & 0.002975 & 16.8 & -20.2 & 37 & 26.2 & 3.90 & 1 \\
SN2006be & IIP & IC 4582 & S? & 0.007497 & 37.7 & -18.8 & 83 & 270.1 & 5.95 & 2,4 \\
SN2006bp & IIP & NGC 3953 & SB(r)bc & 0.003502 & 16.2 & -21.0 & 62 & 33.9 & 112.05 & 1 \\
SN2006dn & Ib & UGC 12188 & S+COMP & 0.017179 & 85.3 & -20.0 & ? & 67.4 & 5.24 & 1 \\
SN2006ei & Ic & NGC 0735 & Sb & 0.015441 & 75.0 & -20.7 & 88 & 275.7 & 10.35 & 1 \\
SN2006F & Ib & NGC 0935 & Scd: & 0.013916 & 57.9 & -21.3 & 53 & 10.5 & 16.99 & 1 \\
SN2006fo & Ib & UGC 02019 & S? & 0.020392 & 97.0 & -20.6 & 48 & 275.7 & 6.03 & 1,2,3 \\
SN2006lc & Ib & NGC 7364 & S0/a & 0.016161 & 77.7 & -21.3 & 52 & 173.4 & 10.46 & 2,4 \\
SN2006lv & Ib & UGC 06517 & Sbc & 0.008330 & 40.7 & -19.0 & 57 & 45.1 & 15.29 & 4 \\
SN2006my & IIP & NGC 4651 & SA(rs)c & 0.002669 & 10.0 & -18.6 & 49 & 230.9 & 35.04 & 2 \\
SN2006ov & IIP & NGC 4303 & SAB(rs)bc & 0.005224 & 16.2 & -20.8 & 18 & 6.6 & 51.94 & 2 \\
SN2007aa & IIP & NGC 4030 & SA(s)bc & 0.004863 & 32.1 & -20.7 & 47 & 41.6 & 91.80 & 1 \\
SN2007ag & Ib & UGC 05392 & Scd: & 0.020711 & 87.3 & -19.6 & 90 & 11.6 & 15.32 & 4 \\
SN2007C & Ib & NGC 4981 & SAB(r)bc & 0.005597 & 20.1 & -19.7 & 45 & 159.1 & 23.43 & 1,2 \\
SN2007fo & Ib & NGC 7714 & SB(s)b: pec & 0.009333 & 44.9 & -20.1 & 45 & 351.3 & 11.92 & 1 \\
SN2007gr & Ic & NGC 1058 & SA(rs)c & 0.001735 & 10.9 & -18.2 & 58 & 303.4 & 28.86 & 1,4 \\
SN2007kj & Ib & NGC 7803 & S0/a & 0.017766 & 87.7 & -21.3 & 60 & 209.9 & 11.42 & 4 \\
SN2007rz & Ic & NGC 1590 & pec & 0.012999 & 52.0 & -20.1 & 28 & 88.2 & 9.22 & 2,4 \\
SN2008bs & Ib & UGC 04085 & S? & 0.024692 & 113.7 & -20.9 & 42 & 215.9 & 11.61 & 1 \\
SN2008cn & IIP & NGC 4603 & SA(s)c: & 0.008647 & 28.3 & -20.9 & 45 & 12.6 & 23.45 & 2 \\
SN2008D & Ib & NGC 2770 & SA(s)c: & 0.006494 & 28.1 & -20.8 & 82 & 325.3 & 67.58 & 1 \\
SN2008dv & Ic & NGC 1343 & SAB(s)b: pec & 0.007375 & 31.9 & -20.0 & 67 & 57.3 & 9.24 & 1 \\
SN2008ew & Ic & IC 1236 & SAB(s)c & 0.020087 & 88.5 & -20.9 & 25 & 312.2 & 12.94 & 3 \\
SN2008gj & Ic & NGC 7321 & SB(r)b & 0.023843 & 115.4 & -21.4 & 49 & 11.8 & 37.60 & 4 \\
SN2008hn & Ib & NGC 2545 & (R)SB(r)ab & 0.011291 & 49.8 & -20.5 & 61 & 9.3 & 18.30 & 1 \\
SN2008X & IIP & NGC 4141 & SBcd: & 0.006354 & 32.6 & -18.3 & 51 & 60.2 & 9.03 & 1 \\
SN2009B & IIP & UGC 04423 & Sd & 0.011701 & 72.2 & -19.6 & 60 & 227.1 & 10.00 & 4 \\
SN2009em & Ic & NGC 0157 & SAB(rs)bc & 0.005500 & 24.2 & -21.2 & 62 & 252.6 & 34.68 & 1 \\
SN2009ga & IIP & NGC 7678 & SAB(rs)c & 0.011638 & 53.0 & -21.6 & 43 & 211.2 & 30.01 & 1 \\
SN2009ha & Ib & PGC 009923 & SA(s)c pec: & 0.014750 & 60.4 & -20.3 & 42 & 335.4 & 15.41 & 1 \\
SN2009ie & IIP & NGC 1093 & SABab? & 0.017612 & 88.5 & -20.9 & 53 & 236.6 & 33.37 & 4 \\
SN2009ii & IIP & UGC 03627 & Sd & 0.020954 & 93.3 & -20.7 & 49 & 166.7 & 21.29 & 4 \\
SN2009iz & Ib & UGC 02175 & SAB(s)bc & 0.014196 & 109.4 & -20.3 & 39 & 320.4 & 18.95 & 4 \\
SN2009je & Ic & UGC 03312 & S & 0.018970 & 76.7 & -20.0 & 71 & 240.4 & 5.27 & 4 \\
SN2009js & IIP & NGC 0918 & SAB(rs)c: & 0.005033 & 14.7 & -18.7 & 58 & 240.4 & 41.89 & 1 \\
SN2009md & IIP & NGC 3389 & SA(s)c & 0.004346 & 25.1 & -20.1 & 66 & 277.9 & 24.09 & 2 \\
SN2009mi & Ic & IC 2151 & SB(s)bc: & 0.010377 & 45.6 & -19.8 & 61 & 65.6 & 28.24 & 2 \\
SN2010do & Ic & NGC 5374 & SB(r)bc? & 0.014483 & 61.7 & -21.2 & 37 & 268.3 & 22.68 & 2 \\
SN2010F & IIP & NGC 3120 & SAB(s)bc: & 0.009300 & 41.8 & -20.3 & 47 & 249.6 & 26.34 & 2 \\
SN2010ie & IIP & NGC 2333 & Sa & 0.015781 & 66.0 & -20.5 & 49 & 91.8 & 29.07 & 4 \\
SN2010io & Ic & UGC 04543 & SAdm & 0.006538 & 30.7 & -18.8 & 45 & 331.8 & 9.52 & 1 \\
SN2010iz & IIP & UGC 03552 & SBcd: & 0.016228 & 63.9 & -19.9 & 71 & 321.6 & 8.07 & 4 \\
SN2010jw & Ic & IC 2394 & SB(s)b & 0.021152 & 87.0 & -19.7 & 31 & 360.0 & 5.81 & 3 \\
SN2010kc & Ib & NGC 7624 & Scd: & 0.014260 & 66.5 & -21.2 & 49 & 289.5 & 3.96 & 1 \\
SN2010ln & Ib & UGC 02685 & SAB(s)b & 0.016802 & 72.0 & -20.8 & 60 & 95.8 & 21.20 & 1 \\
SN2011bh & Ic & NGC 2431 & (R')SB(s)a: & 0.018943 & 81.1 & -20.2 & 27 & 174.0 & 19.81 & 4 \\
SN2011bm & Ic & IC 3918 & ? & 0.022192 & 88.5 & -19.8 & 45 & 56.6 & 5.81 & 4 \\
SN2011cj & IIP & UGC 09356 & S? & 0.007409 & 26.7 & -17.9 & 60 & 30.1 & 8.21 & 4 \\
SN2011ck & IIP & NGC 5425 & Sd & 0.006931 & 36.6 & -19.8 & 77 & 297.7 & 16.07 & 1,4 \\
SN2011cl & IIP & IC 2373 & Scd? & 0.025094 & 111.6 & -19.8 & 36 & 3.8 & 38.48 & 4 \\
SN2011dq & IIP & NGC 0337 & SB(s)d & 0.005504 & 26.2 & -20.6 & 51 & 300.1 & 40.23 & 1 \\
SN2011fl & Ib & IC 1584 & SB? & 0.015818 & 78.7 & -19.7 & 30 & 102.9 & 18.37 & 1 \\
SN2011gd & Ib & NGC 6186 & (R')SB(s)a & 0.009795 & 50.6 & -19.3 & 71 & 65.8 & 2.91 & 4 \\
SN2011it & Ic & PGC 067911 & S? & 0.015888 & 76.7 & -19.8 & 74 & 49.4 & 6.90 & 4 \\
SN2011jk & IIP & UGC 03843 & Scd: & 0.017602 & 70.9 & -19.5 & 71 & 322.4 & 17.31 & 4 \\
SN2012A & IIP & NGC 3239 & IB(s)m & 0.002367 & 13.0 & -18.4 & 47 & 133.7 & 49.59 & 1 \\
SN2012au & Ib & NGC 4790 & SB(rs)c:? & 0.004483 & 17.0 & -18.9 & 59 & 60.7 & 4.06 & 1,2 \\
SN2012bu & IIP & NGC 3449 & SA(s)ab: & 0.010928 & 45.2 & -20.7 & 90 & 128.1 & 45.89 & 2 \\
SN2012C & Ic & NGC 2926 & S? & 0.014540 & 63.2 & -20.0 & 27 & 289.7 & 7.09 & 4 \\
SN2012dx & Ib & PGC 063016 & ? & 0.014844 & 73.2 & -19.6 & 68 & 12.4 & 18.10 & 4 \\
SN2012ec & IIP & NGC 1084 & SA(s)c & 0.004707 & 19.4 & -19.9 & 50 & 358.4 & 15.50 & 1,2 \\
SN2012ej & Ic & IC 2166 & SAB(s)bc & 0.008946 & 39.1 & -20.4 & 63 & 291.9 & 38.45 & 1 \\
SN2012ex & Ib & UGC 00838 & S? & 0.022949 & 109.9 & -21.3 & 42 & 21.1 & 4.82 & 1 \\
SN2012gi & Ic & UGC 11922 & Sc & 0.024190 & 119.0 & -19.2 & 38 & 211.0 & 39.31 & 4 \\
SN2012ho & IIP & MCG -01-57-021 & SB(s)bc & 0.009907 & 48.2 & -20.0 & 52 & 104.8 & 29.30 & 2 \\
SN2013ab & IIP & NGC 5669 & SAB(rs)cd & 0.004563 & 20.1 & -19.4 & 57 & 132.0 & 19.88 & 1 \\
SN2013bu & IIP & NGC 7331 & SA(s)b & 0.002722 & 19.8 & -21.8 & 70 & 204.2 & 55.60 & 1 \\
SN2013ce & IIP & PGC 035860 & ? & 0.021081 & 84.5 & -17.1 & 68 & 324.9 & 9.27 & 4 \\
SN2013cf & IIP & PGC 013858 & (R')SAB(s)a: & 0.019249 & 102.4 & -20.8 & 56 & 226.1 & 6.35 & 4 \\
SN2013dk & Ic & NGC 4038 & SB(s)m & 0.005417 & 19.4 & -20.5 & 52 & 195.4 & 15.46 & 1,2 \\
SN2013ej & IIP & NGC 0628 & SA(s)c & 0.002192 & 9.5 & -20.0 & 20 & 134.3 & 129.22 & 4 \\
SN2013ff & Ic & NGC 2748 & SAbc & 0.004923 & 23.6 & -19.9 & 68 & 215.5 & 25.04 & 1 \\
SN2013ge & Ic & NGC 3287 & SB(s)d & 0.004420 & 24.5 & -19.3 & 75 & 18.6 & 50.75 & 1 \\
SN2013gl & Ib & IC 0701 & ? & 0.020474 & 81.7 & -20.6 & 47 & 306.9 & 3.51 & 2,4 \\
SN2014A & IIP & NGC 5054 & SA(s)bc & 0.005811 & 15.1 & -19.9 & 57 & 55.8 & 15.12 & 1 \\
SN2014az & IIP & NGC 7691 & SAB(rs)bc & 0.013479 & 63.1 & -20.4 & 37 & 308.9 & 29.29 & 4 \\
SN2014bc & IIP & NGC 4258 & SAB(s)bc & 0.001494 & 9.3 & -21.3 & 68 & 144.5 & 3.68 & 1 \\
SN2014bi & IIP & NGC 4096 & SAB(rs)c & 0.001908 & 12.1 & -20.1 & 81 & 20.2 & 54.47 & 1 \\
SN2014C & Ib & NGC 7331 & SA(s)b & 0.002722 & 19.8 & -21.8 & 70 & 140.7 & 31.03 & 1 \\
SN2014cv & IIP & UGC 10123 & Sab & 0.012506 & 59.2 & -19.8 & 90 & 78.5 & 6.03 & 4 \\
SN2014cx & IIP & NGC 0337 & SB(s)d & 0.005504 & 26.2 & -20.6 & 51 & 303.3 & 40.23 & 1 \\
SN2014cy & IIP & NGC 7742 & SA(r)b & 0.005534 & 25.6 & -19.9 & 17 & 21.9 & 11.86 & 2 \\
SN2014dj & Ic & NGC 0317B & SB? & 0.017792 & 90.5 & -22.0 & 65 & 305.6 & 3.59 & 1 \\
SN2014eh & Ic & NGC 6907 & SB(s)bc & 0.010571 & 46.6 & -21.9 & 37 & 223.1 & 55.19 & 2 \\
SN2014ei & Ib & MCG -01-13-050 & Sb pec & 0.014440 & 53.2 & -20.9 & 90 & 262.2 & 15.57 & 1 \\
SN2014L & Ic & NGC 4254 & SA(s)c & 0.008026 & 16.6 & -20.8 & 20 & 220.0 & 20.78 & 2 \\
SN2015ap & Ib & IC 1776 & SB(s)d & 0.011378 & 44.8 & -19.1 & 38 & 240.2 & 32.65 & 2 \\
SN2015bb & Ic & NGC 5772 & SA(r)b: & 0.016245 & 73.7 & -21.2 & 73 & 244.9 & 13.21 & 4 \\
SN2015dj & Ib & NGC 7371 & (R)SA(r)0/a: & 0.008966 & 43.9 & -21.1 & 25 & 50.6 & 24.76 & 1,2 \\
SN2015V & IIP & UGC 11000 & S? & 0.004566 & 23.5 & -18.1 & 70 & 144.6 & 8.36 & 1 \\
SN2016aqf & IIP & NGC 2101 & IB(s)m & 0.004016 & 20.9 & -18.0 & 69 & 266.2 & 2.40 & 2 \\
SN2016B & IIP & CGCG 012-116 & S0? & 0.004336 & 19.4 & -16.4 & 53 & 245.5 & 10.88 & 2 \\
SN2016bau & Ib & NGC 3631 & SA(s)c & 0.003856 & 20.6 & -20.6 & 35 & 293.7 & 37.81 & 1 \\
SN2016blb & IIP & WISEA J113720.57-045443.9 & ? & 0.018300 & 77.7 & -19.6 & ? & 12.1 & 8.59 & 2 \\
SN2016bmi & IIP & IC 4721 & SB(s)cd: & 0.007455 & 16.9 & -19.6 & 75 & 153.6 & 130.68 & 2 \\
SN2016C & IIP & NGC 5247 & SA(s)bc & 0.004523 & 18.1 & -20.5 & 38 & 16.8 & 111.95 & 2 \\
SN2016ccm & IIP & IC 0983 & SB(r)bc & 0.018156 & 70.8 & -21.8 & 31 & 323.7 & 131.93 & 2 \\
SN2016cok & IIP & NGC 3627 & SAB(s)b & 0.002405 & 11.4 & -20.8 & 68 & 118.9 & 67.96 & 2 \\
SN2016css & IIP & ESO 400-G005 & SA(s)c & 0.020001 & 92.9 & -21.9 & 65 & 136.5 & 39.85 & 2 \\
SN2016I & IIP & UGC 09450 & Sdm: & 0.015003 & 60.1 & -18.4 & 90 & 240.6 & 9.01 & 2 \\
SN2016iae & Ic & NGC 1532 & SB(s)b & 0.003468 & 25.7 & -21.4 & 83 & 19.5 & 45.27 & 2 \\
SN2016X & IIP & UGC 08041 & SB(s)d & 0.004416 & 10.0 & -17.7 & 54 & 144.8 & 74.12 & 2 \\
SN2017cio & IIP & UGC 03944 & Scd: & 0.013002 & 55.9 & -19.8 & 72 & 322.6 & 30.91 & 4 \\
SN2017ein & Ic & NGC 3938 & SA(s)c & 0.002695 & 18.7 & -20.5 & 18 & 74.3 & 42.51 & 1 \\
SN2017ewx & Ib & NGC 5418 & SB? & 0.015179 & 61.4 & -20.2 & 69 & 220.3 & 24.38 & 2 \\
SN2017fem & IIP & IC 4452 & E? & 0.014212 & 61.2 & -19.4 & 21 & 211.4 & 2.81 & 2 \\
SN2017fvf & IIP & NGC 1285 & (R')SB(r)b & 0.017512 & 76.2 & -21.3 & 59 & 191.4 & 9.05 & 2 \\
SN2017gat & Ic & LEDA 686296 & ? & 0.023100 & 91.2 & -17.6 & 46 & 49.6 & 3.64 & 2 \\
SN2017gry & IIP & ESO 155-G036 & SO(r?) & 0.019337 & 91.4 & -19.8 & 82 & 153.8 & 7.12 & 2 \\
SN2017hcb & Ib & CGCG 505-019 & ? & 0.016185 & 78.6 & -20.1 & 59 & 187.0 & 6.32 & 1 \\
SN2017hyf & Ib & UGC 01652 & Sd & 0.017142 & 86.0 & -20.0 & 67 & 41.8 & 9.02 & 1 \\
SN2017pn & IIP & LEDA 959170 & ? & 0.014000$^{\star}$  & ? & ? & 62 & 285.1 & 2.89 & 2 \\
SN2017rt & Ic & NGC 3836 & Sb & 0.012208 & 56.1 & -20.5 & 40 & 311.0 & 10.27 & 2 \\
SN2018cew & Ib & NGC 7775 & Scd: & 0.022549 & 108.4 & -21.7 & 34 & 338.8 & 9.09 & 1 \\
SN2018cho & IIP & IC 0004 & S? & 0.016688 & 81.8 & -21.2 & 46 & 324.3 & 10.30 & 2 \\
SN2018ec & Ic & NGC 3256 & pec & 0.009354 & 45.2 & -22.2 & 48 & 323.1 & 9.00 & 2 \\
SN2018gsk & Ic & NGC 1517 & Scd: & 0.011615 & 44.3 & -20.2 & 34 & 225.7 & 7.46 & 1 \\
SN2018hfc & IIP & UGC 06103 & pec & 0.019996 & 82.6 & -20.6 & 32 & 247.9 & 4.68 & 3 \\
SN2018yo & IIP & UGC 07840 & SAB(s)d: & 0.013385 & 59.0 & -18.6 & 58 & 44.0 & 8.63 & 2 \\
SN2019ccm & Ib & UGC 03110 & SBcd: & 0.014950 & 70.5 & -20.8 & 67 & 290.0 & 14.62 & 1 \\
SN2019cvz & IIP & WISEA J163054.20+463524.5 & Sd & 0.018346 & 81.0 & -17.9 & ? & 190.5 & 6.23 & 3 \\
SN2019tls & Ib & NGC 2514 & SB(s)bc: & 0.016177 & 67.9 & -20.6 & 20 & 170.6 & 24.70 & 1 \\
SN2019yvr & Ib & NGC 4666 & SABc: & 0.005047 & 11.7 & -19.4 & 70 & 325.9 & 12.04 & 2 \\
SN2019yxp & Ib & IC 0995 & Sdm: & 0.010424 & 53.1 & -20.2 & 90 & 133.0 & 11.14 & 4 \\
SN2020aaxs & Ib & NGC 5394 & SB(s)b & 0.011501 & 54.1 & -20.7 & 71 & 105.8 & 3.59 & 4 \\
SN2020eai & Ib & UGC 03223 & SBa & 0.015621 & 69.8 & -21.1 & 62 & 247.8 & 21.63 & 1 \\
SN2020jfo & IIP & NGC 4303 & SAB(rs)bc & 0.005224 & 16.2 & -20.8 & 18 & 293.7 & 72.16 & 2 \\
SN2020ksa & Ib & NGC 3478 & SB(rs)bc & 0.022165 & 90.8 & -21.9 & 69 & 37.5 & 9.90 & 3 \\
SN2020lid & Ib & MCG -01-02-001 & SB(s)ab pec? & 0.012372 & 56.6 & -19.9 & 71 & 100.2 & 11.06 & 1 \\
SN2020nac & Ib & UGC 00460 & Scd: & 0.017392 & 88.1 & -21.5 & 56 & 129.7 & 12.73 & 1 \\
SN2020oi & Ic & NGC 4321 & SAB(s)bc & 0.005240 & 12.2 & -20.4 & 24 & 12.6 & 6.62 & 2 \\
SN2020rur & Ic & UGC 03224 & Sb & 0.015644 & 62.2 & -20.8 & 39 & 34.9 & 18.51 & 1 \\
SN2020sgf & Ic & UGC 02700 & SBb? & 0.022009 & 107.8 & -21.3 & 81 & 309.9 & 15.70 & 1 \\
SN2020yyz & IIP & NGC 0976 & SA(rs)c: & 0.014330 & 60.0 & -21.2 & 21 & 170.4 & 10.10 & 4 \\
SN2021aexi & Ic & NGC 7771 & SB(s)a & 0.014460 & 66.4 & -21.8 & 67 & 258.0 & 35.57 & 2 \\
SN2021afuq & Ic & NGC 3256 & pec & 0.009354 & 45.2 & -22.2 & 48 & 56.4 & 11.42 & 2 \\
SN2021kos & Ib & IC 0719 & S0? & 0.006114 & 33.2 & -19.1 & 90 & 74.5 & 7.69 & 2 \\
SN2021ocs & Ic & NGC 7828 & RING & 0.019110 & 89.8 & -20.8 & 90 & 79.7 & 6.08 & 2 \\
SN2021qip & Ib & UGC 01420 & S? & 0.015587 & 75.4 & -20.1 & 62 & 17.4 & 9.67 & 1 \\
SN2021rfs & Ib & UGC 11946 & SAB(s)c & 0.018446 & 92.5 & -22.0 & 54 & 213.2 & 7.57 & 1 \\
SN2021vnw & Ib & UGC 03596 & S0? & 0.017166 & 70.7 & -20.9 & 24 & 280.7 & 9.97 & 1 \\
SN2021zju & Ib & NGC 0009 & Sb: pec & 0.015104 & 70.8 & -19.8 & 68 & 258.5 & 6.72 & 1 \\
SN2022aang & IIP & UGC 04140 & Sbc & 0.015711 & 62.4 & -20.2 & 90 & 307.7 & 6.90 & 4 \\
SN2022adui & Ic & LEDA 176224 & Sb & 0.025908 & 113.9 & -19.6 & 56 & 244.3 & 4.38 & 3 \\
SN2022ibn & Ic & CGCG 130-008 & S? & 0.024157 & 93.0 & -20.0 & 43 & 293.8 & 7.95 & 3 \\
\hline
\multicolumn{11}{l}{$^{\star}$ Redshift of the SN taken from SN catalogues} \\   
\label{tab_targets}
\end{longtable}

\begin{longtable}{@{\extracolsep{\fill}}ccccccc@{}}
\caption{Alphabetically ordered list of metallicities of the observed SNe environments. The columns are target name, target SN type, reference for type classification, J2000.0 target coordinates and measured environment metallicities by N2 and O3N2 method based on the M13 calibration. The uncertainties are dominated by the calibration uncertainties in all cases, which are $\pm$0.16 dex for M13-N2, $\pm$0.18 dex for M13-O3N2 (all uncertainties are 1$\sigma$ values as given in \citealt{marino2013} and and correspond to the uncertainty on the final metallicity). The metallicities for the older PP04 calibration may be calculated from M13 values using equations \ref{m13topp04_n2} and \ref{m13topp04_o3n2}, respectively.} \\ %
\hline \hline
    target & type & type reference & SN RA   & SN Dec & M13-N2 & M13-O3N2 \\
          &       &                & [h m s] & [\textdegree\,\arcmin\,\arcsec] & 12+log(O/H) & 12+log(O/H) \\
\hline
\endfirsthead
\caption[]{(continued)}\\
\hline \hline
    target & type & type reference & SN RA   & SN Dec & M13-N2 & M13-O3N2 \\
          &       &                & [h m s] & [\textdegree\,\arcmin\,\arcsec] & 12+log(O/H) & 12+log(O/H) \\
\hline
\endhead
\tabularnewline
ASASSN14bf & IIP & \cite{atel6249} & 13:58:12.75 & +17:31:53.66 & 8.60 & - \\
ASASSN14iz & IIP & \cite{atel6580} & 20:21:49.84 & -31:17:06.78 & 8.55 & - \\
ASASSN14ma & IIP & \cite{atel6827} & 23:55:09.13 & +10:12:54.20 & 8.35 & 8.30 \\
ASASSN15bb & IIP & \cite{atel6948} & 13:01:06.38 & -36:36:00.17 & 8.28 & 8.25 \\
ASASSN15kz & IIP & \cite{atel7630} & 13:37:18.67 & -28:39:23.55 & 8.23 & 8.24 \\
ASASSN15ng & IIP & \cite{atel8031} & 13:51:31.65 & -48:04:33.73 & 8.39 & 8.33 \\
iPTF13bvn & Ib & \cite{shivvers2019} & 15:00:00.18 & +01:52:53.50 & 8.55 & 8.61 \\
LSQ11jw & Ib & \cite{sanders2012} & 02:04:47.40 & +00:50:06.00 & 8.37 & 8.35 \\
LSQ12fhh & IIP &\cite{atel4461} & 03:16:31.54 & -00:28:03.60 & 8.60 & 8.57 \\
LSQ12fnt & Ic &\cite{atel4505} & 03:32:41.40 & -01:11:11.00 & 8.55 & 8.54 \\
LSQ15xp & IIP & \cite{atel7261} & 11:32:42.79 & -16:44:01.20 & 8.32 & 8.26 \\
PS15afa & IIP & \cite{atel7452} & 10:50:18.15 & -12:06:28.10 & 8.64 & - \\
PS15cer & Ib & \cite{atel8057} & 22:41:14.79 & -21:47:42.10 & 8.46 & 8.41 \\
PTF11ixk & Ic &\cite{atel3531} & 13:21:45.03 & +31:14:04.60 & 8.54 & 8.51 \\
SN1961V & IIP & \cite{utrobin1984} & 02:43:24.00 & +37:21:00.00 & 8.42 & 8.24 \\
SN1983V & Ib & \cite{wheeler1987} & 03:33:31.66 & -36:08:54.89 & 8.53 & 8.57 \\
SN1984L & Ib & \cite{harkness1987} & 02:35:30.55 & -07:09:30.49 & 8.43 & 8.41 \\
SN1985F & Ib & \cite{shivvers2019} & 12:41:33.05 & +41:09:06.30 & 8.36 & 8.30 \\
SN1988L & Ic & \cite{shivvers2019} & 14:06:21.90 & +50:43:40.00 & 8.54 & 8.57 \\
SN1990aa & Ic & \cite{shivvers2019} & 00:52:59.22 & +29:01:48.30 & 8.41 & 8.36 \\
SN1990U & Ib & \cite{shivvers2019} & 23:04:54.92 & +12:18:20.09 & 8.55 & - \\
SN1991ar & Ib & \cite{shivvers2019} & 00:43:56.72 & +01:51:13.39 & 8.48 & 8.43 \\
SN1991N & Ic & \cite{shivvers2019} & 10:38:46.37 & +53:30:04.72 & 8.48 & 8.37 \\
SN1996an & IIP & \cite{tsvetkov2001} & 02:46:00.83 & -07:34:20.32 & 8.49 & 8.52 \\
SN1996aq & Ic & \cite{iauc6454} & 14:22:22.73 & -00:23:24.29 & 8.48 & 8.49 \\
SN1996D & Ic & \cite{iauc6317} & 04:34:00.30 & -08:34:43.98 & 8.56 & 8.46 \\
SN1997dc & Ib & \cite{shivvers2019} & 23:28:28.41 & +22:25:23.02 & 8.53 & 8.57 \\
SN1997X & Ib & \cite{shivvers2019} & 12:48:14.28 & -03:19:58.51 & 8.53 & 8.55 \\
SN1998dt & Ib & \cite{shivvers2019} & 02:28:35.72 & -10:32:59.78 & 8.54 & - \\
SN1999bg & IIP & \cite{poznanski2009} & 12:04:07.30 & +62:30:01.19 & 8.38 & 8.36 \\
SN1999di & Ib & \cite{shivvers2019} & 01:59:54.86 & +23:38:22.70 & 8.56 & 8.59 \\
SN1999eh & Ib & \cite{iauc7282} & 09:09:32.67 & +33:07:16.90 & 8.45 & - \\
SN1999em & IIP & \cite{galbany2016} & 04:41:27.04 & -02:51:45.22 & 8.58 & 8.61 \\
SN1999ex & Ic & \cite{iauc7310} & 22:16:07.27 & -36:50:53.70 & 8.59 & 8.52 \\
SN1999gi & IIP & \cite{leonard2002} & 10:18:16.66 & +41:26:28.21 & 8.49 & 8.63 \\
SN1999gn & IIP & \cite{kuncarayakti2013} & 12:21:57.02 & +04:27:45.61 & 8.56 & 8.53 \\
SN2000de & Ib & \cite{iauc7481} & 12:25:12.17 & +54:30:28.69 & 8.49 & 8.47 \\
SN2000ew & Ib & \cite{shivvers2019} & 11:40:58.52 & +11:27:55.91 & 8.50 & 8.67 \\
SN2000F & Ic & \cite{iauc7360} & 03:12:52.71 & +04:42:34.42 & 8.56 & - \\
SN2000fn & Ib & \cite{shivvers2019} & 08:06:57.79 & +08:00:19.30 & 8.51 & 8.50 \\
SN2001B & Ib & \cite{shivvers2019} & 04:57:19.24 & +78:11:16.51 & 8.51 & 8.49 \\
SN2001ch & Ic & \cite{shivvers2019} & 21:25:59.19 & -03:48:46.80 & 8.13 & 8.14 \\
SN2001ci & Ic & \cite{shivvers2019} & 10:01:57.33 & +55:41:14.60 & 8.57 & 8.50 \\
SN2001du & IIP & \cite{iauc7704} & 03:33:29.10 & -36:08:31.20 & 8.54 & 8.59 \\
SN2001em & Ib & \cite{shivvers2019} & 21:42:23.66 & +12:29:50.89 & 8.54 & - \\
SN2001fv & IIP & \cite{iauc7756} & 11:04:01.66 & +28:01:55.70 & 8.58 & 8.56 \\
SN2001is & Ib & \cite{shivvers2019} & 05:42:09.07 & +69:21:54.79 & 8.50 & 8.39 \\
SN2001X & IIP & \cite{anupama2002} & 15:21:55.45 & +05:03:42.08 & 8.54 & 8.63 \\
SN2002cw & Ib & \cite{iauc7905} & 18:46:02.40 & +32:17:03.19 & 8.61 & - \\
SN2002dz & Ib & \cite{shivvers2019} & 00:13:34.08 & -05:05:45.71 & 8.52 & 8.49 \\
SN2002ee & IIP & \cite{iauc7953} & 14:51:40.55 & +40:36:37.30 & 8.52 & 8.43 \\
SN2002hh & IIP & \cite{pozzo2006} & 20:34:44.29 & +60:07:18.98 & 8.53 & 8.54 \\
SN2002hn & Ic & \cite{shivvers2019} & 08:10:14.95 & +33:57:19.40 & 8.48 & 8.60 \\
SN2002ho & Ic & \cite{shivvers2019} & 12:15:17.97 & +65:58:55.09 & 8.55 & 8.57 \\
SN2002hy & Ib & \cite{shivvers2019} & 10:54:39.18 & -21:03:41.18 & 8.54 & 8.59 \\
SN2002J & Ic & \cite{shivvers2019} & 10:54:38.36 & -21:03:53.60 & 8.55 & 8.58 \\
SN2002ji & Ib & \cite{shivvers2019} & 11:22:53.15 & +16:35:10.00 & 8.53 & 8.53 \\
SN2002jz & Ib & \cite{shivvers2019} & 04:13:12.52 & +13:25:07.28 & 8.23 & 8.22 \\
SN2003bl & IIP & \cite{jones2009} & 13:57:30.65 & +06:05:36.38 & 8.61 & 8.55 \\
SN2003bn & IIP & \cite{jones2009} & 10:02:35.51 & -21:10:54.52 & 8.45 & 8.40 \\
SN2003bp & Ib & \cite{iauc8091} & 08:27:27.81 & +17:17:14.50 & 8.55 & 8.51 \\
SN2003E & IIP & \cite{galbany2016} & 04:39:10.88 & -24:10:36.52 & 8.29 & 8.30 \\
SN2003el & Ic & \cite{iauc8136} & 13:09:46.20 & +28:54:21.60 & 8.58 & 8.61 \\
SN2003gd & IIP & \cite{vandyk2003} & 01:36:42.65 & +15:44:20.90 & 8.49 & 8.43 \\
SN2003gk & Ib & \cite{iauc8164} & 23:01:42.99 & +02:16:08.69 & 8.52 & 8.49 \\
SN2003hg & IIP & \cite{galbany2016} & 23:51:24.13 & +20:06:38.30 & 8.57 & 8.58 \\
SN2003I & Ib & \cite{shivvers2019} & 09:27:29.48 & +03:55:45.59 & 8.47 & 8.41 \\
SN2003ih & Ic & \cite{shivvers2019} & 03:43:57.72 & +39:17:43.91 & 8.59 & 8.58 \\
SN2003is & Ib & \cite{shivvers2019} & 19:21:08.00 & +43:19:35.40 & 8.53 & - \\
SN2003Z & IIP & \cite{spiro2014} & 09:07:32.46 & +60:29:17.48 & 8.54 & 8.37 \\
SN2004A & IIP & \cite{hendry2006} & 16:43:01.90 & +36:50:12.52 & 8.49 & - \\
SN2004am & IIP & \cite{iauc8299} & 09:55:46.61 & +69:40:38.10 & 8.61 & 8.59 \\
SN2004ao & Ib & \cite{shivvers2019} & 17:28:09.35 & +07:24:55.51 & 8.48 & 8.44 \\
SN2004aw & Ic & \cite{shivvers2019} & 11:57:50.24 & +25:15:55.12 & 8.47 & 8.45 \\
SN2004bm & Ic & \cite{shivvers2019} & 10:52:35.33 & +22:56:05.50 & 8.55 & 8.61 \\
SN2004bs & Ib & \cite{shivvers2019} & 10:39:39.30 & +25:19:19.88 & 8.52 & 8.52 \\
SN2004dj & IIP & \cite{iauc8378} & 07:37:17.02 & +65:35:57.80 & 8.41 & 8.38 \\
SN2004dk & Ib & \cite{shivvers2019} & 16:21:48.93 & -02:16:17.29 & 8.38 & 8.48 \\
SN2004dn & Ic & \cite{shivvers2019} & 02:35:37.30 & +37:37:54.19 & 8.38 & 8.29 \\
SN2004eu & Ic & \cite{shivvers2019} & 02:20:34.86 & +41:34:18.52 & 8.51 & 8.53 \\
SN2004fc & IIP & \cite{martinez2022a} & 01:51:03.85 & -09:42:06.91 & 8.52 & 8.53 \\
SN2004ge & Ic & \cite{shivvers2019} & 06:50:00.13 & +25:38:02.00 & 8.54 & 8.58 \\
SN2004gk & Ic & \cite{shivvers2019} & 12:25:33.21 & +12:15:40.10 & 8.47 & 8.45 \\
SN2004gn & Ib & \cite{shivvers2019} & 12:34:12.10 & +02:39:34.42 & 8.50 & 8.62 \\
SN2004gq & Ib & \cite{shivvers2019} & 05:12:04.81 & -15:40:54.19 & 8.55 & 8.56 \\
SN2004gt & Ic & \cite{shivvers2019} & 12:01:50.37 & -18:52:12.68 & 8.53 & 8.53 \\
SN2004gv & Ib & \cite{shivvers2019} & 02:13:37.42 & -00:43:05.81 & 8.50 & 8.55 \\
SN2005ay & IIP & \cite{danziger2006} & 11:52:48.07 & +44:06:18.40 & 8.53 & 8.58 \\
SN2005az & Ic & \cite{modjaz2014} & 13:05:46.97 & +27:44:08.41 & 8.44 & 8.47 \\
SN2005bk & Ic & \cite{shivvers2019} & 16:02:17.04 & +42:54:55.30 & 8.56 & 8.59 \\
SN2005bz & IIP & \cite{atel493} & 18:13:01.73 & +29:41:51.50 & 8.40 & 8.37 \\
SN2005cs & IIP & \cite{danziger2006} & 13:29:52.78 & +47:10:35.69 & 8.43 & - \\
SN2005kl & Ic & \cite{shivvers2019} & 12:24:35.68 & +39:23:03.52 & 8.51 & 8.53 \\
SN2005V & Ic & \cite{harutyunyan2008} & 06:18:38.28 & +78:21:28.80 & 8.59 & 8.54 \\
SN2006be & IIP & \cite{martinez2022a} & 15:45:39.00 & +28:05:19.21 & 8.56 & 8.45 \\
SN2006bp & IIP & \cite{immler2007} & 11:53:55.74 & +52:21:09.40 & 8.53 & 8.58 \\
SN2006dn & Ib & \cite{atel854} & 22:47:37.84 & +39:52:46.81 & 8.44 & 8.39 \\
SN2006ei & Ic & \cite{shivvers2019} & 01:56:37.15 & +34:10:37.42 & 8.61 & - \\
SN2006F & Ib & \cite{iauc8660} & 02:28:11.37 & +19:36:13.50 & 8.57 & 8.59 \\
SN2006fo & Ib & \cite{shivvers2019} & 02:32:38.89 & +00:37:03.00 & 8.53 & 8.53 \\
SN2006lc & Ib & \cite{shivvers2019} & 22:44:24.45 & -00:09:53.89 & 8.58 & 8.57 \\
SN2006lv & Ib & \cite{shivvers2019} & 11:32:03.30 & +36:42:03.60 & 8.49 & 8.50 \\
SN2006my & IIP & \cite{li2007} & 12:43:40.74 & +16:23:14.10 & 8.54 & 8.48 \\
SN2006ov & IIP & \cite{li2007} & 12:21:55.30 & +04:29:16.69 & 8.52 & 8.60 \\
SN2007aa & IIP & \cite{martinez2022a} & 12:00:27.69 & -01:04:51.60 & 8.51 & 8.50 \\
SN2007ag & Ib & \cite{shivvers2019} & 10:01:35.99 & +21:36:42.01 & 8.46 & 8.38 \\
SN2007C & Ib & \cite{shivvers2019} & 13:08:49.30 & -06:47:01.00 & 8.58 & 8.59 \\
SN2007fo & Ib & \cite{shivvers2019} & 23:36:13.98 & +02:09:30.38 & 8.53 & 8.40 \\
SN2007gr & Ic & \cite{shivvers2019} & 02:43:27.98 & +37:20:44.70 & 8.52 & 8.55 \\
SN2007kj & Ib & \cite{shivvers2019} & 00:01:19.58 & +13:06:30.60 & 8.62 & - \\
SN2007rz & Ic & \cite{modjaz2014} & 04:31:10.84 & +07:37:51.49 & 8.54 & 8.61 \\
SN2008bs & Ib & \cite{shivvers2019} & 07:55:18.88 & +53:19:43.39 & 8.59 & 8.54 \\
SN2008cn & IIP & \cite{eliasrosa2009} & 12:40:55.66 & -40:58:12.11 & 8.55 & 8.57 \\
SN2008D & Ib & \cite{shivvers2019} & 09:09:30.65 & +33:08:20.29 & 8.48 & 8.46 \\
SN2008dv & Ic & \cite{shivvers2019} & 03:37:51.43 & +72:34:21.79 & 8.55 & 8.55 \\
SN2008ew & Ic & \cite{shivvers2019} & 16:58:28.92 & +20:02:38.00 & 8.54 & 8.57 \\
SN2008gj & Ic & \cite{shivvers2019} & 22:36:28.57 & +21:37:55.31 & 8.54 & 8.42 \\
SN2008hn & Ib & \cite{shivvers2019} & 08:14:14.25 & +21:21:37.91 & 8.53 & 8.59 \\
SN2008X & IIP & \cite{cbet1300} & 12:09:48.33 & +58:51:01.58 & 8.26 & 8.23 \\
SN2009B & IIP & \cite{cbet1646} & 08:30:51.47 & +74:41:08.09 & 8.39 & 8.34 \\
SN2009em & Ic & \cite{cbet1806} & 00:34:44.53 & -08:23:57.59 & 8.51 & 8.55 \\
SN2009ga & IIP &\cite{cbet1851} & 23:28:26.78 & +22:24:50.62 & 8.53 & 8.54 \\
SN2009ha & Ib & \cite{shivvers2019} & 02:36:58.21 & -05:20:43.19 & 8.52 & 8.51 \\
SN2009ie & IIP & \cite{cbet1915} & 02:48:13.90 & +34:24:52.81 & 8.48 & 8.43 \\
SN2009ii & IIP & \cite{cbet2073} & 07:01:05.58 & +51:15:50.18 & 8.47 & 8.40 \\
SN2009iz & Ib & \cite{shivvers2019} & 02:42:15.41 & +42:23:50.10 & 8.47 & 8.45 \\
SN2009je & Ic & \cite{cbet1953} & 05:28:40.07 & +22:06:46.40 & 8.57 & 8.63 \\
SN2009js & IIP & \cite{cbet1969} & 02:25:48.28 & +18:29:25.80 & 8.54 & - \\
SN2009md & IIP & \cite{cbet2068} & 10:48:26.28 & +12:32:02.80 & 8.45 & 8.42 \\
SN2009mi & Ic & \cite{shivvers2019} & 05:52:38.23 & -17:47:02.51 & 8.46 & 8.41 \\
SN2010do & Ic & \cite{shivvers2019} & 13:57:28.11 & +06:05:48.52 & 8.55 & 8.57 \\
SN2010F & IIP & \cite{cbet2137} & 10:05:21.05 & -34:13:21.00 & 8.53 & 8.54 \\
SN2010ie & IIP & \cite{cbet2477} & 07:08:23.71 & +35:10:11.21 & 8.65 & - \\
SN2010io & Ic & \cite{cbet2489} & 08:43:21.41 & +45:44:17.99 & 8.31 & 8.24 \\
SN2010iz & IIP & \cite{cbet2517} & 06:49:50.94 & +28:22:22.12 & 8.49 & 8.48 \\
SN2010jw & Ic & \cite{cbet2550} & 08:47:06.91 & +28:14:17.41 & 8.55 & 8.48 \\
SN2010kc & Ib & \cite{shivvers2019} & 23:20:22.33 & +27:18:57.82 & 8.51 & 8.60 \\
SN2010ln & Ib & \cite{cbet2609} & 03:20:53.62 & +38:15:11.92 & 8.57 & 8.54 \\
SN2011bh & Ic & \cite{cbet2690} & 07:45:13.62 & +53:04:10.70 & 8.61 & - \\
SN2011bm & Ic & \cite{valenti2012} & 12:56:53.89 & +22:22:28.20 & 8.33 & 8.28 \\
SN2011cj & IIP & \cite{cbet2721} & 14:32:53.81 & +11:35:49.31 & 8.42 & 8.38 \\
SN2011ck & IIP & \cite{cbet2722} & 14:00:46.24 & +48:26:45.38 & 8.44 & 8.42 \\
SN2011cl & IIP & \cite{cbet2724} & 08:26:49.16 &  +20:22:31.80 & 8.38 & 8.34 \\
SN2011dq & IIP & \cite{cbet2749} & 00:59:47.75 & -07:34:20.50 & 8.37 & 8.31 \\
SN2011fl & Ib & \cite{cbet2806} & 00:47:19.93 & +27:49:35.51 & 8.56 & - \\
SN2011gd & Ib & \cite{shivvers2019} & 16:34:25.67 & +21:32:28.39 & 8.56 & 8.61 \\
SN2011it & Ic & \cite{sanders2012} & 22:02:44.45 & +31:41:49.09 & 8.45 & 8.42 \\
SN2011jk & IIP & \cite{cbet2956} & 07:25:43.05 & +20:06:21.71 & 8.20 & 8.26 \\
SN2012A & IIP & \cite{cbet2975} & 10:25:07.39 & +17:09:14.62 & 8.20 & 8.16 \\
SN2012au & Ib & \cite{shivvers2019} & 12:54:52.18 & -10:14:50.21 & 8.49 & 8.49 \\
SN2012bu & IIP & \cite{cbet3088} & 10:52:56.53 & -32:56:07.69 & 8.58 & 8.51 \\
SN2012C & Ic & \cite{cbet2979} & 09:37:30.48 & +32:50:31.49 & 8.55 & 8.53 \\
SN2012dx & Ib & \cite{cbet3196} & 19:17:17.35 & +33:25:48.68 & 8.50 & 8.46 \\
SN2012ec & IIP & \cite{cbet3201} & 02:45:59.88 & -07:34:27.01 & 8.52 & 8.51 \\
SN2012ej & Ic & \cite{cbet3210} & 06:26:51.01 & +59:05:02.62 & 8.49 & 8.48 \\
SN2012ex & Ib & \cite{atel4386} & 01:18:45.97 & +14:59:40.20 & 8.55 & 8.55 \\
SN2012gi & Ic & \cite{cbet3299} & 22:08:52.44 & +40:20:24.00 & 8.45 & 8.44 \\
SN2012ho & IIP & \cite{cbet3338} & 22:40:17.02 & -02:25:34.10 & 8.44 & 8.41 \\
SN2013ab & IIP & \cite{atel4839} & 14:32:44.49 & +09:53:12.30 & 8.53 & - \\
SN2013bu & IIP & \cite{valenti2016} & 22:37:02.17 & +34:24:05.18 & 8.54 & 8.50 \\
SN2013ce & IIP & \cite{cbet3515} & 11:35:50.80 & +34:17:02.69 & 8.09 & 8.12 \\
SN2013cf & IIP & \cite{cbet3516} & 11:34:36.49 & +54:53:24.40 & 8.52 & 8.44 \\
SN2013dk & Ic & \cite{shivvers2019} & 12:01:52.72 & -18:52:18.30 & 8.57 & 8.61 \\
SN2013ej & IIP & \cite{atel5275} & 01:36:48.16 & +15:45:31.00 & 8.50 & 8.54 \\
SN2013ff & Ic & \cite{cbet3647} & 09:13:38.88 & +76:28:10.81 & 8.52 & 8.50 \\
SN2013ge & Ic & \cite{shivvers2019} & 10:34:48.46 & +21:39:41.90 & 8.48 & 8.42 \\
SN2013gl & Ib & \cite{shivvers2019} & 11:31:00.48 & +20:28:10.31 & 8.50 & 8.49 \\
SN2014A & IIP & \cite{cbet3771} & 13:16:59.36 & -16:37:57.00 & 8.53 & 8.57 \\
SN2014az & IIP &\cite{cbet3872} & 23:32:22.84 & +15:51:10.58 & 8.37 & 8.39 \\
SN2014bc & IIP & \cite{atel6159} & 12:18:57.71 & +47:18:11.30 & 8.64 & - \\
SN2014bi & IIP & \cite{atel6192} & 12:06:02.99 & +47:29:33.50 & 8.52 & - \\
SN2014C & Ib & \cite{shivvers2019} & 22:37:05.60 & +34:24:31.90 & 8.65 & - \\
SN2014cv & IIP & \cite{cbet3960} & 15:59:03.59 & +51:18:18.00 & 8.58 & 8.55 \\
SN2014cx & IIP & \cite{atel7084} & 00:59:47.83 & -07:34:18.59 & 8.37 & 8.28 \\
SN2014cy & IIP & \cite{valenti2016} & 23:44:16.03 & +10:46:12.50 & 8.60 & 8.54 \\
SN2014dj & Ic & \cite{shivvers2019} & 00:57:40.18 & +43:47:34.19 & 8.62 & 8.52 \\
SN2014eh & Ic & \cite{shivvers2019} & 20:25:03.86 & -24:49:13.30 & 8.58 & 8.46 \\
SN2014ei & Ib & \cite{shivvers2019} & 05:03:16.39 & -02:56:11.00 & 8.53 & 8.45 \\
SN2014L & Ic & \cite{shivvers2019} & 12:18:48.68 & +14:24:43.49 & 8.48 & 8.61 \\
SN2015ap & Ib & \cite{shivvers2019} & 02:05:13.32 & +06:06:08.39 & 8.27 & 8.22 \\
SN2015bb & Ic & \cite{cbet4211} & 14:51:37.83 & +40:35:51.40 & 8.55 & 8.59 \\
SN2015dj & Ib & \cite{shivvers2019} & 22:46:05.04 & -10:59:48.40 & 8.55 & - \\
SN2015V & IIP & \cite{atel7353} & 17:49:27.05 & +36:08:35.99 & 8.36 & 8.31 \\
SN2016aqf & IIP & \cite{muellerbravo2020} & 05:46:23.91 & -52:05:18.86 & 8.17 & 8.17 \\
SN2016B & IIP & \cite{tnscr2016-11} & 11:55:04.25 & +01:43:06.78 & 8.26 & 8.25 \\
SN2016bau & Ib & \cite{tnscr2016-224} & 11:20:59.02 & +53:10:25.60 & 8.53 & - \\
SN2016blb & IIP & \cite{tnscr2016-274} & 11:37:20.70 & -04:54:36.50 & 8.41 & 8.38 \\
SN2016bmi & IIP & \cite{tnscr2016-296} & 18:34:32.19 & -58:31:44.80 & 8.35 & - \\
SN2016C & IIP & \cite{tnscr2016-9} & 13:38:05.30 & -17:51:15.30 & 8.59 & 8.51 \\
SN2016ccm & IIP & \cite{tnscr2016-341} & 14:09:58.90 & +17:45:48.10 & 8.51 & - \\
SN2016cok & IIP & \cite{tnscr2016-391} & 11:20:19.03 & +12:58:56.64 & 8.51 & 8.57 \\
SN2016css & IIP & \cite{atel9170} & 20:16:53.44 & -36:59:42.00 & 8.45 & 8.39 \\
SN2016I & IIP & \cite{tnscr2016-21} & 14:39:44.73 & +23:23:43.27 & 8.12 & 8.15 \\
SN2016iae & Ic & \cite{tnscr2016-916} & 04:12:05.53 & -32:51:44.53 & 8.56 & 8.52 \\
SN2016X & IIP & \cite{tnscr2016-43} & 12:55:15.50 & +00:05:59.35 & 8.34 & 8.33 \\
SN2017cio & IIP & \cite{tnscr2017-540} & 07:38:34.92 & +37:38:25.16 & 8.35 & 8.30 \\
SN2017ein & Ic & \cite{tnscr2017-599} & 11:52:53.25 & +44:07:26.11 & 8.52 & - \\
SN2017ewx & Ib & \cite{tnscr2017-727} & 14:02:16.53 & +07:40:44.40 & 8.60 & - \\
SN2017fem & IIP & \cite{tnscr2017-921} & 14:32:27.32 & +27:25:36.30 & 8.54 & 8.53 \\
SN2017fvf & IIP & \cite{tnscr2017-921} & 03:17:53.31 & -07:18:00.97 & 8.52 & 8.58 \\
SN2017gat & Ic & \cite{tnscr2017-881} & 23:53:17.03 &  -32:46:32.84 & 8.14 & 8.24 \\
SN2017gry & IIP & \cite{atel10769} & 03:28:07.94 & -56:34:41.99 & 8.55 & 8.59 \\
SN2017hcb & Ib & \cite{tnscr2017-1308} & 02:36:23.72 & +31:42:34.83 & 8.53 & 8.50 \\
SN2017hyf & Ib & \cite{tnscr2017-1309} & 02:09:33.01 & +21:14:56.22 & 8.44 & - \\
SN2017pn & IIP & \cite{atel10032} & 04:46:24.59 & -11:59:18.25 & 8.28 & 8.29 \\
SN2017rt & Ic & \cite{tnscr2017-128} & 11:43:29.26 & -16:47:37.27 & 8.43 & 8.40 \\
SN2018cew & Ib & \cite{tnscr2018-827} & 23:52:24.20 & +28:46:30.07 & 8.51 & 8.52 \\
SN2018cho & IIP & \cite{tnscr2018-827} & 00:13:26.52 & +17:29:19.57 & 8.57 & 8.58 \\
SN2018ec & Ic & \cite{tnscr2018-48} & 10:27:50.77 & -43:54:06.30 & 8.55 & 8.55 \\
SN2018gsk & Ic & \cite{tnscr2018-1437} & 04:09:11.57 & +08:38:50.49 & 8.53 & 8.54 \\
SN2018hfc & IIP & \cite{tnscr2018-1877} & 11:01:58.58 & +45:13:39.14 & 8.59 & 8.54 \\
SN2018yo & IIP & \cite{tnscr2018-280} & 12:41:11.16 & -01:35:19.79 & 8.34 & 8.31 \\
SN2019ccm & Ib & \cite{tnscr2019-551} & 04:41:05.37 & +73:40:23.29 & 8.54 & - \\
SN2019cvz & IIP & \cite{tnscr2019-514} & 16:30:54.08 & +46:35:18.47 & 8.31 & 8.30 \\
SN2019tls & Ib & \cite{tnscr2019-2273} & 08:02:49.92 & +15:48:05.40 & 8.47 & - \\
SN2019yvr & Ib & \cite{tnscr2019-2736} & 12:45:08.14 & -00:27:32.83 & 8.56 & 8.50 \\
SN2019yxp & Ib & \cite{tnscr2020-37} & 14:16:32.13 & +57:48:28.80 & 8.34 & 8.29 \\
SN2020aaxs & Ib & \cite{tnscr2020-3859} & 13:58:33.94 & +37:27:11.52 & 8.64 & 8.60 \\
SN2020eai & Ib & \cite{tnscr2020-809} & 04:59:08.05 & +04:58:21.81 & 8.59 & 8.58 \\
SN2020jfo & IIP & \cite{teja2022} & 12:21:50.48 & +04:28:54.05 & 8.56 & 8.59 \\
SN2020ksa & Ib & \cite{tnscr2020-1908} & 10:59:27.94 & +46:07:28.45 & 8.53 & 8.60 \\
SN2020lid & Ib & \cite{tnscr2020-2074} & 00:20:00.78 & -06:20:04.06 & 8.48 & 8.46 \\
SN2020nac & Ib & \cite{tnscr2020-1891} & 00:43:24.64 & +50:40:28.06 & 8.45 & 8.42 \\
SN2020oi & Ic & \cite{tnscr2020-90} & 12:22:54.93 & +15:49:24.96 & 8.42 & 8.62 \\
SN2020rur & Ic & \cite{tnscr2020-2799} & 04:59:21.80 & +05:37:21.07 & 8.57 & 8.51 \\
SN2020sgf & Ic & \cite{tnscr2020-2811} & 03:22:52.76 & +42:33:22.07 & 8.55 & - \\
SN2020yyz & IIP & \cite{tnscr2020-3801} & 02:34:00.14 & +20:58:26.44 & 8.50 & 8.55 \\
SN2021aexi & Ic & \cite{tnscr2021-4035} & 23:51:22.34 & +20:06:34.90 & 8.59 & 8.57 \\
SN2021afuq & Ic & \cite{tnscr2021-4093} & 10:27:52.15 & -43:54:07.17 & 8.56 & 8.53 \\
SN2021kos & Ib & \cite{tnscr2021-1379} & 11:40:19.01 & +09:00:37.66 & 8.55 & 8.57 \\
SN2021ocs & Ic & \cite{tnscr2021-2064} & 00:06:27.48 & -13:24:56.91 & 8.59 & 8.50 \\
SN2021qip & Ib & \cite{tnscr2021-2187} & 01:56:43.73 & +15:01:02.53 & 8.56 & 8.56 \\
SN2021rfs & Ib & \cite{tnscr2021-2288} & 22:11:38.57 & +46:18:28.26 & 8.52 & 8.63 \\
SN2021vnw & Ib & \cite{tnscr2021-3014} & 06:55:34.87 & +39:45:53.75 & 8.58 & - \\
SN2021zju & Ib & \cite{tnscr2021-3392} & 00:08:54.22 & +23:48:59.76 & 8.46 & 8.36 \\
SN2022aang & IIP & \cite{tnscr2022-3357} & 07:59:21.837 & +18:06:40.92 & 8.39 & 8.31 \\
SN2022adui & Ic & \cite{tnscr2023-5} & 03:27:30.547 & -01:07:41.70 & 8.50 & 8.50 \\
SN2022ibn & Ic & \cite{tnscr2022-1460} & 13:06:14.583 & +25:27:41.21 & 8.47 & 8.43 \\
\hline

\label{tab_results}
\end{longtable}

\newpage
\section*{Appendix B}  \label{appndx_a2}
\setcounter{figure}{0}
\renewcommand{\thefigure}{B.\arabic{figure}}

\begin{minipage}{160mm}
\includegraphics[trim={0cm 0cm 0cm 0cm}, clip, width=1.0\columnwidth, angle=0]{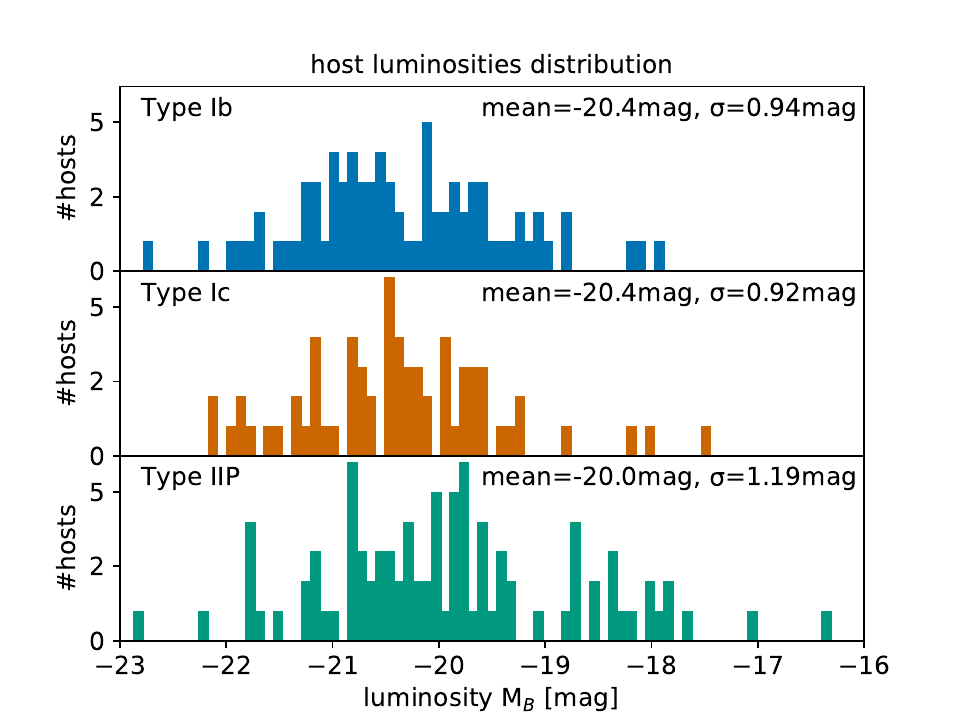}
\captionof{figure}{Distribution of M\textsubscript{B} luminosities of the host galaxies for the different SN Types.}
\label{fig_host_luminosities}

\end{minipage}

\newpage
\section*{Appendix C}  \label{appndx_b}
\setcounter{table}{0}
\renewcommand{\thetable}{C.\arabic{table}}

\begin{longtable}{@{\extracolsep{\fill}}ccC{3cm}C{2cm}ccccccc@{}}
\caption{Supplementary table with host data of archival Type IIb and Type IIn SNe (Section \ref{sub_iib_iin}). Columns are in accordance with Table \ref{tab_targets}.} \\
\hline \hline
 target & type  & host  & host type & redshift & D\textsubscript{L}    &  M\textsubscript{B}   & Incl. &  PA   &d\textsubscript{C} & instr.\\
          &       &   &    &  &  [Mpc]     & [mag] & [\textdegree] & [\textdegree] &  [\arcsec]     &      \\
\hline
\endfirsthead
\caption[]{(continued)}\\
\hline \hline
 target & type  & host  & host type & redshift & D\textsubscript{L}    &  M\textsubscript{B}    &  PA   &d\textsubscript{C} & instr.\\
          &       &   &    &  &  [Mpc]     & [mag] & [\textdegree] &  [\arcsec]     &      \\
\hline
\endhead
\tabularnewline
ASASSN14az & IIb & WISEA J234448.30-020653.1 & ? & 0.006731 & 31.1 & -16.6 & ? & 198.0 & 10.17 & 2 \\
PS15apj & IIb & NGC 6641 & S? & 0.013826 & 62.2 & -20.4 & 30 & 91.0 & 11.88 & 2 \\
SN1987K & IIb & NGC 4651 & SA(rs)c & 0.002669 & 10.0 & -18.6 & 49 & 292.5 & 22.75 & 2 \\
SN1996cr & IIn & ESO 097-G013       & SA(s)b & 0.001448 & 4.4 & -19.1 & 64 & 179.1 & 23.21 & 2 \\
SN1998fa & IIb & UGC 03513 & Scd: & 0.024514 & 112.1 & -20.3 & 67 & 323.2 & 4.88 & 4 \\
SN1998S & IIn & NGC 3877         & SA(s)c: & 0.002985 & 13.9 & -19.8 & 84 & 199.3 & 46.72 & 4 \\
SN2000bg & IIn & NGC 6240         & I0: & 0.024307 & 106.9 & -22.3 & 72 & 220.9 & 16.94 & 2 \\
SN2000cl & IIn & NGC 3318         & SAB(rs)b & 0.009255 & 43.3 & -21.3 & 60 & 143.8 & 10.63 & 2 \\
SN2000P & IIn & NGC 4965         & SAB(s)d & 0.007542 & 25.8 & -19.2 & 35 & 144.1 & 25.92 & 2 \\
SN2002au & IIb & UGC 05100 & SB(s)b & 0.018277 & 84.5 & -20.6 & 70 & 228.9 & 20.58 & 4 \\
SN2003as & IIn & MCG +08-10-007     & ? & 0.023223 & 104.1 & -21.6 & 33 & 96.7 & 10.40 & 4 \\
SN2003G & IIn & IC 0208           & SAbc & 0.011730 & 48.1 & -18.3 & 33 & 29.7 & 11.74 & 4 \\
SN2003ki & IIb & MCG +11-10-034 & ? & 0.025000 & ? & ? & 53 & 327.7 & 11.71 & 4 \\
SN2004C & IIb & NGC 3683 & SB(s)c? & 0.005701 & 41.7 & -20.7 & 69 & 301.8 & 20.54 & 1 \\
SN2004ff & IIb & ESO 552-G040 & SB(s)ab: & 0.022649 & 96.3 & -20.7 & 54 & 260.2 & 12.32 & 2 \\
SN2004gj & IIb & IC 0701 & ? & 0.020474 & 81.7 & -20.6 & 47 & 264.7 & 14.82 & 2,4 \\
SN2005db & IIn & NGC 0214          & SAB(r)c & 0.015134 & 70.6 & -21.8 & 48 & 249.1 & 17.95 & 4 \\
SN2005ip & IIn & NGC 2906         & Scd: & 0.007148 & 36.9 & -20.5 & 55 & 12.0 & 14.32 & 4 \\
SN2005kd & IIn & CGCG 327-013 & ? & 0.015094 & 71.9 & -20.7 & 79 & 4.3 & 5.63 & 4 \\
SN2006am & IIn & NGC 5630         & Sdm: & 0.008918 & 37.3 & -20.0 & 90 & 43.5 & 10.33 & 4 \\
SN2006jd & IIn & UGC 04179         & SBb & 0.018556 & 81.8 & -19.3 & 51 & 95.7 & 23.97 & 4 \\
SN2007ay & IIb & UGC 04310 & SA(s)m & 0.014527 & 62.9 & -18.6 & 28 & 220.2 & 19.75 & 4 \\
SN2007bb & IIn & UGC 03627         & Sd & 0.020954 & 93.3 & -20.7 & 49 & 121.1 & 26.30 & 4 \\
SN2008aq & IIb & PGC 043458 & SB(s)m & 0.007972 & 29.2 & -20.0 & 90 & 161.9 & 47.89 & 2 \\
SN2008ax & IIb & NGC 4490 & SB(s)d & 0.001885 & 14.0 & -20.8 & 90 & 114.7 & 56.26 & 3 \\
SN2008B & IIn & NGC 5829         & SA(s)c & 0.018973 & 85.3 & -20.4 & 41 & 73.2 & 23.58 & 4 \\
SN2008bo & IIb & NGC 6643 & SA(rs)c & 0.004950 & 27.4 & -21.1 & 63 & 65.1 & 35.19 & 4 \\
SN2008ie & IIb & NGC 1070 & Sb & 0.013636 & 58.1 & -21.3 & 33 & 300.2 & 25.42 & 4 \\
SN2008J & IIn & MCG -02-07-033     & SBbc? & 0.015874 & 73.1 & -20.5 & 70 & 256.4 & 4.71 & 2 \\
SN2008V & IIb & NGC 1591 & SB(r)ab & 0.013719 & 65.1 & -20.9 & 57 & 53.7 & 12.47 & 2 \\
SN2009K & IIb & NGC 1620 & SAB(rs)bc & 0.011715 & 41.8 & -21.4 & 81 & 279.2 & 8.81 & 2 \\
SN2010bt & IIn & NGC 7130         & Sa & 0.016151 & 68.4 & -21.5 & 34 & 144.3 & 14.76 & 2 \\
SN2011an & IIn & UGC 04139         & SA(s)c & 0.016301 & 64.5 & -20.7 & 47 & 126.8 & 14.56 & 3 \\
SN2011ei & IIb & NGC 6925 & SA(s)bc & 0.009303 & 29.4 & -21.1 & 84 & 43.4 & 37.98 & 2 \\
SN2011fh & IIn & NGC 4806         & SB(s)c? & 0.008032 & 27.0 & -19.1 & 33 & 53.1 & 26.27 & 2 \\
SN2011jg & IIb & UGC 10331 & S & 0.014914 & 69.3 & -19.5 & 83 & 315.1 & 24.52 & 4 \\
SN2012an & IIb & NGC 6373 & SAB(s)c & 0.011061 & 53.9 & -18.5 & 49 & 33.9 & 29.65 & 4 \\
SN2012cd & IIb & PGC 046714 & Sb & 0.011807 & 58.8 & -19.3 & 49 & 126.8 & 29.89 & 4 \\
SN2012P & IIb & NGC 5806 & SAB(s)b & 0.004493 & 23.1 & -19.9 & 60 & 257.4 & 19.67 & 1,2 \\
SN2013fc & IIn & ESO 154-G010       & (R’)SB(r)a & 0.018633 & 80.3 & -21.3 & 35 & 119.9 & 2.43 & 2 \\
SN2013ha & IIn & MCG +11-08-025     & S? & 0.013599 & 58.6 & -19.8 & 53 & 113.7 & 4.51 & 4 \\
SN2014ds & IIb & NGC 2536 & SB(rs)c & 0.013715 & 58.3 & -19.4 & 53 & 76.8 & 7.39 & 3 \\
SN2014ee & IIn & UGC 04132         & Sbc & 0.017372 & 75.6 & -21.3 & 77 & 230.9 & 20.63 & 4 \\
SN2014es & IIn & MCG -01-24-012  & SAB(rs)c: & 0.019650 & 85.4 & -20.8 & 67 & 140.5 & 15.42 & 2 \\
SN2015bf & IIn & NGC 7653 & Sb & 0.014200 & 68.2 & -21.1 & 31 & 346.5 & 20.46 & 2,4 \\
SN2016bas & IIb & ESO 163-G011 & SB(s)b? & 0.009433 & 46.4 & -19.9 & 71 & 179.5 & 20.20 & 2 \\
SN2016blt & IIb & ESO 271-G026 & (R')SB0\textasciicircum+(rs) & 0.016361 & 81.2 & -20.6 & 62 & 358.7 & 13.10 & 2 \\
SN2016gkg & IIb & NGC 0613 & SB(rs)bc & 0.004940 & 31.1 & -21.8 & 36 & 212.2 & 92.35 & 2 \\
SN2017gfh & IIb & PGC 1311562 & ? & 0.024477 & 118.9 & -19.7 & 52 & 309.8 & 7.55 & 2 \\
SN2017ixz & IIb & LEDA 1788023 & ? & 0.023592 & 102.3 & -18.3 & 54 & 132.3 & 14.13 & 3 \\
SN2017mw & IIb & ESO 316-G007 & Sa-b & 0.011661 & 53.4 & -19.7 & 70 & 6.4 & 7.06 & 2 \\
SN2018ddr & IIb & UGC 08896 & S? & 0.014623 & 62.0 & -19.6 & 84 & 321.3 & 2.14 & 2 \\
SN2019pqo & IIb & NGC 5980 & S & 0.013649 & 57.9 & -20.7 & 76 & 184.7 & 12.34 & 4 \\
SN2020tlf & IIn & NGC 5731 & S? & 0.008382 & 42.0 & -19.1 & 82 & 128.0 & 11.32 & 4 \\
SN2021aefs & IIn & NGC 3836 & Sb & 0.012208 & 56.1 & -20.5 & 40 & 194.8 & 6.19 & 2 \\
SN2021bxu & IIb & ESO 478-G006 & Sbc & 0.017792 & 81.8 & -21.7 & 58 & 291.8 & 24.01 & 2 \\
SN2021lwd & IIb & NGC 2596 & Sb & 0.019807 & 84.2 & -21.2 & 74 & 15.3 & 11.40 & 4 \\
\hline
\label{tab_iibiin_targets}
\end{longtable}

\begin{longtable}{@{\extracolsep{\fill}}ccccccc@{}}
\caption{Supplementary table with metallicities of archival Type IIb and Type IIn SNe environments (Section \ref{sub_iib_iin}). Columns are in accordance with Table \ref{tab_results}} \\ %
\hline \hline
    target & type & type reference & SN RA   & SN Dec & M13-N2 & M13-O3N2 \\
          &       &                & [h m s] & [\textdegree\,\arcmin\,\arcsec] & 12+log(O/H) & 12+log(O/H) \\
\hline
\endfirsthead
\caption[]{(continued)}\\
\hline \hline
    target & type & type reference & SN RA   & SN Dec & M13-N2 & M13-O3N2 \\
          &       &                & [h m s] & [\textdegree\,\arcmin\,\arcsec] & 12+log(O/H) & 12+log(O/H) \\
\hline
\endhead
\tabularnewline
ASASSN14az & IIb & \cite{shivvers2019} & 23:44:48.00 & -02:07:03.17 & 8.14 & 8.24 \\
PS15apj & IIb & \cite{atel7519} & 18:28:58.23 & +22:54:10.60 & 8.55 & 8.50 \\
SN1987K & IIb & \cite{shivvers2019} & 12:43:41.17 & +16:23:44.92 & 8.54 & 8.58 \\
SN1996cr & IIn & \cite{ransome2021} & 14:13:10.010 & -65:20:44.41 & 8.62 & 8.53 \\
SN1998fa & IIb & \cite{shivvers2019} & 06:42:51.51 & +41:25:18.91 & 8.44 & 8.39 \\
SN1998S & IIn & \cite{ransome2021} & 11:46:06.180 & +47:28:55.49 & 8.54 & 8.56 \\
SN2000bg & IIn & \cite{iauc7394} & 16:52:58.130 & +02:23:50.50 & 8.71 & 8.51 \\
SN2000cl & IIn & \cite{iauc7782} & 10:37:16.070 & -41:37:47.78 & 8.53 & 8.56 \\
SN2000P & IIn & \cite{ransome2021} & 13:07:10.529 & -28:14:02.51 & 8.46 & 8.45 \\
SN2002au & IIb & \cite{shivvers2019} & 09:34:37.60 & +05:50:15.68 & 8.54 & 8.44 \\
SN2003as & IIn & \cite{iauc8078} & 05:28:45.809 & +49:52:59.09 & 8.55 & 8.55 \\
SN2003G & IIn & \cite{ransome2021} & 02:08:28.130 & +06:23:51.90 & 8.50 & 8.65 \\
SN2003ki & IIb & \cite{shivvers2019} & 07:51:33.24 & +63:55:51.60 & 8.46 & 8.43 \\
SN2004C & IIb & \cite{shivvers2019} & 11:27:29.72 & +56:52:48.22 & 8.52 & 8.52 \\
SN2004ff & IIb & \cite{modjaz2014} & 04:58:46.19 & -21:34:12.00 & 8.64 & - \\
SN2004gj & IIb & \cite{shivvers2019} & 11:30:59.63 & +20:28:06.82 & 8.45 & 8.41 \\
SN2005db & IIn & \cite{ransome2021} & 00:41:26.791 & +25:29:51.61 & 8.52 & 8.63 \\
SN2005ip & IIn & \cite{ransome2021} & 09:32:06.420 & +08:26:44.41 & 8.52 & 8.47 \\
SN2005kd & IIn & \cite{atel656} & 04:03:16.879 & +71:43:18.91 & 8.35 & 8.31 \\
SN2006am & IIn & \cite{ransome2021} & 14:27:37.241 & +41:15:35.39 & 8.36 & 8.31 \\
SN2006jd & IIn & \cite{ransome2021} & 08:02:07.430 & +00:48:31.50 & 8.34 & 8.33 \\
SN2007ay & IIb & \cite{shivvers2019} & 08:17:14.85 & +01:12:06.91 & 8.45 & 8.42 \\
SN2007bb & IIn & \cite{cbet916} & 07:01:07.459 & +51:15:57.31 & 8.44 & 8.34 \\
SN2008aq & IIb & \cite{shivvers2019} & 12:50:30.42 & -10:52:01.42 & 8.30 & 8.20 \\
SN2008ax & IIb & \cite{shivvers2019} & 12:30:40.80 & +41:38:14.50 & 8.41 & 8.37 \\
SN2008B & IIn & \cite{ransome2021} & 15:02:43.649 & +23:20:07.80 & 8.51 & 8.47 \\
SN2008bo & IIb & \cite{shivvers2019} & 18:19:54.41 & +74:34:20.89 & 8.52 & 8.47 \\
SN2008ie & IIb & \cite{shivvers2019} & 02:43:20.80 & +04:58:19.09 & 8.59 & - \\
SN2008J & IIn & \cite{ransome2021} & 02:34:24.199 & -10:50:38.51 & 8.52 & 8.59 \\
SN2008V & IIb & \cite{cbet1257} & 04:29:31.30 & -26:42:39.82 & 8.53 & 8.54 \\
SN2009K & IIb & \cite{cbet1703} & 04:36:36.77 & -00:08:35.59 & 8.56 & - \\
SN2010bt & IIn & \cite{ransome2021} & 21:48:20.220 & -34:57:16.49 & 8.56 & 8.63 \\
SN2011an & IIn & \cite{ransome2021} & 07:59:24.420 & +16:25:08.18 & 8.41 & 8.38 \\
SN2011ei & IIb & \cite{atel3528} & 20:34:22.62 & -31:58:23.59 & 8.53 & 8.40 \\
SN2011fh & IIn & \cite{pessi2022} & 12:56:14.009 & -29:29:54.82 & 8.37 & 8.31 \\
SN2011jg & IIb & \cite{cbet2952} & 16:17:18.87 & +59:19:29.78 & 8.40 & 8.35 \\
SN2012an & IIb & \cite{cbet3035} & 17:24:10.23 & +59:00:06.91 & 8.31 & 8.29 \\
SN2012cd & IIb & \cite{cbet3107} & 13:22:35.25 & +54:48:47.70 & 8.53 & 8.40 \\
SN2012P & IIb & \cite{fremling2016} & 14:59:59.12 & +01:53:24.40 & 8.51 & 8.65 \\
SN2013fc & IIn & \cite{ransome2021} & 02:45:08.959 & -55:44:27.31 & 8.61 & 8.56 \\
SN2013ha & IIn &\cite{cbet3745} & 06:15:49.850 & +66:50:19.39 & 8.55 & 8.47 \\
SN2014ds & IIb & \cite{shivvers2019} & 08:11:16.45 & +25:10:47.39 & 8.47 & 8.44 \\
SN2014ee & IIn & \cite{ransome2021} & 07:59:11.779 & +32:54:39.89 & 8.47 & 8.37 \\
SN2014es & IIn & \cite{ransome2021} & 09:20:46.910 & -08:03:34.00 & 8.53 & 8.53 \\
SN2015bf & IIn & \cite{ransome2021} & 23:24:49.030 & +15:16:52.00 & 8.25 & 8.19 \\
SN2016bas & IIb & \cite{tnscr2016-219} & 07:38:05.53 & -55:11:47.00 & 8.52 & 8.53 \\
SN2016blt & IIb & \cite{tnscr2016-296} & 14:15:45.76 & -47:38:15.00 & 8.56 & - \\
SN2016gkg & IIb & \cite{tnscr2016-699} & 01:34:14.40 & -29:26:24.22 & 8.59 & - \\
SN2017gfh & IIb & \cite{tnscr2017-894} & 20:03:27.41 & +06:59:27.73 & 8.73 & - \\
SN2017ixz & IIb & \cite{tnscr2017-1464} & 07:47:03.07 & +26:46:25.79 & 8.38 & 8.38 \\
SN2017mw & IIb & \cite{tnscr2017-1029} & 09:57:20.97 & -41:35:20.98 & 8.30 & 8.24 \\
SN2018ddr & IIb & \cite{tnscr2018-988} & 13:58:38.47 & +07:13:01.27 & 8.48 & 8.48 \\
SN2019pqo & IIb & \cite{tnscr2019-1952} & 15:41:30.33 & +15:47:03.30 & 8.55 & 8.56 \\
SN2020tlf & IIn & \cite{tnscr2020-2839} & 14:40:10.020 & +42:46:39.43 & 8.50 & 8.49 \\
SN2021aefs & IIn & \cite{tnscr2021-4138} & 11:43:29.690 & -16:47:49.99 & 8.67 & - \\
SN2021bxu & IIb & \cite{tnscr2021-447} & 02:09:16.46 & -23:24:45.40 & 8.57 & 8.56 \\
SN2021lwd & IIb & \cite{tnscr2021-2534} & 08:27:26.73 & +17:17:13.70 & 8.53 & 8.50 \\
\hline

\label{tab_iibiin_results}
\end{longtable}

\newpage
\section*{Appendix D}  \label{appndx_D}
\setcounter{figure}{0}
\renewcommand{\thefigure}{D.\arabic{figure}}
\includegraphics[trim={0cm 0cm 0cm 0cm}, clip, width=1.0\columnwidth, angle=0]{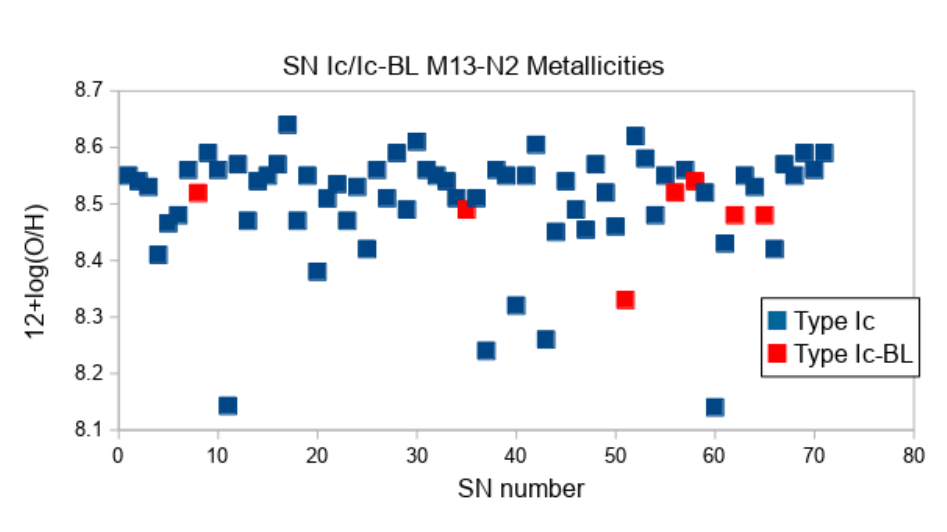}
\captionof{figure}{Environment metallicities for our Type Ic and Ic-BL determined using the M13-N2 calibration.}
\label{fig_ic_bl}





\bsp	
\label{lastpage}
\end{document}